\documentstyle[12pt,subeqnarray,epic,eepicemu,epsf]{article}

\begin{document}

\def\theequation{\thesection.\arabic{equation}}

\newcommand{\taum}{\tau_{int,\,\vec{\cal M}}}
\newcommand{\taux}{\tau_{int,\,{\cal M}^2}}
\newcommand{\tauA}{\tau_{int,\,A}}
\newcommand{\taue}{\tau_{int,\,{\cal E}}}
\newcommand{\taudele}{\tau_{int,\,({\cal E}-\overline{E})^2}}

\newcommand{\plotdot}{\makebox(0,0){$\bullet$}}
\newcommand{\plotcross}{\makebox(0,0){{\Large $\times$}}}

\newcommand{\plota}{\makebox(0,0){$\circ$}}      
\newcommand{\plotb}{\makebox(0,0){$\star$}}      
\newcommand{\plotc}{\makebox(0,0){$\bullet$}}    
\newcommand{\plotd}{\makebox(0,0){{\scriptsize $+$}}}       
\newcommand{\plote}{\makebox(0,0){{\scriptsize $\times$}}}  
\newcommand{\plotf}{\makebox(0,0){$\ast$}}       

\newcommand{\plotA}{\makebox(0,0){$\triangleleft$}}   
\newcommand{\plotB}{\makebox(0,0){$\triangleright$}}  
\newcommand{\plotC}{\makebox(0,0){$\diamond$}}        
\newcommand{\plotD}{\makebox(0,0){{\scriptsize $\oplus$}}} 
\newcommand{\plotE}{\makebox(0,0){{\scriptsize $\otimes$}}}
\newcommand{\plotF}{\makebox(0,0){{\scriptsize $\ominus$}}}

\def\reff#1{(\ref{#1})}
\newcommand{\csd}{critical slowing-down}
\newcommand{\be}{\begin{equation}}
\newcommand{\ee}{\end{equation}}
\newcommand{\<}{\langle}
\renewcommand{\>}{\rangle}
\newcommand{\half}{ {{1 \over 2 }}}
\newcommand{\quarter}{ {{1 \over 4 }}}
\newcommand{\fourth}{\quarter}
\newcommand{\eighth}{ {{1 \over 8 }}}
\newcommand{\sixteenth}{ {{1 \over 16 }}}
\def\var{ \hbox{var} }
\newcommand{\HB}{ {\hbox{{\scriptsize\em HB}\/}} }
\newcommand{\MGMC}{ {\hbox{{\scriptsize\em MGMC}\/}} }
\newcommand{\gtilde}{ {\widetilde{G}} }
\newcommand{\longto}{\longrightarrow}

\def\spose#1{\hbox to 0pt{#1\hss}}
\def\ltapprox{\mathrel{\spose{\lower 3pt\hbox{$\mathchar"218$}}
 \raise 2.0pt\hbox{$\mathchar"13C$}}}
\def\gtapprox{\mathrel{\spose{\lower 3pt\hbox{$\mathchar"218$}}
 \raise 2.0pt\hbox{$\mathchar"13E$}}}

\newcommand{\scra}{{\cal A}}
\newcommand{\scrb}{{\cal B}}
\newcommand{\scrc}{{\cal C}}
\newcommand{\scrd}{{\cal D}}
\newcommand{\scre}{{\cal E}}
\newcommand{\scrf}{{\cal F}}
\newcommand{\scrg}{{\cal G}}
\newcommand{\scrh}{{\cal H}}
\newcommand{\scrk}{{\cal K}}
\newcommand{\scrl}{{\cal L}}
\newcommand{\scrm}{{\cal M}}
\newcommand{\scrmvec}{\vec{\cal M}}
\newcommand{\scrp}{{\cal P}}
\newcommand{\scrr}{{\cal R}}
\newcommand{\scrs}{{\cal S}}
\newcommand{\scru}{{\cal U}}

\def\bsigma{\mbox{\protect\boldmath $\sigma$}}
\renewcommand{\Re}{\mathop{\rm Re}\nolimits}
\newcommand{\tr}{\mathop{\rm tr}\nolimits}
\newcommand{\CP}{ \hbox{\it CP\/} }

\font\specialroman=msym10 scaled\magstep1  
\font\sevenspecialroman=msym7              
\def\zed{\hbox{\specialroman Z}}
\def\szed{\hbox{\sevenspecialroman Z}}
\def\R{\hbox{\specialroman R}}
\def\sR{\hbox{\sevenspecialroman R}}
\def\N{\hbox{\specialroman N}}
\def\C{\hbox{\specialroman C}}
\renewcommand{\emptyset}{\hbox{\specialroman ?}}

%
%
\newenvironment{sarray}{
          \textfont0=\scriptfont0
          \scriptfont0=\scriptscriptfont0
          \textfont1=\scriptfont1
          \scriptfont1=\scriptscriptfont1
          \textfont2=\scriptfont2
          \scriptfont2=\scriptscriptfont2
          \textfont3=\scriptfont3
          \scriptfont3=\scriptscriptfont3
        \renewcommand{\arraystretch}{0.7}
        \begin{array}{l}}{\end{array}}

\newenvironment{scarray}{
          \textfont0=\scriptfont0
          \scriptfont0=\scriptscriptfont0
          \textfont1=\scriptfont1
          \scriptfont1=\scriptscriptfont1
          \textfont2=\scriptfont2
          \scriptfont2=\scriptscriptfont2
          \textfont3=\scriptfont3
          \scriptfont3=\scriptscriptfont3
        \renewcommand{\arraystretch}{0.7}
        \begin{array}{c}}{\end{array}}

\title{\vspace*{-2cm} Multi-Grid Monte Carlo \\
   \hbox{\hspace*{-1.3cm}
         III. Two-Dimensional $O(4)$-Symmetric Nonlinear $\sigma$-Model
        }
      }
\author{
  \\[-0.3cm]
  \small Robert G. Edwards            \\[-0.2cm]
  \small\it Supercomputer Computations Research Institute \\[-0.2cm]
  \small\it Florida State University  \\[-0.2cm]
  \small\it Tallahassee, FL 32306 USA \\[-0.2cm]
  \small Internet:  {\tt EDWARDS@MAILER.SCRI.FSU.EDU} \\[-0.2cm]
    \\[-0.3cm] \and
  \small  Sabino Jos\'e Ferreira\thanks{Address after September 1, 1991:
             Departamento de F\'{\i}sica -- ICEx,
             Universidade Federal de Minas Gerais,
             Caixa Postal 702, Belo Horizonte MG 30161, BRAZIL.
             Bitnet: {\tt SABINO@BRUFMG.BITNET}}  \\[-0.2cm]
  \small\it Department of Physics     \\[-0.2cm]
  \small\it New York University       \\[-0.2cm]
  \small\it 4 Washington Place        \\[-0.2cm]
  \small\it New York, NY 10003 USA    \\[-0.2cm]
  \small Internet:  {\tt FERREIRA@MAFALDA.PHYSICS.NYU.EDU} \\[-0.2cm]
    \\[-0.3cm] \and
  \small  Jonathan Goodman\thanks{Address from September 1, 1991 to
             June 30, 1992: Department of Mathematics,
             Stanford University, Stanford CA 94305, USA.
             Internet: {\tt GOODMAN@CAUCHY.STANFORD.EDU}}   \\[-0.2cm]
  \small\it Courant Institute of Mathematical Sciences \\[-0.2cm]
  \small\it New York University       \\[-0.2cm]
  \small\it 251 Mercer St.            \\[-0.2cm]
  \small\it New York, NY 10012 USA    \\[-0.2cm]
  \small Internet:  {\tt GOODMAN@NYU.EDU} \\[-0.2cm]
   \\[-0.3cm] \and
  \small  Alan D. Sokal               \\[-0.2cm]
  \small\it Department of Physics     \\[-0.2cm]
  \small\it New York University       \\[-0.2cm]
  \small\it 4 Washington Place        \\[-0.2cm]
  \small\it New York, NY 10003 USA    \\[-0.2cm]
  \small Internet:  {\tt SOKAL@ACF3.NYU.EDU} \\[-0.2cm]
  {\protect\makebox[5in]{\quad}}  
   \\
}

\vspace{0.2cm}
\maketitle
\thispagestyle{empty}   
\clearpage

\begin{abstract}
We study the dynamic critical behavior
of the multi-grid Monte Carlo (MGMC) algorithm
with piecewise-constant interpolation
applied to the two-dimensional $O(4)$-symmetric nonlinear $\sigma$-model
[= $SU(2)$ principal chiral model],
on lattices up to $256 \times 256$.
We find a dynamic critical exponent
$z_{int,{\cal M}^2} = 0.60 \pm 0.07$ for the W-cycle
and $z_{int,{\cal M}^2} = 1.13 \pm 0.11$ for the V-cycle,
compared to $z_{int,{\cal M}^2} = 2.0 \pm 0.15$ for the single-site
heat-bath algorithm
(subjective 68\% confidence intervals).
Thus, for this asymptotically free model,
critical slowing-down is greatly reduced compared to local algorithms,
but not completely eliminated.
For a $256 \times 256$ lattice, W-cycle MGMC is about 35 times
as efficient as a single-site heat-bath algorithm.
\end{abstract}

\clearpage

%
%

\section{Introduction}
\setcounter{equation}{0}
\vspace{-0.3cm}\quad\par

By now it is widely recognized
\cite{Sokal_Lausanne,Adler_LAT88,Wolff_LAT89,Sokal_LAT90}
that better simulation algorithms,
with strongly reduced critical slowing-down,
are needed for high-precision Monte Carlo studies of
statistical-mechanical systems near critical points
and of quantum field theories (such as QCD) near the continuum limit.
One promising class of such algorithms is {\em multi-grid Monte Carlo} (MGMC)
\cite{Goodman:Multigrid0,Goodman:Multigrid1,Goodman:Multigrid2}.
In paper I of this series \cite{Goodman:Multigrid1},
we explained the conceptual foundations of the MGMC method,
and proved the absence of critical slowing-down for
Gaussian (free-field) MGMC.
In paper II \cite{Goodman:Multigrid2},
we applied MGMC to the two-dimensional ferromagnetic $XY$ model,
and analyzed the dynamic critical behavior in the
high-temperature (vortex) and low-temperature (spin-wave) phases
(see also \cite{Hulsebos_91}).

In the present paper we apply MGMC to the two-dimensional $O(N)$-symmetric
nonlinear $\sigma$-model (also called $N$-vector model) with $N=4$,
defined by
\be
  H(\bsigma)   \;=\;   - \beta_{O(4)}  \sum_{\< xx' \>}
                               \bsigma_x \cdot \bsigma_{x'}    \;,
 \label{eqn1}
\ee
where each spin $\bsigma_x$ is a unit vector in $\R^4$,
$\< xx' \>$ runs over all nearest-neighbor pairs,
and $\beta_{O(4)} > 0$ is the inverse temperature in the $O(4)$ normalization.
Since the sphere $S^3$ is isometric (as a Riemannian manifold) to the
group $SU(2)$ via the map
\be
  g \;=\; \left(\! \begin{array}{cc}
                      \sigma^0 + i\sigma^3   &   \sigma^2 + i\sigma^1   \\
                     -\sigma^2 + i\sigma^1   &   \sigma^0 - i\sigma^3
                   \end{array}
          \!\right)   \;,
 \label{su2}
\ee
this model can also be thought of as the $SU(2)$ principal chiral model
\be
  H(g)   \;=\;   - \beta_{SU(2)}  \sum_{\< xx' \>} \Re\tr (g_x^\dagger g_{x'})
   \;,
 \label{eqn1a}
\ee
with $\beta_{SU(2)} = \beta_{O(4)}/2$.
Our principal aim is to make a careful study of the
dynamic critical behavior of the MGMC algorithm for this
asymptotically free model,
and in particular to extract the dynamic critical exponent(s).
We also take advantage of the high efficiency of the MGMC algorithm
--- about 35 times that of a single-site heat-bath algorithm,
on a $256 \times 256$ lattice ---
to make a detailed comparison of the static scaling behavior
with the asymptotic-freedom predictions.

Two-dimensional $N$-vector models
are of direct interest in condensed-matter physics,
and they are important ``toy models'' in elementary-particle physics
by virtue of their close similarity to four-dimensional gauge theories.
They have therefore been extensively studied
by rigorous \cite{Mermin_67,McBryan_77,Pfister_81,Bricmont_77a,Messager_78},
exact-but-nonrigorous 
 \cite{Zamolodchikov_79,Polyakov-Wiegmann_83,Wiegmann_85,%
Hasenfratz-Niedermayer_1,Hasenfratz-Niedermayer_2,Hasenfratz-Niedermayer_3},
renormalization-group
 \cite{Polyakov_75,Brezin_76,Amit_80,Parisi_80,Shigemitsu_81,Bernreuther_86,%
Falcioni_86,Jolicoeur_88a,Jolicoeur_88b,Luscher-Weisz_unpub},
high-temperature expansion \cite{Butera_88,Butera_89,Butera_90},
$1/N$ expansion
\cite{Muller_85,Flyvbjerg_89,Drouffe_90,Flyvbjerg_90-91,Flyvbjerg_90,%
Campostrini_90ab}
and Monte Carlo
 \cite{Heller_88a,Heller_88b,Wolff_89a,Edwards_89,%
Hasenbusch_90,Wolff_90,Hasenfratz_90,Wolff_O4_O8,Hasenbusch_LAT90,%
Apostolakis_91,CEPS_swwo4c2}
methods.\footnote{
   This is a {\em very}\/ incomplete set of references,
   but a significant fraction of the relevant articles
   can be tracked down starting from these.
   Also, studies dealing solely with the $XY$ case are cited in
   \cite{Goodman:Multigrid2} and therefore omitted here.
}
It is known rigorously that every infinite-volume Gibbs measure
is invariant under global spin rotations \cite{Pfister_81}
and in particular has zero spontaneous magnetization \cite{Mermin_67};
it is believed, though not yet proven,
that at each $\beta$ there is a {\em unique}\/
infinite-volume Gibbs measure.\footnote{
   In the $XY$ case it has been proven that there is a unique
   {\em translation-invariant}\/ infinite-volume Gibbs measure
   \cite{Bricmont_77a,Messager_78}.
}

Renormalization-group calculations in the low-temperature expansion
($\equiv$ weak-coupling perturbation theory)
\cite{Polyakov_75,Brezin_76}
suggest that the $N$-vector models with $N > 2$ are {\em asymptotically free}\/
--- i.e.\ that the only critical point is at $\beta = \infty$ ---
and that the correlation length $\xi$ and susceptibility $\chi$ behave as
\begin{eqnarray}
\xi(\beta)    & = &   C_\xi \, \beta^{-1/(N-2)} e^{2\pi\beta/(N-2)} \,
         \left[ 1 + {a_1 \over \beta} + {a_2 \over \beta^2} + \cdots \right]
                                           \label{O(N)_xi_predicted}  \\[0.2cm]
\chi(\beta)   & = &   C_\chi \, \beta^{-(N+1)/(N-2)} e^{4\pi\beta/(N-2)} \,
         \left[ 1 + {b_1 \over \beta} + {b_2 \over \beta^2} + \cdots \right]
                                           \label{O(N)_chi_predicted}
\end{eqnarray}
as $\beta \to \infty$.\footnote{
   In this paper the susceptibility is defined as
   $\chi = \sum_x \< \bsigma_0 \cdot \bsigma_x \>$.
   This definition is employed also in
\cite{Falcioni_86,Butera_88,Muller_85,Drouffe_90,Flyvbjerg_90-91,%
Heller_88b,Edwards_89}.
   By contrast, the standard thermodynamic definition
   is $\chi_{thermo} = \beta \chi_{this\ paper}$;
   the extra factor of $\beta$ arises because the inverse temperature $\beta$
   is introduced as an overall factor multiplying also the magnetic field.
   The thermodynamic definition is employed in
   \cite{Brezin_76,Jolicoeur_88b}.
   This difference in notation must be borne in mind when comparing
   results in the literature!
}
Here the $a_k$ and $b_k$ are nonuniversal constants (depending on $N$)
that can be computed in weak-coupling perturbation theory on the lattice
at $k+2$ loops;  in particular, $a_1$ and $b_1$ have been computed
for the nearest-neighbor action \reff{eqn1} and two variant actions
\cite{Falcioni_86,Luscher-Weisz_unpub}.\footnote{
   Reference \protect\cite{Falcioni_86} employs, as we do,
   the {\em second-moment}\/ definition of the correlation length:
   compare equation (12a) of \protect\cite{Falcioni_86} with
   \reff{def_xi}/\reff{def_xiprime} below.
   Several other papers (e.g.\ 
\protect\cite{Wolff_90,Hasenfratz_90,Apostolakis_91,Hasenfratz-Niedermayer_1,%
Hasenfratz-Niedermayer_2,Hasenfratz-Niedermayer_3,Wolff_O4_O8})
   emphasize, by contrast, the {\em exponential}\/ correlation length
   (= inverse mass gap).
   The coefficients $a_1,a_2,\ldots$ and $C_\xi$
   need not be the same for the two quantities.
}
On the other hand, $C_\xi$ and $C_\chi$ are
nonuniversal constants (depending on $N$)
that have to be computed by some non-perturbative method.
For the exponential correlation length (= inverse mass gap),
the coefficient 
$C_\xi \equiv 2^{-5/2} [(N-2)/(2\pi e^{\pi/2})]^{1/(N-2)}
 (m/\Lambda_{\overline{\hbox{\scriptsize MS}}})^{-1}$
has\footnote{
   This relation between $C_\xi$ and
   $m/\Lambda_{\overline{\hbox{\scriptsize MS}}}$
   is based on the result
   $\Lambda_{\overline{\hbox{\scriptsize MS}}}/\Lambda_{lat}
    = 2^{5/2} e^{\pi/2(N-2)}$
   from lattice perturbation theory
   \protect\cite{Parisi_80,Shigemitsu_81}.
   Please note that equation (3) of \protect\cite{Wolff_O4_O8}
   suffers from a typographical error:
   the division sign should be multiplication.
}
been computed exactly using the Bethe Ansatz
\cite{Hasenfratz-Niedermayer_1,Hasenfratz-Niedermayer_2,%
Hasenfratz-Niedermayer_3}:
the result is
\be
   {m \over \Lambda_{\overline{\hbox{\scriptsize MS}}} }
   \;=\;
   \left( {8 \over e} \right) ^{1/(N-2)} \,
   {1 \over \Gamma\biggl(1 + {1 \over N-2} \biggr) }
   \;.
 \label{exact_Cxi}
\ee

The asymptotic-freedom predictions
\reff{O(N)_xi_predicted}/\reff{O(N)_chi_predicted}
have not yet been proven rigorously
(but see \cite{Ito_87a,Ito_87b,Ito_90a,Ito_90b} for some progress),
and indeed they have been questioned by one group
\cite{Patrascioiu_87,Seiler_88,Patrascioiu_89}.
However, they are consistent with Monte Carlo studies
for $N=3$ \cite{Wolff_90,Hasenfratz_90,Apostolakis_91},
$N=4$ \cite{Heller_88b,Edwards_89,Wolff_O4_O8}
and $N=8$ \cite{Wolff_O4_O8}.
More precisely, for $N=3$ it is found that for 
$1.5 \ltapprox \beta \ltapprox 2.05$
(corresponding to $10 \ltapprox \xi \ltapprox 300$),
the quantity $C_\xi$ defined by \reff{O(N)_xi_predicted}
varies by about 20\%
--- contrary to the prediction that it should be constant ---
and also differs by about 20\% from the predicted exact value \reff{exact_Cxi}.
It appears, therefore, that {\em quantitative}\/ asymptotic scaling
has not yet set in at $\xi \sim 100$,
even though the qualitative behavior is correct.
For $N=4,8$, however, the agreement with \reff{O(N)_xi_predicted} is better;
this suggests that as $N$ increases, asymptotic scaling sets in at smaller
values of $\xi$ \cite{Wolff_O4_O8}.

A very similar picture has been obtained by Pad\'e extrapolation
of high-temperature series \cite{Butera_88,Butera_89,Butera_90}.
It is found that the critical singularity in the complex $\beta$-plane,
which lies at real $\beta$ when $-2 \le N \le 2$,
bifurcates for $N > 2$ into a pair of complex-conjugate singularities,
which tend as $N \to \infty$ to
$\widetilde{\beta} \equiv \beta/N \approx 0.32162 (1 \pm i)$.
In particular, for $N=3$, this first pair of singularities is located
at $\beta \approx 1.73 \pm 0.42 i$ \cite[Figure 2]{Butera_90},
extremely close to the real axis
and to the region in which the Monte Carlo simulations
are currently being performed
($\beta=1.73$ corresponds to $\xi \approx 42$).
By contrast, for $N=4$ the first singularity has moved to
$\beta \approx 2.27 \pm 0.88i$,
i.e.\ farther away from the real axis and also towards smaller $\xi$
($\beta=2.27$ corresponds to $\xi \approx 17$).
For $N=8$ the first singularity is located at
$\beta \approx 3.78 \pm 2.38i$
($\beta=3.78$ corresponds to $\xi \approx 5$).\footnote{
   We thank Paolo Butera for providing us with precise numerical values
   of the singularity locations first reported in \cite[Figure 2]{Butera_90}.
}
It is thus very plausible that these complex singularities are responsible
for preventing asymptotic scaling for $\xi \sim 100$ at $N=3$.

Let us remark that by eliminating $\beta$ from
\reff{O(N)_xi_predicted}/\reff{O(N)_chi_predicted},
we obtain
\be
 \label{O(N)_chi-versus-xi_predicted}
   \chi   \;=\;   C_{\chi\xi} \, \xi^2 \, (\log\xi)^{-(N-1)/(N-2)} \,
    \left[ 1 + c_{11} {\log\log\xi \over \log\xi} + c_{01} {1 \over \log\xi}
             + O \!\left( {\log^2 \log\xi \over \log^2 \xi} \right)   \right]
   \;,
\ee
where
\begin{subeqnarray}
   c_{11}   & = &  - {N-1  \over  (N-2)^2}           \\[0.2cm]
   c_{01}   & = &
  {N-1  \over  (N-2)^2} \left[ \log {2\pi \over N-2} + (N-2) \log C_\xi \right]
  \,+\, {2\pi \over N-2} (b_1 - 2a_1)                
\end{subeqnarray}

In finite volume (with periodic boundary conditions),
the usual finite-size-scaling Ansatz gives,
up to corrections of order $\beta^{-1}$,
\begin{eqnarray}
\xi(\beta,L)   & = &   \beta^{-1/(N-2)} \, e^{2\pi\beta/(N-2)} \,
                                           f_\xi(\xi_\infty/L)
                                    \label{O(N)_xi_FSS_predicted}    \\[0.2cm]
\chi(\beta,L)  & = &  \beta^{-(N+1)/(N-2)} \, e^{4\pi\beta/(N-2)} \,
                                           f_\chi(\xi_\infty/L)
                                    \label{O(N)_chi_FSS_predicted}
\end{eqnarray}
where $f_\xi$ and $f_\chi$ are finite-size-scaling functions,
and $\xi_\infty \equiv \xi(\beta,\infty)$.
Note that $\lim_{x \to 0} f_\xi (x) = C_\xi > 0$
and likewise for $f_\chi$.
One can also eliminate $\beta$ from
\reff{O(N)_xi_FSS_predicted}/\reff{O(N)_chi_FSS_predicted},
yielding
\be
 \label{O(N)_chi-versus-xi_FSS_predicted}
   \chi(\beta,L)   \;=\;   \xi(\beta,L)^2 \, [\log\xi(\beta,L)]^{-(N-1)/(N-2)}
       \, f_{\chi\xi}(\xi_\infty/L)
\ee
up to corrections of order $\log\log\xi/\log\xi$.
The Ans\"atze
\reff{O(N)_xi_FSS_predicted}--\reff{O(N)_chi-versus-xi_FSS_predicted}
can be written in numerous equivalent forms, for example,
\begin{eqnarray}
\xi(\beta,L)   & = &   \beta^{-1/(N-2)} \, e^{2\pi\beta/(N-2)} \,
                                           \widetilde{f}_\xi(\xi/L)  
                                    \label{O(N)_xi_FSS_predicted2}   \\[0.2cm]
\chi(\beta,L)  & = &  \beta^{-(N+1)/(N-2)} \, e^{4\pi\beta/(N-2)} \,
                                           \widetilde{f}_\chi(\xi/L)
                                    \label{O(N)_chi_FSS_predicted2}  \\[0.2cm]
\chi(\beta,L)  & = &  \xi^2 \, (\log\xi)^{-(N-1)/(N-2)} \,
                                           \widetilde{f}_{\chi\xi}(\xi/L)
                                    \label{O(N)_chi-versus-xi_FSS_predicted2}
\end{eqnarray}
where $\xi \equiv \xi(\beta,L)$.

\section{MGMC Algorithm for Nonlinear $\sigma$-Models}   \label{sec2}
\setcounter{equation}{0}
\vspace{-0.3cm}\quad\par

The physical causes of critical slowing-down in the traditional (local)
Monte Carlo algorithms, and the motivation for collective-mode algorithms,
have been discussed previously
\cite{Sokal_Lausanne,Goodman:Multigrid1,Sokal_LAT90}.
In particular, paper I \cite{Goodman:Multigrid1} provides an extensive
treatment of the physics behind MGMC and the details of its
implementation, as well as a comparison with other types of collective-mode
algorithms (Fourier acceleration and auxiliary-variable algorithms).
Here we confine ourselves to a brief review of the structure of the MGMC
algorithm as applied to a nonlinear $\sigma$-model.
We begin by considering a principal chiral model, i.e.\ a $\sigma$-model
taking values in some {\em group}\/ $G \subset U(N)$.
At the end of this section we shall make some brief remarks about the
more general case of a $\sigma$-model taking values
in a homogeneous space $M = G/H$.

Consider a principal chiral model with values in $G \subset U(N)$,
defined on a $d$-dimensional lattice of linear size $L$;
for simplicity let us use a hypercubic lattice with
periodic boundary conditions.
The Hamiltonian is given by
\be
  H(g)   \;=\;   - \beta  \sum_{\< xx' \>} \Re\tr (g_x^\dagger g_{x'})   \;,
 \label{eqn1b}
\ee
and the Gibbs probability distribution is
\be
  d \mu (g)   \;=\;   Z^{-1}\exp[ -H(g)] \; \prod_x d g_x  \;,
 \label{eqn2}
\ee
where $dg$ is Haar measure on $G$.
Our goal is to devise a stochastic updating procedure that
leaves invariant the probability distribution (\ref{eqn2})
and has an autocorrelation time $\tau$ which is as small as possible.

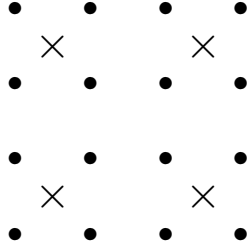
\begin{figure}
\unitlength = 1.00mm
\begin{center}
\begin{normalsize}
\begin{picture}(40,40)(0,0)
\thicklines
\matrixput(0,0)(10,0){4}(0,10){4}{\plotdot}
\matrixput(5,5)(20,0){2}(0,20){2}{\plotcross}
\end{picture}
\end{normalsize}
\vspace{0.4in}
\caption[fig1]{
   Staggered coarsening (factor-of-2) in dimension $d=2$.
   Dots ($\bullet$) are fine-grid sites,
   crosses ($\times$) are coarse-grid sites.
}
\label{factor_2_coarsening_b}
\end{center}
\end{figure}

The MGMC algorithm for this model is defined as follows
\cite[Sections IV and VII]{Goodman:Multigrid1}:
We first introduce,
in addition to the original lattice $\Omega$,
a sequence of {\em coarse grids}\/
$\Omega_M\equiv\Omega, \Omega_{M-1}, \Omega_{M-2}, \ldots, \Omega_0$
of linear size $L, L/2, L/4, \ldots, L/2^M$, respectively.\footnote{
   Clearly $L$ must be a multiple of $2^M$.  It is therefore
   most convenient for $L$ to be a power of 2, or at least
   a power of 2 times a small integer.
}
These coarse grids will play a role in intermediate stages
of the multi-grid algorithm.
Each site $y$ of a grid $\Omega_{l-1}$ is considered to be associated
with a $2^d$ block $B_y$ in the next-finer grid $\Omega_l$ 
(Figure \ref{factor_2_coarsening_b}).
We define, therefore, an {\em interpolation operator}\/ $p_{l,l-1}$
that maps a field configuration $g^{(l-1)}$ on grid $\Omega_{l-1}$
into one on grid $\Omega_l$, using
{\em piecewise-constant interpolation}\/:\footnote{
   This update had been proposed earlier, in a ``unigrid'' context,
   by Meyer-Ortmanns \protect\cite{Meyer-Ortmanns_85}.
}

\be
    (p_{l,l-1} g^{(l-1)})_x  \;=\;  g^{(l-1)}_y
                                      \quad \mbox{for } x \in B_y \;.
\ee
Each coarse-grid field is to be thought of as a {\em correction}\/
or {\em update}\/ to the next-finer-grid field,
acting by left multiplication:
that is, given a ``fine-grid'' field $g^{(l,old)}$
and a ``coarse-grid'' field $g^{(l-1)}$,
we define an updated ``fine-grid'' field
$g^{(l,new)} \equiv (p_{l,l-1} g^{(l-1)}) g^{(l,old)}$,
or in detail
\be
   g^{(l,new)}_x  \;=\;  g^{(l-1)}_y g^{(l,old)}_x
                                  \quad \mbox{for } x \in B_y \;.
 \label{interp}
\ee

On each grid $\Omega_l$, we consider Hamiltonians $H_l$ of the form
\begin{equation}
  H_l(g)   \;=\;   - \sum_{\< xx' \>}
                         \Re\tr (\alpha^{(l)}_{xx'} g_x^\dagger g_{x'})   \;,
 \label{eq:new_ham}
\end{equation}
where $\alpha^{(l)}_{xx'}$ is an $N \times N$ complex matrix;
note that $\alpha^{(l)}_{xx'}$ is not necessarily unitary or
a multiple of a unitary matrix, and it can vary from one bond $\< xx' \>$
to another.
On the finest grid ($l=M$) we have $\alpha^{(M)}_{xx'} = \beta I$
for all $\< xx' \>$, but on coarser grids more general Hamiltonians of the form
\reff{eq:new_ham} will be generated [see \reff{eq:coarse_ham2} below].

For each Hamiltonian $H_l$ of the form (\ref{eq:new_ham}),
we assume that there is given 
a {\em basic stochastic iteration\/}
$g \,\rightarrow\, \scrg_l(g,H_l)$
that updates the field configuration $g$ (which lives on grid $\Omega_l$)
in such a way as to leave invariant
the probability distribution $\sim \exp[-H_l(g)]$.
For concreteness, we shall take $\scrg_l$
to be the single-site heat-bath algorithm for the Hamiltonian $H_l$.
Finally, we let $\gamma$ be a positive integer,
called the {\em cycle control parameter\/};
it will control the number of times that the coarse grids are visited.

The multi-grid Monte Carlo (MGMC) algorithm is then
defined recursively as follows:
\begin{quote}
\begin{tabbing}
\qquad \= \qquad \= \qquad \= \qquad \= \+ \kill 
{\bf procedure} $mgmc (l,g,H_l)$ \\
{\bf comment} This algorithm updates a field configuration $g$ in such a \\
   \quad way as to leave invariant the Gibbs distribution
                                                  $\exp[-H_l(g)] \, dg$. \\
\\
{\bf for} $j=1$ {\bf until} $m_1$ {\bf do} $g \leftarrow {\cal G}_l(g, H_l)$ \\

{\bf if} $l>0$ \= {\bf then} \\
\>{\em compute} $H_{l-1}(\;\cdot\;) \equiv
                 H_l ( (p_{l,l-1} \,\cdot\,) g )$  \\
\>$h \leftarrow I$ \\
\>{\bf for} $j=1$ {\bf until} $\gamma$ {\bf do} $mgmc(l-1,h,H_{l-1})$ \\
\>$g \leftarrow (p_{l,l-1} h) g$  \\
{\bf endif} \\

{\bf for} $j=1$ {\bf until} $m_2$ {\bf do} $g \leftarrow {\cal G}_l(g, H_l)$ \\

{\bf end}
\end{tabbing}
\end{quote}
\bigskip

We see, then, that a
single MGMC iteration at level $l$ consists of three steps:
\medskip

1)  {\em Pre-sweep.}  We apply a few ($m_1$) iterations of the basic
stochastic iteration (i.e.\ single-site heat-bath) at level $l$.
This produces a new field configuration $g$ that differs
significantly from the old one in its short-wavelength
components, but not (unfortunately!) in its long-wavelength components.

2)  {\em Coarse-grid correction.}  We compute the coarse-grid Hamiltonian
$H_{l-1}(h) \equiv H_l( (p_{l,l-1} h) g )$.
As explained earlier, the coarse-grid field $h$ should be thought of as a
correction to $g$.
We initialize the coarse-grid-correction field $h$ to the identity matrix,
and make $\gamma$ updates of this field using
MGMC itself (at level $l-1$).
We then transfer the updated field $h$ back to grid $\Omega_l$
using the interpolation operator $p_{l,l-1}$,
and use it to correct the field $g$.
The goal of this coarse-grid correction step is to
update rapidly the long-wavelength components of $g$.

3)  {\em Post-sweep.}  We apply, for good measure,
a few ($m_2$) more iterations of the basic stochastic iteration
at level $l$.

\bigskip

Let us now look more closely at some aspects of the MGMC algorithm:
\bigskip

{\em Recursive definition.}
We have defined the MGMC algorithm recursively ---
in the coarse-grid correction step the procedure
{\em mgmc} calls itself ---
because that is by far the simplest and clearest approach.
However, the MGMC algorithm can be implemented in {\em any} computer
programming language, including those (like Fortran)
that forbid recursive subroutine calls,
although the control structure may be a bit more complicated.
See \cite[Section 4.1]{Hackbusch_85}
for Fortran and Algol implementations
of the multi-grid control structure.
Our codes are available on request (preferably via electronic mail).
\bigskip

{\em Computation of $H_{l-1}$.}
Assume that the Hamiltonian $H_l$ on level $l$
is of the form \reff{eq:new_ham}.
Then a simple computation shows that the coarse-grid Hamiltonian
$H_{l-1}(h) \equiv H_l((p_{l,l-1}h)g)$
is\footnote{
   Note that equation (4.9) of \cite{Goodman:Multigrid1} contains a
   slight error: the sum should be restricted to {\em nearest-neighbor}\/
   pairs $\< xx' \>$ satisfying $x \in B_y$, $x' \in B_{y'}$.
   A more serious error in \cite{Goodman:Multigrid1} arises from our
   attempt to allow there an arbitrary compact but
   {\em not-necessarily-unitary}\/ group $G \subset GL(N,\C)$.
   In such a case the nearest-neighbor pairs $\< xx' \>$ with
   $x$ and $x'$ belonging to the {\em same}\/ block $B_y$
   make a non-constant contribution to the coarse-grid Hamiltonian,
   of the form $\Re\tr (\alpha^{(l-1)}_{yy} h_y^\dagger h_y)$.
   It is thus necessary to generalize the Hamiltonian \reff{eq:new_ham}
   to include both nearest-neighbor pairs $\< xx' \>$
   {\em and}\/ ``diagonal'' terms $x=x'$;
   such a form is then preserved under coarsening.
   We thank Tereza Mendes for pointing out these errors.
}
\begin{equation}
H_{l-1}(h)   \;=\;   - \sum_{\< yy' \>}
                         \Re\tr (\alpha^{(l-1)}_{yy'} h_y^\dagger h_{y'})
      \,+\, {\rm const}   \;,
 \label{eq:coarse_ham}
\end{equation}
where
\be
\alpha^{(l-1)}_{yy'}   \;=\;
   \sum_{\begin{scarray}
             \< xx' \>   \\
             x\in B_y    \\
             x'\in B_{y'}
         \end{scarray}
        }
   g_{x'} \alpha^{(l)}_{xx'} g^\dagger_x   \;.
 \label{eq:coarse_ham2}
\ee
Note that the coarse-grid Hamiltonian $H_{l-1}$ has the same
functional form as the ``fine-grid'' Hamiltonian $H_l$: it is specified
by the coefficients $\{\alpha^{(l-1)}_{yy'}\}$.
The step ``{\em compute} $H_{l-1}$'' therefore means to compute these
coefficients.
Note also the importance of generalizing the Hamiltonian from \reff{eqn1b}
to \reff{eq:new_ham}:
even if the fine-grid coefficients are $\alpha^{(M)}_{xx'} = \beta I$
with $\beta > 0$ independent of $\< xx' \>$, the coarse-grid coefficients
$\alpha^{(l)}_{xx'}$ for $l < M$, defined by \reff{eq:coarse_ham2},
will be $\beta$ times arbitrary finite sums of elements of $G$,
considered in the space ${\sf M}_N$ of all complex $N \times N$ matrices.
For $G = SU(N)$ with $N \ge 3$, the finite sums of $G$ are dense in ${\sf M}_N$.
The group $G = SU(2)$ is a special case:
every linear combination of $SU(2)$ matrices (with real coefficients)
is a nonnegative multiple of an $SU(2)$ matrix, as can be seen from \reff{su2}.
Such a matrix can therefore be represented in the form \reff{su2}
but without the constraint $\bsigma^2 = 1$.

Note also an important qualitative distinction between the
original Hamiltonian \reff{eqn1b} and the
``frustrated'' Hamiltonian \reff{eq:new_ham}\footnote{
   The Hamiltonian \reff{eq:new_ham} can also be thought of as a
   principal chiral model in a fixed background
   ``dielectric gauge field'' \cite{Mack_dielectric}.
}:
the Hamiltonian \reff{eqn1b} has a $G \times G$ global symmetry
$g_x \to h g_x h'$, while the Hamiltonian \reff{eq:new_ham}
has only the $G$ symmetry of {\em left}\/ multiplication $g_x \to h g_x$.
(If one {\em right}\/-multiplies $g_x$ and $g_{x'}$ by an element $h'$
that does not comute with $\alpha^{(l)}_{xx'}$, the energy changes.)
Fortunately, it is only the {\em left}\/ symmetry that is exploited in
the passage \reff{interp} to yet coarser grids.
(The coarse-grid-correction move should be a symmetry {\em within}\/
each block, and contribute an energy only {\em between}\/ blocks.)
Of course, this fortunate circumstance is not a coincidence:
from the ``unigrid'' interpretation
\cite{McCormick-Ruge_unigrid,Goodman:Multigrid1}
one sees that left-multiplication on a very coarse grid
acts simply as left-multiplication on a very large block on the fine grid
--- which is of course a symmetry within the block.

It should also be emphasized that the coefficients in 
the coarse-grid Hamiltonian \reff{eq:coarse_ham}
depend implicitly on the current fine-grid field configuration $g$.
Finally, let us emphasize that in the MGMC algorithm,
unlike the corresponding deterministic MG algorithm,
the coarse-grid Hamiltonian {\em must} (as far as we can tell)
be defined by the ``variational formula''
$H_{l-1}(h) \equiv H_l((p_{l,l-1}h)g)$;
only in this way does the algorithm leave invariant
the correct Gibbs distribution \reff{eqn2}.\footnote{
   This point has caused some confusion in the literature;
   for further discussion and references,
   see \cite[Section X.A]{Goodman:Multigrid1}.
}
\bigskip

{\em Choice of $\scrg_l$.}
The basic stochastic iteration $\scrg_l(\;\cdot\;,H_l)$
can, in principle, be {\em any} stochastic updating procedure
that leaves invariant the probability distribution
$\sim \exp[-H_l(g)]$.
It could, for example, be a single-site (possibly multi-hit)
Metropolis algorithm,
or a single-site quasi-heat-bath \cite{Fredenhagen_87} algorithm.
Indeed, a quasi-heat-bath algorithm may turn out to be the most
efficient one in practice (taking into account CPU time).
However, for conceptual and experimental purposes,
a pure heat-bath algorithm is far preferable,
as it is a clearly defined object with no free parameters.
By contrast, the effectiveness of a Metropolis (or quasi-heat-bath)
algorithm can vary dramatically as a function of the
hit size (or underlying Von Neumann proposal algorithm) and
the Hamiltonian $H_l$.
This is a particularly severe problem in MGMC,
as the coefficients in $H_l$ are {\em random}
and highly variable.
We have therefore used a pure heat-bath algorithm
(with red-black ordering)
in all our experiments.
Technical details of this algorithm can be found in Appendix \ref{appendix_A}.
\bigskip

{\em Pre-sweeps vs.\ post-sweeps.}
The balance between pre-sweeps and post-sweeps
is not very crucial;  only the total $m_1 + m_2$ seems to matter much.
Indeed, either the pre-sweep or the post-sweep (but not both!)
could be eliminated entirely.
Increasing $m_1$ and $m_2$ improves the performance
of the MGMC iteration, at the expense of increased
computational labor per iteration.
In this paper we have always taken $m_1 = m_2 = 1$,
but we make no claim that this is optimal.
\bigskip

{\em Cycle control parameter $\gamma$ and computational labor.}
Clearly, one iteration of the multi-grid algorithm at level $M$
comprises one visit to grid $M$, $\gamma$ visits to grid $M - 1$,
$\gamma^2$ visits to grid $M - 2$, and so on.
Thus, $\gamma$ determines the degree of emphasis placed on
the coarse-grid updates.  ($\gamma = 0$ would
correspond to the pure heat-bath iteration on the finest
grid alone.)

We can now estimate the computational labor required for
one iteration of the multi-grid algorithm.  Each visit to
a given grid involves $m_{1} + m_{2}$ heat-bath
sweeps on that grid, plus the work involved in computing the
coarse-grid Hamiltonian and in carrying out the interpolation
back to the fine-grid.
The work involved in all these operations
is proportional to the number of lattice points
on that grid.  Let $W_{l}$ be the work required for these
operations on grid $l$.  Then, for grids defined by a factor-of-2
coarsening in $d$ dimensions, we have
\be
   W_{l} \;\approx\; 2^{-d(M-l)}  W_{M} \;,
 \label{labor1}
\ee
so that the total work for one multi-grid iteration is
\begin{eqnarray}
  work(MG)  &   =   &  \sum_{l=M}^0  \gamma^{M-l}  W_{l}         \nonumber \\
            &\approx&  W_{M} \sum_{l=M}^0  (\gamma 2^{-d})^{M-l} \nonumber \\
            &  \le  &  W_{M}  (1 - \gamma 2^{-d} )^{-1}
   					\qquad \hbox{if } \gamma < 2^d  \;.
  \label{labor}
\end{eqnarray}
Thus, provided that $\gamma < 2^d$, the work required
for one entire multi-grid iteration is no more than
$(1 - \gamma 2^{-d} )^{-1}$ times the work required
for $m_{1} + m_{2}$ heat-bath iterations
(plus a little auxiliary computation) on the finest
grid alone --- {\em irrespective of the total number of levels}.
The most common choices are $\gamma = 1$ (the V-cycle)
and $\gamma = 2$ (the W-cycle),
but we shall also consider $\gamma = 3$ (the ``super-W-cycle'' or S-cycle).
\bigskip

{\em MGMC vs.\ renormalization group.}
We wish to emphasize that although the multi-grid method
and the block-spin renormalization group (RG)
are based on very similar {\em philosophies} ---
dealing with a single length scale at a time ---
they are in fact {\em very different}.
In particular, the conditional coarse-grid Hamiltonian $H_{l-1}$
employed in the MGMC method is {\em not} the same
as the renormalized Hamiltonian given by a block-spin RG transformation.
The RG transformation computes the {\em marginal},
not the conditional, distribution of the block means ---
that is, it {\em integrates} over the complementary degrees
of freedom, while the MGMC method {\em fixes} these degrees of freedom
at their current (random) values.  Our conditional Hamiltonian
is given by an explicit finite expression,
while the marginal (RG) Hamiltonian cannot be computed in closed form.
The failure to appreciate these distinctions has led to
some confusion in the literature;  for discussion,
see \cite[Section X.A]{Goodman:Multigrid1}.
\bigskip

{\em Behavior of Gaussian MGMC.}
An analogous MGMC algorithm can be devised for the Gaussian model (free field)
\be
   H(\varphi) \;=\;   {A \over 2}  \sum_{\< xx' \>} (\varphi_x -\varphi_{x'})^2
                    + {B \over 2}  \sum_x  \varphi_x^2   \;,
 \label{eq2.10}
\ee
and its behavior can be predicted analytically.
A simple Fock-space argument
\cite{Goodman:Multigrid0,Goodman:Multigrid1}
shows that the autocorrelation time of the Gaussian MGMC algorithm
is {\em equal}\/ to the convergence time of the corresponding
deterministic MG algorithm for solving Laplace's equation.
The deterministic MG algorithm is known to behave as follows:
\begin{itemize}
 \item  Critical slowing-down is completely eliminated
   for $\gamma \ge 1$ (i.e.\ at least a V-cycle)
   with piecewise-linear or smoother interpolation \cite{Mandel_87},
   and for $\gamma \ge 2$ (i.e.\ at least a W-cycle)
   with piecewise-constant interpolation \cite{GS_unpublished}.
 \item  For a V-cycle with piecewise-constant interpolation,
   the exact behavior is not known, but it is strongly suspected
   \cite[section 4.7]{Brandt_86} and observed in preliminary numerical tests
   \cite{Sokal_unpublished,Linn_91} that the critical slowing-down
   is {\em not}\/ eliminated, and that the dynamic critical exponent is
   in fact $z=1$.
\end{itemize}
The behavior of Gaussian MGMC will be a useful standard of comparison
for the behavior of MGMC in asymptotically free continuous-spin models.

{\bf Remark.}
Parisi \cite{Parisi_84}
and Mack and collaborators
\cite{Mack_87,Mack-Meyer_90,Hasenbusch_LAT90,Hasenbusch-Meyer_91}
have pointed out that an update of amplitude $A$
on a block of linear size $\ell_B$ costs an energy
$\Delta E \sim \ell_B^{d-2} A^2$ if a smooth
(i.e.\ once-continuously-differentiable) kernel is employed,
while it costs a much larger energy $\Delta E \sim \ell_B^{d-1} A^2$ 
in the case of a piecewise-constant kernel.
Thus, the typical update amplitude is $A \sim \ell_B^{-(d-2)/2}$
in the case of a smooth kernel --- and this is the ``correct'' amplitude for an
excitation on scale $\sim \ell_B$ in the Gaussian model ---
while the amplitude is only $A \sim \ell_B^{-(d-1)/2}$
in the case of piecewise-constant (step-function) kernel,
i.e.\ a factor $\sqrt{\ell_B}$ ``too small''.
(This behavior is confirmed numerically \cite{Hasenbusch_LAT90}.)
 From these facts one might expect that MGMC with piecewise-constant
interpolation would {\em not}\/ succeed in eliminating critical slowing-down
for the Gaussian model.
However, this conclusion is only partially correct:
as explained above, the rigorous analysis \cite{GS_unpublished}
shows that piecewise-constant MGMC {\em does}\/ eliminate critical slowing-down
provided that at least a W-cycle is employed ($\gamma \ge 2$).
Apparently what is happening is that the multiple updates per iteration
--- $\ell_B^{\log_2 \gamma}$ of them on length scale $\ell_B$ ---
compensate for the unduly small amplitude of each individual update.
Unfortunately, we are unable to make this heuristic argument quantitative,
because we do not know to what extent these multiple updates are correlated.
In particular, we are unable to translate the rigorous proof into
a heuristic argument in a way that is quantitatively correct
(i.e.\ yields the correct critical exponent $z$ as a function of $\gamma$).

\bigskip

Now let us make a few further remarks:
\bigskip

{\em MGMC for general $\sigma$-models.}\/
Thus far we have considered the MGMC algorithm for a
$\sigma$-model taking values in some {\em group}\/ $G \subset U(N)$.
Now let us consider a $\sigma$-model taking values
in some manifold $M$ on which a compact group
$G \subset U(N)$ acts transitively.
Then, by fixing a reference configuration
$\overline{\varphi} = \{ \overline{\varphi}_x \} _{x \in \Omega} \in M^\Omega$,
we can define a (many-to-one) map of $G^\Omega$ onto $M^\Omega$ by
\be
   \{ g_x \}   \;\longto\;  \{ g_x \overline{\varphi}_x \}    \;.
\ee
In particular, Haar measure on $G^\Omega$ is mapped onto Haar measure on
$M^\Omega$.
Likewise, all observables $F$ on $M^\Omega$ --- including the Hamiltonian ---
can be ``lifted'' to $G^\Omega$:
\be
   \widetilde{F}(g)  \;\equiv\;   F(g \overline{\varphi})   \;.
\ee
Then the expectation value of any observable $F(\varphi)$
with respect to the Hamiltonian $H(\varphi)$ [and Haar measure on $M^\Omega$]
is equal to the expectation value of $\widetilde{F}(g)$ with respect to
the Hamiltonian $\widetilde{H}(g)$ [and Haar measure on $G^\Omega$]:
\be
   \int F(\varphi) \, e^{-H(\varphi)} \, \prod_x d\varphi_x   \;=\;
   \int \widetilde{F}(g) \, e^{-\widetilde{H}(g)} \, \prod_x dg_x   \;,
\ee
where $d\varphi_x$ and $dg_x$ are normalized Haar measure on
$M$ and $G$, respectively.
Thus, the $\sigma$-model taking values in $M$ can be ``lifted''
to a $\sigma$-model taking values in $G$.

Consider, for example, the $N$-vector model:
here $M$ is the unit sphere in $\R^N$, $G$ is $SO(N)$ or $O(N)$,
and the Hamiltonian is
\be
   H(\varphi)   \;=\;
     - \sum_{\< xx' \>}  a_{xx'} \, \varphi_x^T \varphi_{x'}
\ee
where $a_{xx'}$ is a real number
(usually $a_{xx'} = \beta$ for all ${\< xx' \>}$).
Then the ``lifted'' Hamiltonian is
\be
  \widetilde{H}(g)   \;=\;   - \sum_{\< xx' \>}
                               \tr (\alpha_{xx'} g_x^T g_{x'})   \;,
\ee
where
\be
   \alpha_{xx'}   \;=\;
     a_{xx'} \, \overline{\varphi}_{x'} \, \overline{\varphi}_x^T
\ee
is an $N \times N$ real matrix of rank 1.
This Hamiltonian is of the form \reff{eq:new_ham},
and so can be handled by the MGMC method described previously.

In this paper we have chosen to study the $SU(2)$ principal chiral model
[= $N$-vector model with $N=4$] largely because it is easy to implement
a heat-bath algorithm on $SU(2) \simeq S^3$
(see Appendix \ref{appendix_A}).
The same MGMC principles can be applied to an $N$-vector model
with arbitrary $N$, but one then needs to devise a heat-bath algorithm
on $SO(N)$ or $O(N)$.
  
\bigskip

{\em MGMC with smooth interpolation.}\/
Mack and collaborators
\cite{Mack_87,Mack-Meyer_90,Hasenbusch_LAT90,Hasenbusch-Meyer_91}
have advocated the use of a smooth
interpolation (e.g.\ piecewise-linear or better)
in place of piecewise-constant.
For fields taking values in a {\em linear}\/ state space,
this is straightforward to do;
but for fields taking values in a {\em nonlinear}\/ manifold
(such as a nonlinear $\sigma$-model),
it is notably awkward:
how should one define a fractional power of a group element?
The solution proposed by Hasenbusch, Meyer and Mack
\cite{Hasenbusch_LAT90}
is to define the coarse-grid update by
\be
   g_x^{(new)}   \;=\;   \exp(i A_x \Psi) \, g_x^{(old)}   \;,
 \label{Mack_update}
\ee
where $A_x$ is a suitable smooth kernel
and $\Psi$ is an element of the Lie algebra of $G$.
(More precisely, they take $A_x$ to be the longest-wavelength mode
of the Laplace operator on an $\ell_B \times \ell_B$ block
with Dirichlet boundary conditions.)
Note that here the updates are made in the ``unigrid'' style
\cite{McCormick-Ruge_unigrid}:
the only field in the problem is the original (finest-grid) field;
and the updates \reff{Mack_update} are performed successively
on the $1 \times 1$ blocks (in some order),
then on the $2 \times 2$ blocks, then on the $4 \times 4$ blocks, etc.
For such an update the labor is proportional to the number of
{\em fine-grid}\/ spins involved in the update,
i.e.\ \reff{labor1} is replaced by
\be
   W_{l} \;\approx\; W_{M} \;,
\ee
so that the total work for one unigrid iteration is
\begin{eqnarray}
  work(UG)  &  =   &  W_M \sum_{l=M}^0  \gamma^{M-l}        \nonumber \\[2mm]
            & \sim &  \cases{ M W_M          & if $\gamma = 1$  \cr
                              \gamma^M W_M   & if $\gamma > 1$  \cr
                            }                               \nonumber \\[2mm]
            & \sim &  \cases{ L^d \log L     & if $\gamma = 1$  \cr
                              L^{d + \log_2 \gamma}   & if $\gamma > 1$  \cr
                            }
\end{eqnarray}
Thus, the computational complexity per iteration is tolerable only for a
V-cycle.
Note also that the energy $H(g^{(new)})$ is an extremely complicated
function of the Lie-algebra element $\Psi$,
so it is not feasible to choose $\Psi$ by a heat-bath method;
rather, $\Psi$ is chosen by the Metropolis prescription
with a suitable hit size $\epsilon \sim \ell_B^{-(d-2)/2}$.

It is now natural to ask:  what does one gain by the use of a smooth kernel
in place of a piecewise-constant kernel?
In the Gaussian case the answer was given above:
with a smooth kernel critical slowing-down is eliminated already for
a V-cycle\footnote{
   Here ``smooth'' refers to an interpolation which is piecewise-linear
   or better, i.e.\ one in which linear functions are exactly interpolated.
   It is worth noting that the Hasenbusch-Meyer-Mack kernels
   \cite{Mack_87,Mack-Meyer_90,Hasenbusch_LAT90}
   do {\em not}\/ satisfy this property;
   indeed, not even a constant function can be exactly interpolated.
   (That is, the constant function cannot be
   represented as a linear combination of the kernels $A_x$ corresponding
   to non-overlapping blocks of the same size, with the exception of the cases
   $\ell_B = 1$ and $\ell_B = 2$.)  Therefore, we do not know whether
   critical slowing-down is completely eliminated (for Gaussians) in the
   Hasenbusch-Meyer-Mack V-cycle;  we suspect that it may not be.
   In view of the identities in \cite[Section VIII]{Goodman:Multigrid1},
   this problem could be studied numerically by measuring the convergence rate
   of the corresponding {\em deterministic}\/ MG algorithm for solving
   Laplace's equation.
   (Note, however, that a more recent work by Hasenbusch and Meyer
   \cite{Hasenbusch-Meyer_91} uses piecewise-linear interpolation.)
},
while with a piecewise-constant kernel a W-cycle is required.
However, this fact does not give any advantage to the smooth kernel,
since in the piecewise-constant case we can readily implement
the W-cycle in a CPU time of order volume, provided that $d>1$
[see \reff{labor}].
In the non-Gaussian case it is less clear what will happen,
but we are unable to see why a smooth interpolation should
handle strongly nonlinear excitations (such as meson-meson scattering
or glueball bound states)
any better than a piecewise-constant interpolation.
We return to these points in Section \ref{sec4.2}.

\section{Numerical Results} \label{sec:results}
\setcounter{equation}{0}

\subsection{Quantities to be Measured}
\vspace{-0.3cm}\quad\par

We begin by defining the quantities ---
autocorrelation functions and autocorrelation times ---
that characterize the Monte Carlo dynamics.
Let $A$ be an observable
(i.e.\ a function of the field configuration $g$).
We are interested in the evolution of $A$ in Monte Carlo time,
and more particularly in the rate at which the system ``loses memory''
of the past.
We define, therefore, the {\em unnormalized autocorrelation function}\footnote{
   In the mathematics and statistics literature, this is called the
   {\em autocovariance function}\/.
}
\be
  C_{AA}(t)  \;=\;   \< A_s A_{s+t} \>   -  \< A \> ^2  \,,
\ee
where expectations are taken {\em in equilibrium\/}.
The corresponding {\em normalized autocorrelation function\/} is
\be
  \rho_{AA}(t)  \;=\;  C_{AA}(t) / C_{AA}(0) \,.
\ee
We then define the {\em integrated autocorrelation time}
\begin{eqnarray}
\tau_{int,A}  & =&
 \half \sum_{{t} \,=\, - \infty}^{\infty} \rho_{AA} (t)\nonumber\\
 &=&  \half \ +\  \sum_{{t} \,=\, 1}^{\infty} \  \rho_{AA} (t)
\end{eqnarray}
[The factor of $\half$ is purely a matter of convention;  it is
inserted so that $\tau_{int,A} \approx\ \tau$ if
$\rho_{AA}(t) \approx e^{-|t|/ \tau}$ with $\tau \gg 1$.] 
The integrated autocorrelation time controls the statistical error
in Monte Carlo measurements of $\< A \>$.  More precisely,
the sample mean
\begin{equation}
\bar A \ \ \equiv\ \ {1 \over n }\  \sum_{t=1}^n \ A_t
\end{equation}
has variance
\begin{eqnarray}
\var( \bar A )  &= &
  {1 \over n^2} \ \sum_{r,s=1}^n \ C_{AA} (r-s) \nonumber \\
 &=& {1 \over n }\ \sum_{{t} \,=\, -(n-1)}^{n-1}
  (1 -  {{|t| \over n }} ) C_{AA} (t) \label{var_observa}  \\
 &\approx&  {1 \over n }\ (2 \tau_{int,A} ) \ C_{AA} (0)
   \qquad {\rm for}\ n\gg \tau \label{var_observb} 
\end{eqnarray}
Thus, the variance of $\bar{A}$ is a factor $2 \tau_{int,A}$
larger than it would be if the $\{ A_t \}$ were
statistically independent.
Stated differently, the number of ``effectively independent samples''
in a run of length $n$ is roughly $n/2 \tau_{int,A}$.
The autocorrelation time $\tau_{int,A}$ (for interesting observables $A$)
is therefore a ``figure of (de)merit'' of a Monte Carlo algorithm.


We shall measure static quantities (expectations) and dynamic quantities
(autocorrelation times) for the following observables:
\begin{eqnarray}
  \scrmvec   &=&  \sum_x \bsigma_x                        \\[0.2cm]
  \scrm^2    &=&  \left( \sum_x \bsigma_x \right)^2       \\[0.2cm]
  \scrf      &=&  \half \left[
                           \left| \sum_x e^{2\pi i x_1/L} \bsigma_x \right| ^2
                           \,+\,
                           \left| \sum_x e^{2\pi i x_2/L} \bsigma_x \right| ^2
                        \right]                           \\[0.2cm]
  \scre      &=&  \sum_{\< xx' \>}  \bsigma_x \cdot \bsigma_{x'}
\end{eqnarray}
The mean values of these observables give information on different aspects
of the 2-point function
\begin{subeqnarray}
   G(x)         & = &   \< \bsigma_0 \cdot \bsigma_x \>        \\[0.1cm]
   \gtilde(p)   & = &   \sum_x e^{ip\cdot x} \< \bsigma_0 \cdot \bsigma_x \>
\end{subeqnarray}
In particular, we are interested in the {\em susceptibility}\/
\be
  \chi  \;=\;  \gtilde(0)   \;=\;  V^{-1} \< \scrm^2 \>
\ee
and the analogous quantity at the smallest nonzero momentum
\be
  F  \;=\;  \gtilde(p) | _{|p| = 2\pi/L}  \;=\;   V^{-1} \< \scrf \>    \;.
\ee
By combining these we can obtain the {\em (second-moment) correlation length}\/
\be
 \label{def_xi}
  \xi   \;=\;     \left( { \displaystyle  (\chi/F) - 1
                           \over
                           \displaystyle  4 \sin^2 (\pi/L)
                         }
                  \right) ^{1/2}
\ee
(see the Remark below).
Finally, we have the (negative) {\em energy}\/
\be
  E   \;=\;  2G(x) | _{|x| = 1}   \;=\;   V^{-1}  \< \scre \>   \;.
\ee
Here $V=L^2$ is the number of sites in the lattice.

\medskip

{\bf Remark.}
The definition \reff{def_xi} is sometimes summarized \cite{Edwards_89}
by saying that we are fitting $\gtilde(p)$ to the Ansatz
\be
 \label{bad-form}
  \gtilde(p) = Z \left[ \xi^{-2} + 4 \sum_{i=1}^d \sin^2 (p_i/2) \right] ^{-1}
\ee
at $p=0$ and $|p| = 2\pi/L$.
This is of course true; but it is important to emphasize
that we are {\em not}\/ assuming that $\gtilde(p)$
really has the free-field form \reff{bad-form} at general $p$
(of course it doesn't).
Rather, we simply use this form to motivate {\em one}\/ reasonable definition
of the second-moment correlation length $\xi$ on a finite lattice.
Another definition --- slightly different but equally reasonable --- would be
\begin{eqnarray}
 \label{def_xiprime}
  \xi'     &=&    \left( {1 \over 2d} \,
                         { \displaystyle \sum_x
                              \left( \sum_{i=1}^d (L/\pi)^2 \sin^2 (\pi x_i/L)
                              \right)  G(x)
                           \over
                           \displaystyle  \sum_x G(x)
                         }
                  \right) ^{1/2}                            \nonumber \\[0.2cm]
   &=&  {L^2 \over 4\pi^2} \left( 1 - {F \over \chi} \right)   \;.
\end{eqnarray}
The two definitions coincide in the infinite-volume limit $L \to\infty$.
Finally, let us emphasize that {\em neither}\/ of these quantities is equal
to the exponential correlation length (= 1/mass gap)
\be
   \xi_{exp}   \;=\;   \lim\limits_{|x| \to \infty}  {-|x|  \over  \log G(x)}
   \;.
\ee
However, $\xi$ (or $\xi'$) and $\xi_{exp}$ are believed to scale
in the same way as $\beta \to\infty$.

\medskip

The integrated autocorrelation time $\tau_{int,A}$ can be estimated
by standard procedures of statistical time-series analysis
\cite{Priestley_81,Anderson_71}.
These procedures also give statistically valid {\em error bars}\/
on $\< A \>$ and $\tau_{int,A}$.
For more details, see \cite[Appendix C]{Madras_88}.
In this paper we have used a self-consistent truncation window of width
$c \tau_{int,A}$, where $c$ is a constant.
In most cases we have used $c=6$;  this choice is reasonable whenever the
autocorrelation function $\rho_{AA}(t)$ decays roughly exponentially.
However, for the energy ($A = \scre$) we noticed that $\rho_{AA}(t)$
has a slowly-decaying component with very small amplitude,
and taking $c=6$ led to a systematic underestimate of $\tau_{int,\scre}$.
Therefore, for $A = \scre$ we took $c=20$ for the heat-bath,
$c=15$ for the V-cycle, and $c=10$ for the W-cycle and super-W-cycle;
this choice seemed to give a reasonable window width.

In setting error bars on $\xi$ we have used the triangle inequality;
such error bars are overly conservative, but we did not feel it was
worth the trouble to
measure the cross-correlations between $\scrm^2$ and $\scrf$.

\subsection{Summary of our Runs} \label{section:MGMCO4_results}
\vspace{-0.3cm}\quad\par

We performed extensive runs using the MGMC algorithm with $\gamma = 0,1,2,3$
(heat-bath, V-cycle, W-cycle and S-cycle, respectively)
and $m_1 = m_2 = 1$, on lattices of sizes $L = 32, 64, 128, 256$.
We made an extremely detailed effort to study the W-cycle
and a moderately detailed effort to study the V-cycle;
only a few runs were made to study, for comparison purposes,
the heat-bath and the S-cycle.
In all cases the coarsest grid is taken to be $2 \times 2$.
All runs used a random initial configuration (``hot start'').
The results of these computations are shown in
Tables~\ref{mgmco4_data_32}, \ref{mgmco4_data_64}, \ref{mgmco4_data_128}
and \ref{mgmco4_data_256}.
Here $\beta \equiv \beta_{O(4)} = 2\beta_{SU(2)}$.
The heat-bath runs use the MGMC program
but with a cycle control parameter $\gamma = 0$, so
each ``iteration'' consists of {\em two}\/ heat-bath sweeps.
%
%
\begin{table}
\addtolength{\tabcolsep}{-1.0mm}
%
\protect\footnotesize
\begin{center}
\begin{tabular}{|c|c r r|r@{\ (}r r@{(}r r@{\ (}r|r@{\ (}r r@{\ (}r|} \hline
\multicolumn{14}{|c|}{$d=2$ $O(4)$ model at $L=32$} \\ \hline
  &  &  &  & \multicolumn{6}{|c}{\ } & \multicolumn{4}{|c|}{\ }   \\ 
$\beta$  &  Type  &  Sweeps  &  Disc  &
  \multicolumn{2}{|c}{$\chi$}  &  \multicolumn{2}{c}{$\xi$}  &
     \multicolumn{2}{c|}{$E$}  &
  \multicolumn{2}{|c}{$\taum$} &  \multicolumn{2}{c|}{$\taux$}    \\
\hline
2.00 & HB & 500000 & 10000 &   92.9 &  0.4)
     &  7.61 & 0.03) & 1.15499 & 0.00006) &  66.19 &   3.78)
     & 20.68 & 0.66)\\ 
2.00 &  W & 100000 &  5000 &   93.7 &  0.3)
     &  7.67 & 0.02) & 1.15525 & 0.00010) &   1.19 &   0.02)
     &  2.64 & 0.07)\\ 
\hline
2.10 & HB & 500000 & 10000 &  134.6 &  0.5)
     &  9.63 & 0.03) & 1.20400 & 0.00005) & 108.48 &   7.92)
     & 23.23 & 0.79)\\ 
2.10 &  W & 100000 &  5000 &  134.3 &  0.4)
     &  9.61 & 0.03) & 1.20399 & 0.00009) &   1.33 &   0.03)
     &  2.74 & 0.07)\\ 
\hline
2.20 & HB & 500000 & 10000 &  178.4 &  0.6)
     & 11.57 & 0.04) & 1.24789 & 0.00005) & 145.37 &  12.29)
     & 23.97 & 0.82)\\ 
2.20 &  W & 100000 &  5000 &  178.5 &  0.4)
     & 11.60 & 0.03) & 1.24779 & 0.00008) &   1.46 &   0.03)
     &  2.66 & 0.07)\\ 
\hline
2.30 & HB & 500000 & 10000 &  221.2 &  0.5)
     & 13.43 & 0.04) & 1.28684 & 0.00004) & 213.38 &  21.88)
     & 20.37 & 0.65)\\ 
2.30 &  W & 100000 &  5000 &  221.1 &  0.4)
     & 13.41 & 0.03) & 1.28691 & 0.00008) &   1.57 &   0.03)
     &  2.19 & 0.05)\\ 
\hline
2.40 &  W & 100000 &  5000 &  258.5 &  0.3)
     & 14.94 & 0.03) & 1.32155 & 0.00007) &   1.69 &   0.04)
     &  1.94 & 0.04)\\ 
\hline
\end{tabular}
\end{center}
\caption[tab1]{
Data for the two-dimensional $O(4)$ model on a $32 \times 32$ lattice.
``Type'' is the method used (heat-bath, V-cycle, W-cycle or super-W-cycle).
``Sweeps'' is the number of MGMC iterations performed;
a heat-bath ``iteration'' here is equal to two usual heat-bath sweeps.
``Disc'' is the number of iterations discarded prior to beginning the analysis.
Standard error is shown in parentheses.
}
\label{mgmco4_data_32}
\end{table}
%
%
%
\begin{table}
\addtolength{\tabcolsep}{-1.0mm}
%
\protect\footnotesize
\begin{center}
\begin{tabular}{|c|c r r|r@{\ (}r r@{(}r r@{\ (}r|r@{\ (}r r@{\ (}r|} \hline
\multicolumn{14}{|c|}{$d=2$ $O(4)$ model at $L=64$} \\ \hline
  &  &  &  & \multicolumn{6}{|c}{\ } & \multicolumn{4}{|c|}{\ }   \\ 
$\beta$  &  Type  &  Sweeps  &  Disc  &
  \multicolumn{2}{|c}{$\chi$}  &  \multicolumn{2}{c}{$\xi$}  &
     \multicolumn{2}{c|}{$E$}  &
  \multicolumn{2}{|c}{$\taum$} &  \multicolumn{2}{c|}{$\taux$}    \\
\hline
2.10 &  V & 200000 & 10000 &  161.2 &  0.9)
     & 10.45 & 0.04) & 1.20182 & 0.00003) &  13.47 &   0.56)
     &  8.49 & 0.28)\\ 
2.10 &  W & 200000 &  5000 &  160.0 &  0.6)
     & 10.39 & 0.03) & 1.20187 & 0.00003) &   1.24 &   0.02)
     &  3.44 & 0.07)\\ 
\hline
2.15 &  W & 200000 &  5000 &  202.9 &  0.7)
     & 11.93 & 0.03) & 1.22410 & 0.00003) &   1.32 &   0.02)
     &  3.88 & 0.09)\\ 
\hline
2.20 & HB & 500000 & 10000 &  257.2 &  2.3)
     & 13.74 & 0.09) & 1.24532 & 0.00002) & 217.47 &  22.53)
     & 72.96 & 4.37)\\ 
2.20 &  V & 200000 & 10000 &  258.2 &  1.5)
     & 13.78 & 0.06) & 1.24527 & 0.00003) &  18.92 &   0.93)
     & 11.30 & 0.43)\\ 
2.20 &  W & 200000 &  5000 &  258.3 &  0.9)
     & 13.79 & 0.04) & 1.24533 & 0.00003) &   1.38 &   0.02)
     &  4.39 & 0.10)\\ 
2.20 &  S & 100000 &  5000 &  258.6 &  1.1)
     & 13.83 & 0.04) & 1.24534 & 0.00004) &   0.51 &   0.01)
     &  2.89 & 0.08)\\ 
\hline
2.22 &  W & 200000 &  5000 &  283.3 &  1.0)
     & 14.58 & 0.04) & 1.25349 & 0.00003) &   1.42 &   0.02)
     &  4.58 & 0.11)\\ 
\hline
2.24 &  W & 200000 &  5000 &  309.5 &  1.0)
     & 15.38 & 0.04) & 1.26149 & 0.00003) &   1.45 &   0.02)
     &  4.58 & 0.11)\\ 
\hline
2.26 &  W & 200000 &  5000 &  337.4 &  1.1)
     & 16.20 & 0.04) & 1.26932 & 0.00003) &   1.48 &   0.02)
     &  4.71 & 0.12)\\ 
\hline
2.28 &  W & 200000 &  5000 &  364.7 &  1.2)
     & 16.93 & 0.04) & 1.27692 & 0.00003) &   1.51 &   0.02)
     &  4.79 & 0.12)\\ 
\hline
2.30 & HB & 500000 & 10000 &  393.5 &  3.3)
     & 17.71 & 0.12) & 1.28442 & 0.00002) & 366.26 &  49.44)
     & 90.60 & 6.04)\\ 
2.30 &  V & 200000 & 10000 &  395.8 &  2.0)
     & 17.83 & 0.07) & 1.28441 & 0.00003) &  23.84 &   1.31)
     & 12.39 & 0.49)\\ 
2.30 &  W & 200000 &  5000 &  395.3 &  1.2)
     & 17.82 & 0.04) & 1.28442 & 0.00003) &   1.52 &   0.02)
     &  5.00 & 0.13)\\ 
2.30 &  S & 100000 &  5000 &  394.1 &  1.4)
     & 17.79 & 0.05) & 1.28445 & 0.00004) &   0.52 &   0.01)
     &  3.18 & 0.09)\\ 
\hline
2.35 &  W & 200000 &  5000 &  471.5 &  1.3)
     & 19.87 & 0.05) & 1.30244 & 0.00003) &   1.60 &   0.02)
     &  5.02 & 0.13)\\ 
\hline
2.40 & HB & 500000 & 10000 &  559.1 &  3.6)
     & 22.24 & 0.14) & 1.31957 & 0.00002) & 545.55 &  90.06)
     & 93.08 & 6.29)\\ 
2.40 &  V & 200000 & 10000 &  549.3 &  2.1)
     & 21.86 & 0.08) & 1.31945 & 0.00003) &  29.87 &   1.84)
     & 12.27 & 0.49)\\ 
2.40 &  W & 200000 &  5000 &  550.0 &  1.3)
     & 21.89 & 0.05) & 1.31942 & 0.00002) &   1.66 &   0.02)
     &  4.83 & 0.12)\\ 
2.40 &  S & 100000 &  5000 &  551.9 &  1.5)
     & 21.91 & 0.06) & 1.31956 & 0.00003) &   0.52 &   0.01)
     &  3.04 & 0.09)\\ 
\hline
2.43 &  W & 200000 &  5000 &  601.5 &  1.3)
     & 23.22 & 0.05) & 1.32931 & 0.00002) &   1.71 &   0.03)
     &  4.39 & 0.10)\\ 
\hline
2.45 &  W & 200000 &  5000 &  632.5 &  1.2)
     & 23.97 & 0.05) & 1.33563 & 0.00002) &   1.74 &   0.03)
     &  4.21 & 0.10)\\ 
\hline
2.50 & HB & 500000 & 10000 &  706.6 &  3.8)
     & 25.79 & 0.16) & 1.35092 & 0.00002) & 713.40 & 133.84)
     & 97.46 & 6.74)\\ 
2.50 &  V & 200000 & 10000 &  707.8 &  1.9)
     & 25.81 & 0.08) & 1.35092 & 0.00002) &  35.21 &   2.29)
     & 10.14 & 0.36)\\ 
2.50 &  W & 200000 &  5000 &  706.4 &  1.2)
     & 25.75 & 0.05) & 1.35093 & 0.00002) &   1.79 &   0.03)
     &  4.00 & 0.09)\\ 
\hline
2.55 &  W & 200000 &  5000 &  779.0 &  1.1)
     & 27.49 & 0.05) & 1.36545 & 0.00002) &   1.84 &   0.03)
     &  3.45 & 0.07)\\ 
\hline
2.60 &  W & 200000 &  5000 &  848.0 &  1.1)
     & 29.08 & 0.05) & 1.37928 & 0.00002) &   1.92 &   0.03)
     &  3.27 & 0.07)\\ 
\hline
\end{tabular}
\end{center}
\caption[tab2]{
Data for the two-dimensional $O(4)$ model on a $64 \times 64$ lattice.
``Type'' is the method used (heat-bath, V-cycle, W-cycle or super-W-cycle).
``Sweeps'' is the number of MGMC iterations performed;
a heat-bath ``iteration'' here is equal to two usual heat-bath sweeps.
``Disc'' is the number of iterations discarded prior to beginning the analysis.
Standard error is shown in parentheses.
}
\label{mgmco4_data_64}
\end{table}

%
%
%
\begin{table}
\addtolength{\tabcolsep}{-1.0mm}
%
\protect\footnotesize
\begin{center}
\begin{tabular}{|c|c r r|r@{\ (}r r@{(}r r@{\ (}r|r@{\ (}r r@{\ (}r|} \hline
\multicolumn{14}{|c|}{$d=2$ $O(4)$ model at $L=128$} \\ \hline
  &  &  &  & \multicolumn{6}{|c}{\ } & \multicolumn{4}{|c|}{\ }   \\ 
$\beta$  &  Type  &  Sweeps  &  Disc  &
  \multicolumn{2}{|c}{$\chi$}  &  \multicolumn{2}{c}{$\xi$}  &
     \multicolumn{2}{c|}{$E$}  &
  \multicolumn{2}{|c}{$\taum$} &  \multicolumn{2}{c|}{$\taux$}    \\
\hline
2.10 &  W & 200000 &  5000 &  161.0 &  0.6)
     & 10.31 & 0.03) & 1.20177 & 0.00002) &   1.15 &   0.01)
     &  2.55 & 0.05)\\ 
\hline
2.20 &  V & 200000 & 10000 &  269.7 &  2.0)
     & 14.16 & 0.07) & 1.24508 & 0.00002) &  20.64 &   1.06)
     & 12.50 & 0.50)\\ 
2.20 &  W & 200000 &  5000 &  267.7 &  1.1)
     & 13.99 & 0.04) & 1.24509 & 0.00001) &   1.30 &   0.02)
     &  3.57 & 0.08)\\ 
\hline
2.25 &  W & 200000 &  5000 &  346.3 &  1.5)
     & 16.26 & 0.05) & 1.26504 & 0.00001) &   1.39 &   0.02)
     &  4.34 & 0.10)\\ 
\hline
2.30 &  V & 200000 & 10000 &  452.4 &  3.9)
     & 19.07 & 0.11) & 1.28394 & 0.00001) &  28.36 &   1.70)
     & 18.79 & 0.92)\\ 
2.30 &  W & 200000 &  5000 &  449.0 &  2.0)
     & 18.83 & 0.06) & 1.28395 & 0.00001) &   1.47 &   0.02)
     &  5.08 & 0.13)\\ 
\hline
2.35 &  W & 200000 &  5000 &  580.0 &  2.7)
     & 21.81 & 0.07) & 1.30186 & 0.00001) &   1.54 &   0.02)
     &  6.02 & 0.17)\\ 
\hline
2.40 &  V & 200000 & 10000 &  747.0 &  6.8)
     & 25.24 & 0.16) & 1.31885 & 0.00001) &  39.87 &   2.84)
     & 26.50 & 1.54)\\ 
2.40 &  W & 200000 &  5000 &  744.1 &  3.4)
     & 25.12 & 0.08) & 1.31887 & 0.00001) &   1.61 &   0.02)
     &  6.56 & 0.19)\\ 
\hline
2.45 &  W & 200000 &  5000 &  952.9 &  4.3)
     & 29.08 & 0.10) & 1.33499 & 0.00001) &   1.70 &   0.03)
     &  7.61 & 0.24)\\ 
\hline
2.50 &  V & 200000 & 10000 & 1172.5 &  9.8)
     & 32.60 & 0.21) & 1.35028 & 0.00001) &  53.35 &   4.40)
     & 30.20 & 1.87)\\ 
2.50 &  W & 200000 &  5000 & 1195.9 &  4.9)
     & 33.34 & 0.11) & 1.35029 & 0.00001) &   1.75 &   0.03)
     &  7.91 & 0.25)\\ 
2.50 &  S & 100000 &  5000 & 1193.5 &  5.4)
     & 33.21 & 0.11) & 1.35031 & 0.00002) &   0.50 &   0.01)
     &  4.48 & 0.15)\\ 
\hline
2.55 &  W & 200000 &  5000 & 1463.6 &  5.3)
     & 37.66 & 0.11) & 1.36488 & 0.00001) &   1.82 &   0.03)
     &  7.78 & 0.24)\\ 
\hline
2.60 &  V & 200000 & 10000 & 1745.9 & 11.1)
     & 41.83 & 0.24) & 1.37874 & 0.00001) &  68.42 &   6.37)
     & 30.01 & 1.86)\\ 
2.60 &  W & 200000 &  5000 & 1723.2 &  5.6)
     & 41.33 & 0.12) & 1.37875 & 0.00001) &   1.89 &   0.03)
     &  7.72 & 0.24)\\ 
\hline
2.65 &  W & 200000 &  5000 & 2003.5 &  5.6)
     & 45.39 & 0.12) & 1.39195 & 0.00001) &   1.95 &   0.03)
     &  7.21 & 0.22)\\ 
\hline
2.70 &  V & 200000 & 10000 & 2290.0 & 10.8)
     & 49.53 & 0.25) & 1.40459 & 0.00001) &  73.51 &   7.10)
     & 25.73 & 1.47)\\ 
2.70 &  W & 200000 &  5000 & 2294.7 &  5.2)
     & 49.61 & 0.12) & 1.40460 & 0.00001) &   1.99 &   0.03)
     &  6.23 & 0.18)\\ 
\hline
2.75 &  W & 200000 &  5000 & 2561.3 &  5.0)
     & 53.26 & 0.12) & 1.41666 & 0.00001) &   2.06 &   0.03)
     &  5.94 & 0.16)\\ 
\hline
2.80 &  W & 200000 &  5000 & 2813.8 &  4.6)
     & 56.54 & 0.11) & 1.42821 & 0.00001) &   2.14 &   0.04)
     &  5.12 & 0.13)\\ 
\hline
\end{tabular}
\end{center}
\caption[tab3]{
Data for the two-dimensional $O(4)$ model on a $128 \times 128$ lattice.
``Type'' is the method used (heat-bath, V-cycle, W-cycle or super-W-cycle).
``Sweeps'' is the number of MGMC iterations performed;
a heat-bath ``iteration'' here is equal to two usual heat-bath sweeps.
``Disc'' is the number of iterations discarded prior to beginning the analysis.
Standard error is shown in parentheses.
}
\label{mgmco4_data_128}
\end{table}

%
%
%
\begin{table}
\addtolength{\tabcolsep}{-1.0mm}
%
\protect\footnotesize
\begin{center}
\begin{tabular}{|c|c r r|r@{\ (}r r@{(}r r@{\ (}r|r@{\ (}r r@{\ (}r|} \hline
\multicolumn{14}{|c|}{$d=2$ $O(4)$ model at $L=256$} \\ \hline
  &  &  &  & \multicolumn{6}{|c}{\ } & \multicolumn{4}{|c|}{\ }   \\ 
$\beta$  &  Type  &  Sweeps  &  Disc  &
  \multicolumn{2}{|c}{$\chi$}  &  \multicolumn{2}{c}{$\xi$}  &
     \multicolumn{2}{c|}{$E$}  &
  \multicolumn{2}{|c}{$\taum$} &  \multicolumn{2}{c|}{$\taux$}    \\
\hline
2.50 &  V & 200000 & 10000 & 1316.2 & 16.6)
     & 34.80 & 0.30) & 1.35021 & 0.00001) &  65.79 &   6.01)
     & 37.72 & 2.61)\\ 
2.50 &  W & 100000 &  5000 & 1319.0 & 10.2)
     & 35.04 & 0.19) & 1.35021 & 0.00001) &   1.66 &   0.04)
     &  7.13 & 0.31)\\ 
\hline
2.60 &  V & 200000 & 10000 & 2240.7 & 30.1)
     & 46.77 & 0.44) & 1.37859 & 0.00001) &  86.32 &   9.04)
     & 53.02 & 4.35)\\ 
2.60 &  W & 200000 &  5000 & 2229.4 & 12.8)
     & 46.64 & 0.19) & 1.37860 & 0.00001) &   1.84 &   0.03)
     &  9.80 & 0.34)\\ 
\hline
2.70 &  V & 200000 & 10000 & 3648.7 & 45.2)
     & 61.76 & 0.59) & 1.40444 & 0.00001) & 113.19 &  13.59)
     & 62.62 & 5.58)\\ 
2.70 &  W & 200000 &  5000 & 3646.8 & 19.3)
     & 61.92 & 0.25) & 1.40443 & 0.00000) &   1.98 &   0.03)
     & 11.55 & 0.44)\\ 
\hline
2.80 &  V & 200000 & 10000 & 5519.2 & 54.3)
     & 79.39 & 0.71) & 1.42806 & 0.00000) & 141.24 &  18.91)
     & 64.76 & 5.86)\\ 
2.80 &  W & 200000 &  5000 & 5448.7 & 23.4)
     & 78.33 & 0.29) & 1.42806 & 0.00000) &   2.12 &   0.04)
     & 11.59 & 0.44)\\ 
\hline
3.00 &  W & 100000 &  5000 & 9300.0 & 30.2)
     & 108.21 & 0.41) & 1.46979 & 0.00001) &   2.31 &   0.06)
     &  8.24 & 0.38)\\ 
\hline
\end{tabular}
\end{center}
\caption[tab4]{
Data for the two-dimensional $O(4)$ model on a $256 \times 256$ lattice.
``Type'' is the method used (heat-bath, V-cycle, W-cycle or super-W-cycle).
``Sweeps'' is the number of MGMC iterations performed;
a heat-bath ``iteration'' here is equal to two usual heat-bath sweeps.
``Disc'' is the number of iterations discarded prior to beginning the analysis.
Standard error is shown in parentheses.
}
\label{mgmco4_data_256}
\end{table}

In the heat-bath and V-cycle algorithms, the slowest mode (of those we study)
is the total magnetization $\scrmvec$:
global spin rotations are very slow.
In the W-cycle and S-cycle algorithms, by contrast, the autocorrelation time
$\taum$ is negligible:  the MGMC algorithm performs global spin rotations
very efficiently.
Indeed, if we had taken our coarsest grid to be $1 \times 1$
instead of $2 \times 2$, then the sequence $\{ \scrmvec_t \}$
would be completely uncorrelated,
i.e.\ $\taum = \half$ for all $\beta,L$.

Of greater physical interest is the squared magnetization $\scrm^2$,
which corresponds to {\em relative}\/ rotations of the spins
in different parts of the lattice;
this mode is very slow (though not the slowest)
in the heat-bath and V-cycle algorithms,
and it is essentially the slowest mode of the W-cycle and S-cycle algorithms.
For all these algorithms
the autocorrelation time $\taux$ has the same qualitative behavior:
as a function of $\beta$ it first rises to a peak and then falls;
the location of this peak shifts towards $\beta=\infty$ as $L$ increases;
and the height of this peak grows as $L$ increases.
In Section \ref{subsection:finite-size-scaling_dynamic}
we shall make a detailed finite-size-scaling analysis of these data.

The autocorrelation time of the energy, $\taue$,
is uniformly small: less than about 2.5 for the heat bath,
and less than about 1 for the other cycles.
So we do not bother to report it here.

%
%
\begin{table}
 \protect\footnotesize
\begin{center}
\begin{tabular}{|r|c|r@{\ (}r r@{\ (}r r@{\ (}c|} \hline
\multicolumn{8}{|c|}{Merged $O(4)$-Model Static Data} \\ \hline
$L$&$\beta$
  &\multicolumn{2}{c}{$\chi$}
  &\multicolumn{2}{c}{$\xi$}
  &\multicolumn{2}{c|}{$E$}  \\   \hline
 32 & 1.62 &   21.0 &  0.1) &  3.09 & 0.02) & 0.93417 & 0.00054)\\ 
 32 & 1.70 &   27.8 &  0.0) &  3.65 & 0.01) & 0.98418 & 0.00011)\\ 
 32 & 1.80 &   41.3 &  0.0) &  4.67 & 0.01) & 1.04456 & 0.00007)\\ 
 32 & 1.90 &   62.1 &  0.1) &  5.96 & 0.01) & 1.10161 & 0.00007)\\ 
 32 & 2.00 &   93.7 &  0.1) &  7.68 & 0.01) & 1.15509 & 0.00004)\\ 
 32 & 2.10 &  134.6 &  0.2) &  9.63 & 0.01) & 1.20398 & 0.00004)\\ 
 32 & 2.20 &  178.8 &  0.2) & 11.60 & 0.02) & 1.24787 & 0.00004)\\ 
 32 & 2.30 &  221.1 &  0.3) & 13.42 & 0.02) & 1.28685 & 0.00004)\\ 
 32 & 2.40 &  258.5 &  0.3) & 14.94 & 0.03) & 1.32155 & 0.00007)\\ 
 32 & 2.60 &  322.4 &  1.0) & 17.50 & 0.10) & \multicolumn{2}{c|}{\ }\\ 
 32 & 3.00 &  418.0 &  0.9) & 21.30 & 0.10) & \multicolumn{2}{c|}{\ }\\ 
 32 & 3.40 &  488.1 &  0.8) & 24.30 & 0.10) & \multicolumn{2}{c|}{\ }\\ 
\hline
 64 & 1.70 &   27.9 &  0.2) &  3.65 & 0.02) & \multicolumn{2}{c|}{\ }\\ 
 64 & 1.80 &   41.2 &  0.3) &  4.60 & 0.10) & \multicolumn{2}{c|}{\ }\\ 
 64 & 1.90 &   64.1 &  0.5) &  6.10 & 0.10) & \multicolumn{2}{c|}{\ }\\ 
 64 & 2.00 &   99.4 &  0.2) &  7.83 & 0.01) & 1.15378 & 0.00004)\\ 
 64 & 2.10 &  160.5 &  0.3) & 10.40 & 0.01) & 1.20184 & 0.00002)\\ 
 64 & 2.15 &  202.9 &  0.7) & 11.93 & 0.03) & 1.22410 & 0.00003)\\ 
 64 & 2.20 &  257.9 &  0.4) & 13.77 & 0.02) & 1.24532 & 0.00001)\\ 
 64 & 2.22 &  283.3 &  1.0) & 14.58 & 0.04) & 1.25349 & 0.00003)\\ 
 64 & 2.24 &  309.5 &  1.0) & 15.38 & 0.04) & 1.26149 & 0.00003)\\ 
 64 & 2.26 &  337.4 &  1.1) & 16.20 & 0.04) & 1.26932 & 0.00003)\\ 
 64 & 2.28 &  364.7 &  1.2) & 16.93 & 0.04) & 1.27692 & 0.00003)\\ 
 64 & 2.30 &  395.0 &  0.7) & 17.81 & 0.03) & 1.28442 & 0.00001)\\ 
 64 & 2.35 &  471.5 &  1.3) & 19.87 & 0.05) & 1.30244 & 0.00003)\\ 
 64 & 2.40 &  551.6 &  0.8) & 21.92 & 0.03) & 1.31950 & 0.00001)\\ 
 64 & 2.43 &  601.5 &  1.3) & 23.22 & 0.05) & 1.32931 & 0.00002)\\ 
 64 & 2.45 &  632.5 &  1.2) & 23.97 & 0.05) & 1.33563 & 0.00002)\\ 
 64 & 2.50 &  706.7 &  0.9) & 25.77 & 0.04) & 1.35092 & 0.00001)\\ 
 64 & 2.55 &  779.0 &  1.1) & 27.49 & 0.05) & 1.36545 & 0.00002)\\ 
 64 & 2.60 &  847.9 &  1.0) & 29.06 & 0.04) & 1.37928 & 0.00002)\\ 
 64 & 3.00 & 1286.6 &  3.0) & 38.40 & 0.20) & \multicolumn{2}{c|}{\ }\\ 
 64 & 3.40 & 1611.5 &  2.9) & 45.50 & 0.20) & \multicolumn{2}{c|}{\ }\\ 
 64 & 3.80 & 1865.7 &  2.7) & 51.30 & 0.20) & \multicolumn{2}{c|}{\ }\\ 
\hline
\end{tabular}
\end{center}
\caption[tab5]{
   Best estimates of susceptibility, correlation length and energy
   for the $O(4)$ model on $32\times 32$ and $64\times 64$ lattices,
   from Tables \protect\ref{mgmco4_data_32}, \protect\ref{mgmco4_data_64}
   and \protect\cite{Edwards_89,CEPS_swwo4c2}.
   Standard error is shown in parentheses.
}
\label{o4_merged_table1}
\end{table}

\begin{table}
 \protect\footnotesize
\begin{center}
\begin{tabular}{|r|c|r@{\ (}r r@{\ (}r r@{\ (}c|} \hline
\multicolumn{8}{|c|}{Merged $O(4)$-Model Static Data} \\ \hline
$L$&$\beta$
  &\multicolumn{2}{c}{$\chi$}
  &\multicolumn{2}{c}{$\xi$}
  &\multicolumn{2}{c|}{$E$}  \\   \hline
128 & 2.00 &   99.3 &  0.8) &  7.80 & 0.10) & \multicolumn{2}{c|}{\ }\\ 
128 & 2.10 &  161.1 &  0.5) & 10.32 & 0.03) & 1.20177 & 0.00002)\\ 
128 & 2.20 &  267.9 &  0.9) & 14.02 & 0.03) & 1.24509 & 0.00001)\\ 
128 & 2.25 &  346.3 &  1.5) & 16.26 & 0.05) & 1.26504 & 0.00001)\\ 
128 & 2.30 &  450.6 &  1.6) & 18.91 & 0.05) & 1.28394 & 0.00001)\\ 
128 & 2.35 &  580.0 &  2.7) & 21.81 & 0.07) & 1.30186 & 0.00001)\\ 
128 & 2.40 &  745.1 &  2.7) & 25.15 & 0.07) & 1.31886 & 0.00001)\\ 
128 & 2.45 &  952.9 &  4.3) & 29.08 & 0.10) & 1.33499 & 0.00001)\\ 
128 & 2.50 & 1192.1 &  3.4) & 33.19 & 0.07) & 1.35029 & 0.00001)\\ 
128 & 2.55 & 1463.6 &  5.3) & 37.66 & 0.11) & 1.36488 & 0.00001)\\ 
128 & 2.60 & 1726.8 &  4.6) & 41.38 & 0.09) & 1.37874 & 0.00001)\\ 
128 & 2.65 & 2003.5 &  5.6) & 45.39 & 0.12) & 1.39195 & 0.00001)\\ 
128 & 2.70 & 2293.8 &  4.7) & 49.60 & 0.11) & 1.40460 & 0.00001)\\ 
128 & 2.75 & 2561.3 &  5.0) & 53.26 & 0.12) & 1.41666 & 0.00001)\\ 
128 & 2.80 & 2813.8 &  4.6) & 56.54 & 0.11) & 1.42821 & 0.00001)\\ 
128 & 3.00 & 3715.8 & 10.6) & 67.40 & 0.30) & \multicolumn{2}{c|}{\ }\\ 
128 & 3.40 & 5124.0 & 10.5) & 82.50 & 0.40) & \multicolumn{2}{c|}{\ }\\ 
\hline
256 & 2.20 &  267.8 &  2.0) & 13.70 & 0.10) & \multicolumn{2}{c|}{\ }\\ 
256 & 2.40 &  765.3 &  6.3) & 25.50 & 0.20) & \multicolumn{2}{c|}{\ }\\ 
256 & 2.50 & 1318.2 &  8.7) & 34.97 & 0.16) & 1.35021 & 0.00001)\\ 
256 & 2.60 & 2231.1 & 11.8) & 46.66 & 0.17) & 1.37860 & 0.00001)\\ 
256 & 2.70 & 3647.1 & 17.7) & 61.90 & 0.23) & 1.40443 & 0.00001)\\ 
256 & 2.80 & 5459.7 & 21.5) & 78.48 & 0.27) & 1.42806 & 0.00001)\\ 
256 & 3.00 & 9300.0 & 30.2) & 108.21 & 0.41) & 1.46979 & 0.00001)\\ 
\hline
\end{tabular}
\end{center}
\caption[tab6]{
   Best estimates of susceptibility, correlation length and energy
   for the $O(4)$ model on $128\times 128$ and $256\times 256$ lattices,
   from Tables \protect\ref{mgmco4_data_128}, \protect\ref{mgmco4_data_256}
   and \protect\cite{Edwards_89,CEPS_swwo4c2}.
   Standard error is shown in parentheses.
}
\label{o4_merged_table2}
\end{table}

In Tables~\ref{o4_merged_table1} and \ref{o4_merged_table2}
we summarize, for the convenience of the reader,
our best estimates of the static quantities $\chi$, $\xi$ and $E$
on lattices of size $L = 32, 64, 128, 256$.
These estimates come from merging all of the data in
Tables~\ref{mgmco4_data_32}--\ref{mgmco4_data_256}
together with the data from our Wolff-Swendsen-Wang runs
\cite{Edwards_89,CEPS_swwo4c2}.\footnote{
   The $L=256$, $\beta = 2.60, 2.80$ data from \cite{Edwards_89}
   are unreliable due to an observed metastability,
   and the $L=32,64$, $\beta = 2.20$, $N_{hit} = 1$ data
   from \cite{CEPS_swwo4c2} are unreliable due to possibly insufficient
   equilibration;
   these data points are {\em not}\/ included in the means.
   In all other cases, $\chi$, $\xi$ and $E$ from the different algorithms
   agree within error bars;  this is strong evidence that all the programs
   are correct!
}
These data are consistent with those of Heller \cite{Heller_88b}
and Wolff \cite{Wolff_O4_O8}.
We use these merged data in our finite-size-scaling analyses
whenever $\chi$ and/or $\xi$ is required.

\subsection{Finite-Size-Scaling Analysis: Static Quantities}
\label{subsection:finite-size-scaling_static}
\vspace{-0.3cm}\quad\par

In this subsection and the next, we make our conclusions more quantitative
by performing a finite-size-scaling analysis.

\begin{figure}
\epsfxsize=\textwidth
\epsffile{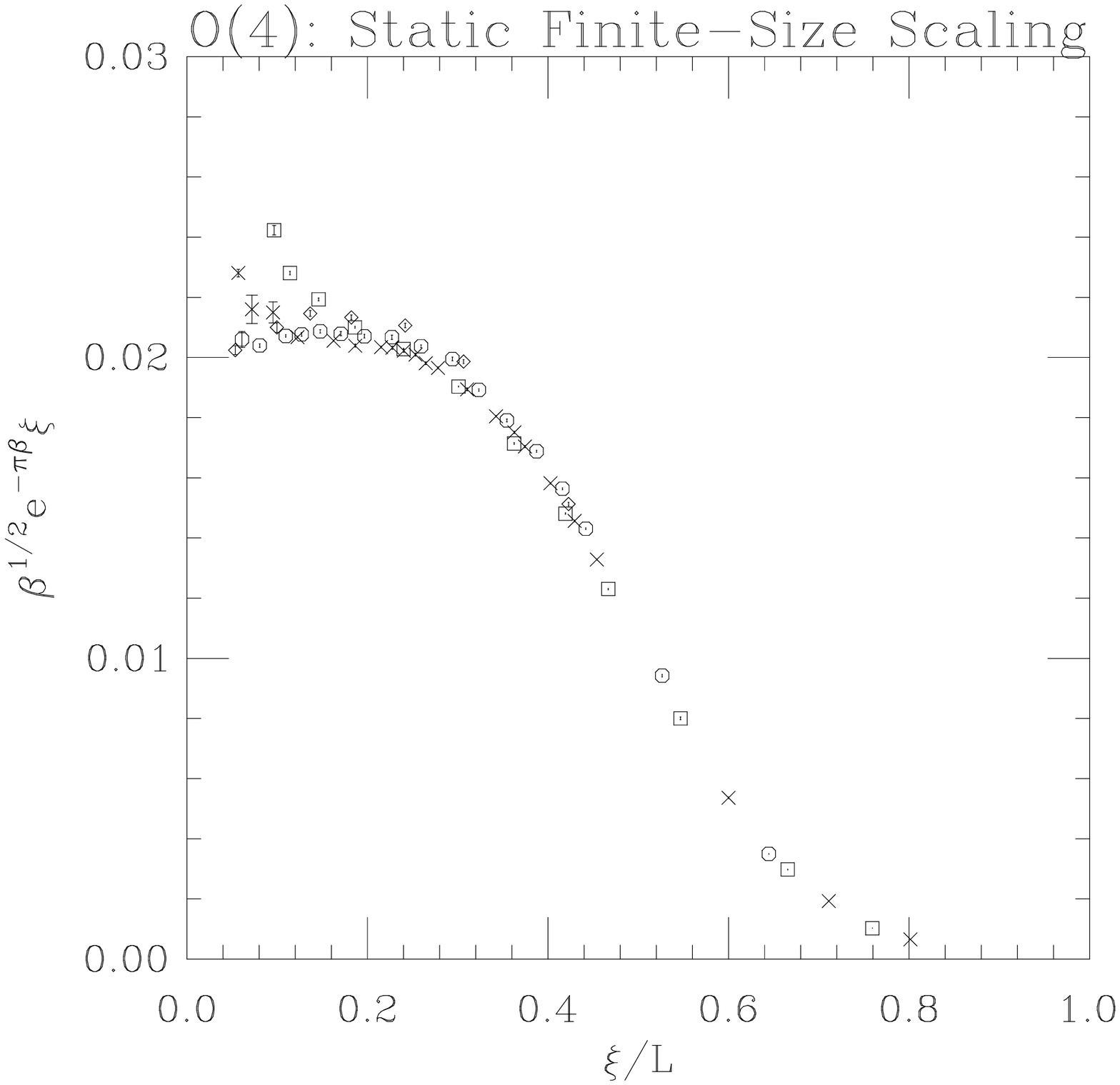}
\caption[fig2]{
   Finite-size-scaling plot of $\beta^{1/2} e^{-\pi\beta} \xi$
   versus $\xi/L$, for lattice sizes $L=32$ ($\Box$),
   64 ($\times$), 128 ($\bigcirc$), 256 ($\Diamond$).
   Error bars are in most cases invisible.
}
\label{mgmco4_xi_scaling} 
\end{figure}
\begin{figure}
\epsfxsize=\textwidth
\epsffile{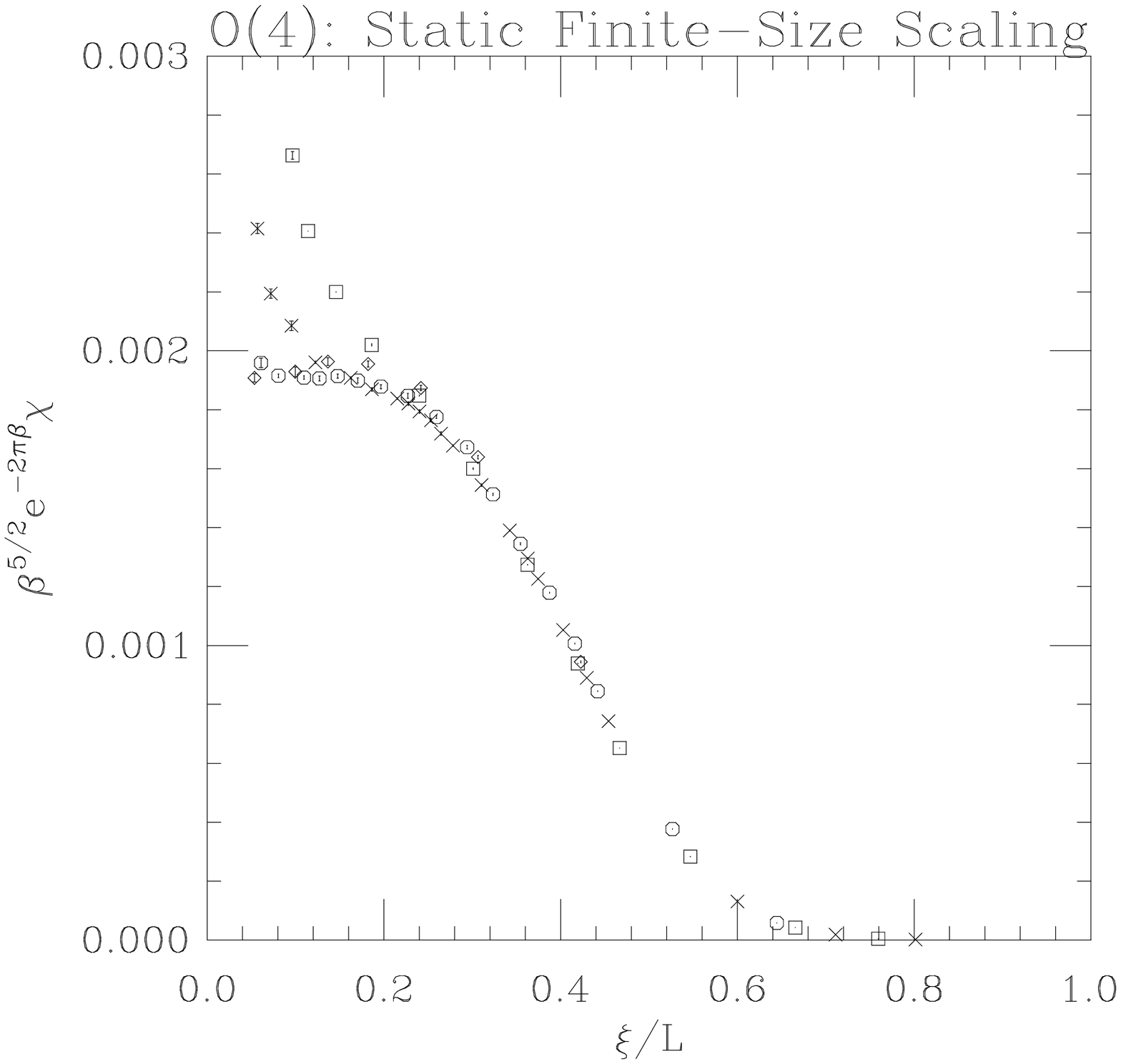}
\caption[fig3]{
   Finite-size-scaling plot of $\beta^{5/2} e^{-2\pi\beta} \chi$
   versus $\xi/L$, for lattice sizes $L=32$ ($\Box$),
   64 ($\times$), 128 ($\bigcirc$), 256 ($\Diamond$).
   Error bars are in most cases invisible.
}
\label{mgmco4_chi_scaling} 

\end{figure}

We fit $\xi(\beta,L)$ to the two-loop asymptotic-freedom
finite-size-scaling prediction \reff{O(N)_xi_FSS_predicted2}
by plotting $\xi \beta^{1/2} e^{-\pi\beta}$ versus $\xi/L$
(Figure~\ref{mgmco4_xi_scaling}).
Likewise, we fit $\chi(\beta,L)$ to the prediction
\reff{O(N)_chi_FSS_predicted2}
by plotting $\chi \beta^{5/2} e^{-2\pi\beta}$ versus $\xi/L$
(Figure~\ref{mgmco4_chi_scaling}).
In both cases the points fall quite nicely on a single curve,
with the exception of the data points at $\beta \ltapprox 2.0$
($\xi \ltapprox 8$).
Note also that the scaling curves drop precipitously for $\xi/L \gtapprox 0.4$
(corresponding to $\xi_\infty/L \gtapprox 0.5$);
thus, only those points with $\xi/L \ltapprox 0.4$ bear much
relation to the infinite-volume physics.
By extrapolating the scaling curves to $\xi/L = 0$,
we can estimate the constants $C_\xi$ and $C_\chi$ defined in
\reff{O(N)_xi_predicted}/\reff{O(N)_chi_predicted}:
\begin{subeqnarray}
   C_{\xi\hbox{\scriptsize (2-loop)}}   & = &  (1.93 \pm 0.10) \times 10^{-2} \\
   C_{\chi\hbox{\scriptsize (2-loop)}}  & = &  (1.88 \pm 0.08) \times 10^{-3}
\end{subeqnarray}
(subjective 68\% confidence intervals).
Unfortunately, this value of $C_\xi$ cannot be directly compared with
the exact value
$C_\xi = 2.349109\ldots \times 10^{-2}$ predicted by the Bethe Ansatz
[cf.\ \reff{exact_Cxi}],
because we use the second-moment definition of $\xi$
[cf.\ \reff{def_xi}]
while the Bethe-Ansatz work
\cite{Hasenfratz-Niedermayer_1,Hasenfratz-Niedermayer_2,%
Hasenfratz-Niedermayer_3}
uses the exponential definition (= inverse mass gap).
However, it is found empirically that the two definitions of $\xi$
agree to within less than 1\% ---
compare our Tables~\ref{o4_merged_table1} and \ref{o4_merged_table2}
with Table~1 of \cite{Wolff_O4_O8} ---
and it would be very interesting to understand {\em why}\/ this is so!

Finally, it should be noted from
Tables~\ref{o4_merged_table1} and \ref{o4_merged_table2}
that the finite-size corrections to $\xi$ and $\chi$ at fixed $\beta$
are of order 1--2\% when $L \approx 6\xi$ ($\approx 6 \xi_\infty$);
this finite-size effect should be taken into account in
future high-precision work.

The three-loop correction terms $a_1$ and $b_1$
in \reff{O(N)_xi_predicted}/\reff{O(N)_chi_predicted}
have been computed by Falcioni and Treves \cite{Falcioni_86}:
they are
\begin{subeqnarray}
   a_1   & = &   - {1 \over 2\pi(N-2)}  \left[ 1 - (N-2) h_1 - h_2 \right]
                                                                    \\[0.2cm]
   b_1   & = &   - {1 \over 2\pi(N-2)}
                      \left[ 2 - 2(N-2) h_1 - 2h_2 + (N-1)(1-k) \right]
\end{subeqnarray}
where $h_1$, $h_2$ and $k$ can be computed by evaluating lattice Feynman
diagrams;  for the nearest-neighbor action \reff{eqn1} they are
$h_1 \approx -0.09$, $h_2 \approx 0.52$, $k \approx 1.57$.
Thus, for $N=4$ we have $a_1 \approx -0.053$ and $b_1 \approx 0.031$.
As already noted by Falcioni and Treves for the case $N=3$,
these corrections are too small to affect much the quality of the fit
in Figures~\ref{mgmco4_xi_scaling} and \ref{mgmco4_chi_scaling}.
However, they do affect the estimates of $C_\xi$ and $C_\chi$
by a few percent:
\begin{subeqnarray}
   C_{\xi\hbox{\scriptsize (3-loop)}}   & \approx &
      C_{\xi\hbox{\scriptsize (2-loop)}}  \times
      \left( 1  +  {0.053 \over \beta_{av}} \right)     \\
   C_{\chi\hbox{\scriptsize (3-loop)}}  & \approx &
      C_{\chi\hbox{\scriptsize (2-loop)}}  \times
      \left( 1  -  {0.031 \over \beta_{av}} \right)
\end{subeqnarray}
where $\beta_{av} \approx 2.4$ is an average value of $\beta$ among those
data points contributing to the estimates of $C_\xi$ and $C_\chi$
(i.e.\ $L=128,256$ and $\xi/L \ltapprox 0.2$).
Hence our best estimates for $C_\xi$ and $C_\chi$ are
\begin{subeqnarray}
   C_{\xi\hbox{\scriptsize (3-loop)}}   & = &  (1.97 \pm 0.10) \times 10^{-2} \\
   C_{\chi\hbox{\scriptsize (3-loop)}}  & = &  (1.86 \pm 0.08) \times 10^{-3}
\end{subeqnarray}
(subjective 68\% confidence intervals).
Unless the higher-order correction coefficients
$a_2,a_3,\ldots,$  $b_2,b_3,\ldots$
turn out to be surprisingly large,
the deviations from scaling observed for $\xi \ltapprox 8$
will have to be ascribed to nonperturbative effects.

\begin{figure}
\epsfxsize=\textwidth
\epsffile{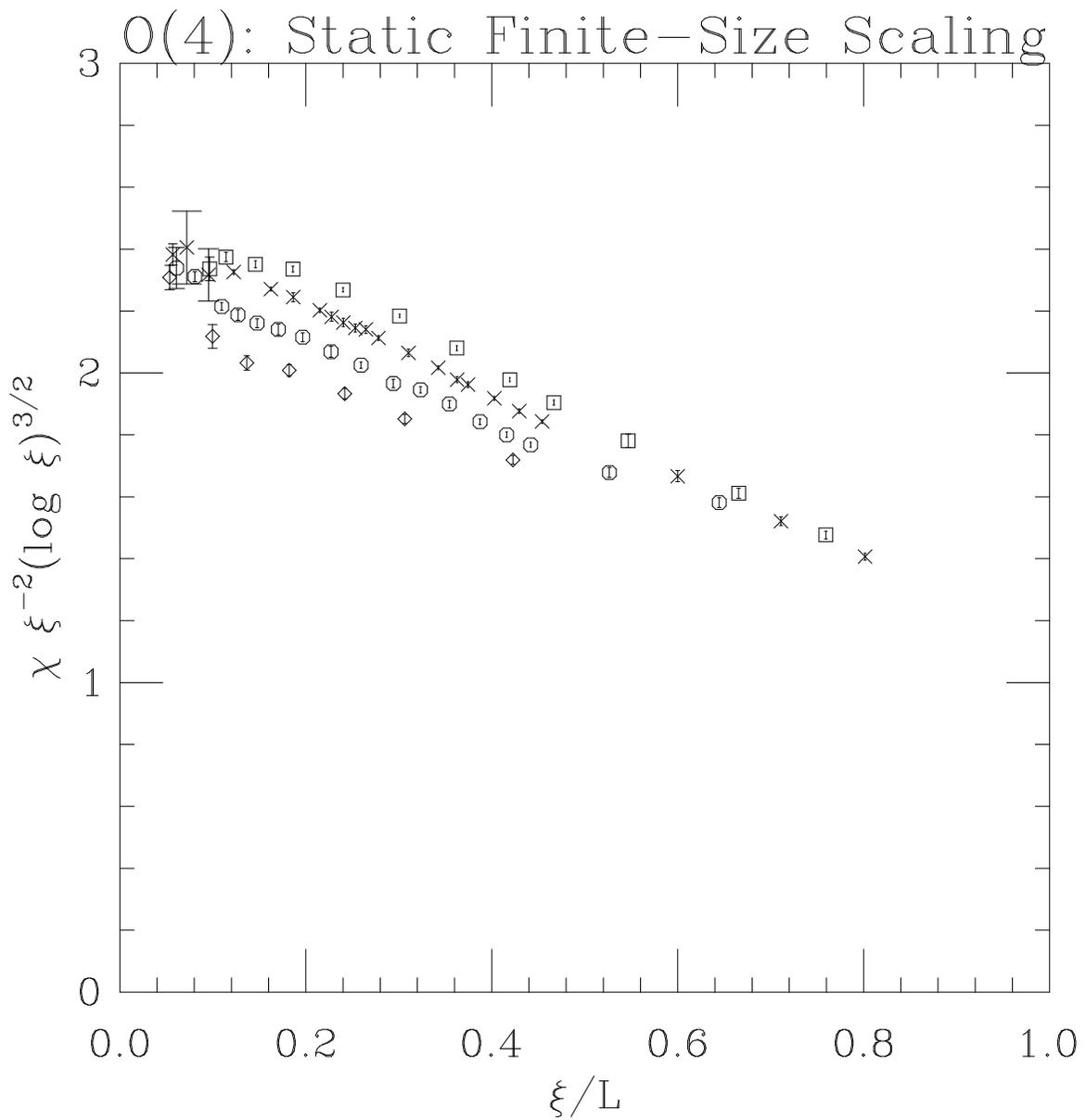}
\caption[fig4]{
   Finite-size-scaling plot of $\chi \xi^{-2} (\log\xi)^{3/2}$ versus $\xi/L$,
   for lattice sizes $L=32$ ($\Box$),
   64 ($\times$), 128 ($\bigcirc$), 256 ($\Diamond$).
   Error bars are in most cases invisible.
}
\label{mgmco4_chi-versus-xi_scaling}
\end{figure}

We also tried to fit $\chi$ and $\xi$ to the prediction
\reff{O(N)_chi-versus-xi_FSS_predicted2}
by plotting $\chi \xi^{-2} (\log\xi)^{3/2}$ versus $\xi/L$.
However, the fit was rather poor
(see Figure \ref{mgmco4_chi-versus-xi_scaling});
and it was not noticeably improved by including the universal correction
term $c_{11} \log\log\xi/\log\xi$ from \reff{O(N)_chi-versus-xi_predicted}.
Apparently the trouble is that $|\log C_\xi|$ is comparable in magnitude
to $\log\xi$ for the range of temperatures considered here
(both are $\approx 4$), so the correction term $c_{01}/\log\xi$
is quite significant (and higher-order corrections may also be significant).

In any case, our data show quite good agreement ---
as good as one has a right to expect for this range of $\beta$ ---
with the two-loop and three-loop asymptotic-freedom predictions.

Recently, however, Patrascioiu, Seiler and collaborators
\cite{Patrascioiu_87,Patrascioiu_89} have argued
on theoretical grounds
that non-Abelian $\sigma$-models in two dimensions should have a
conventional power-law critical point at {\em finite}\/ $\beta$ ---
in sharp contrast to the conventional wisdom
\reff{O(N)_xi_predicted}--\reff{O(N)_chi-versus-xi_FSS_predicted2}.
Moreover, they have claimed to find evidence of such a critical point
in a Monte Carlo study \cite{Seiler_88} of the dodecahedron model
(a discrete-spin model that is expected to fall in the universality
class of the $O(4)$-symmetric nonlinear $\sigma$-model)
on lattices of size up to $L=80$:
their best estimates of the critical exponents are $\nu \approx 2.02$
and $\gamma \approx 3.29$ (with uncertainty of order $\pm 0.1$)
and thus $2-\eta = \gamma/\nu \approx 1.627$.
We therefore attempted to fit our data for $\xi$ and $\chi$
to the finite-size-scaling forms
\begin{eqnarray}
   \xi(\beta,L)   & \sim &
      L  \, f_\xi \!\left( (\beta-\beta_c) L^{1/\nu} \right)
                                           \label{2nd_order_FSS_xi}    \\
   \chi(\beta,L)  & \sim &
      L^{\gamma/\nu} \,  f_\chi \!\left( (\beta-\beta_c) L^{1/\nu} \right)
                                           \label{2nd_order_FSS_chi}
\end{eqnarray}
We also fit our data to the Ansatz
\be
 \label{2nd_order_FSS_chi-versus-xi}
   \chi(\beta,L)   \;\sim\;   \xi^{\gamma/\nu} \, f_{\chi\xi}(\xi/L)
\ee
[where $\xi \equiv \xi(\beta,L)$ and
 $\lim_{x\to 0} f_{\chi\xi}(x) = C_{\chi\xi} > 0$],
which follows from \reff{2nd_order_FSS_xi}/\reff{2nd_order_FSS_chi}
by eliminating $\beta$.  This latter Ansatz does not test the existence of a
power-law critical point at finite $\beta$, but rather tests
the value of the exponent $\gamma/\nu$ without needing to estimate $\beta_c$.

\begin{figure}
\epsfxsize=\textwidth
\epsffile{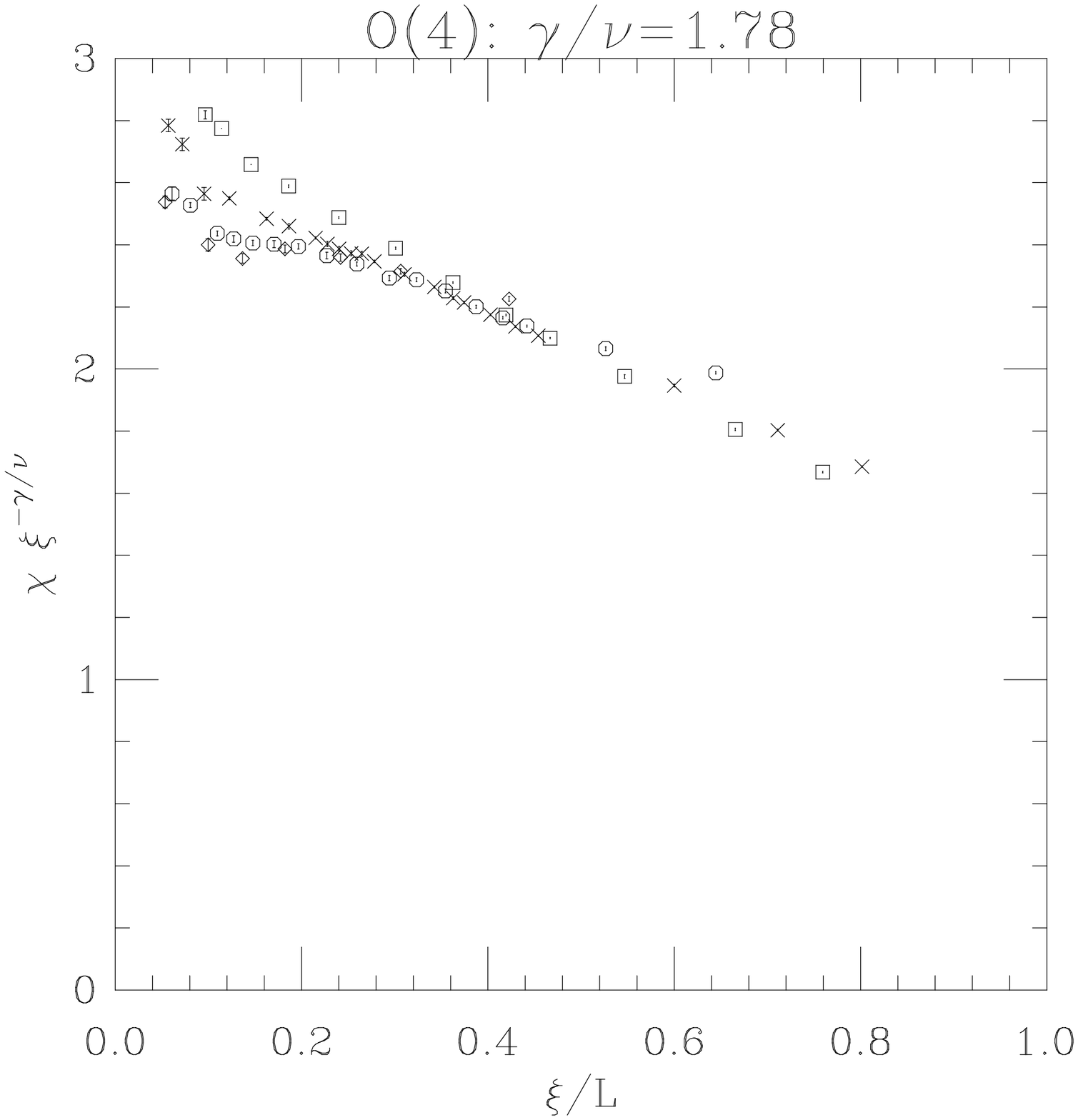}
\caption[fig5]{
   Finite-size-scaling plot of $\chi \xi^{-\gamma/\nu}$ versus $\xi/L$,
   for lattice sizes $L=32$ ($\Box$),
   64 ($\times$), 128 ($\bigcirc$), 256 ($\Diamond$).
   Here $\gamma/\nu = 1.78$.
   Error bars are in most cases invisible.
}
\label{mgmco4_chi-versus-xi_seiler_scaling_1.78} 
\end{figure}

First we attempted a fit to \reff{2nd_order_FSS_chi-versus-xi}
by plotting $\chi \xi^{-\gamma/\nu}$ versus $\xi/L$
for a variety of values of $\gamma/\nu$.
We were unable to obtain a satisfactory fit for any choice of $\gamma/\nu$.
In particular, the Patrascioiu-Seiler value $\gamma/\nu = 1.627$
leads to an extremely poor fit.
The least poor fit was found at $\gamma/\nu \approx 1.78$;
it is shown in Figure~\ref{mgmco4_chi-versus-xi_seiler_scaling_1.78}. 
However, we do not feel entitled to interpret the inadequacy of these fits
as evidence against the Ansatz \reff{2nd_order_FSS_chi-versus-xi},
because the fit to the corresponding asymptotic-freedom Ansatz
\reff{O(N)_chi-versus-xi_FSS_predicted2}
was equally poor.

\begin{figure}
\begin{center}
\epsfxsize=0.5\textwidth
\leavevmode\epsffile{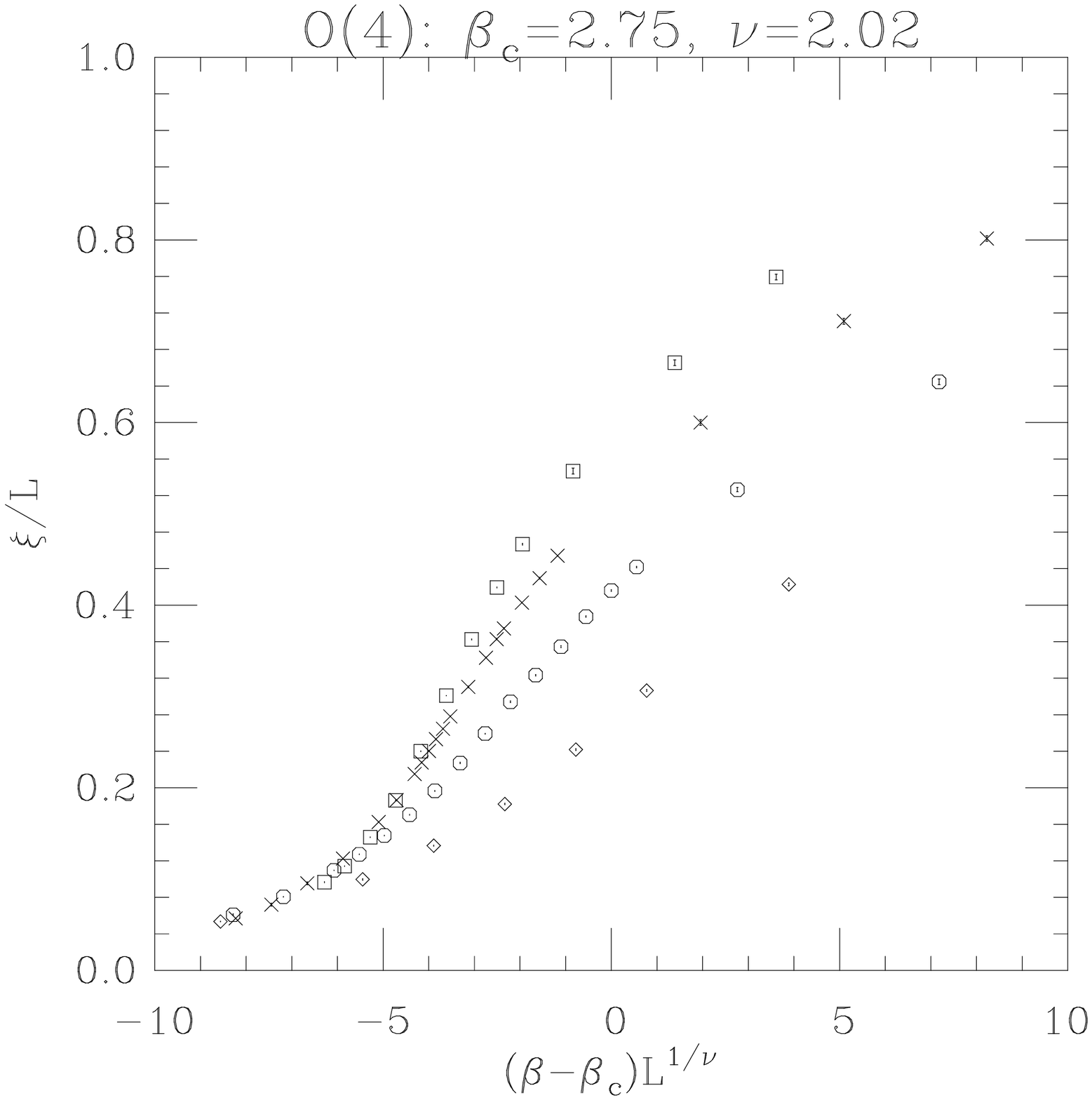} \qquad\qquad (a)   \\
\vspace{2.5cm}
\epsfxsize=0.5\textwidth
\leavevmode\epsffile{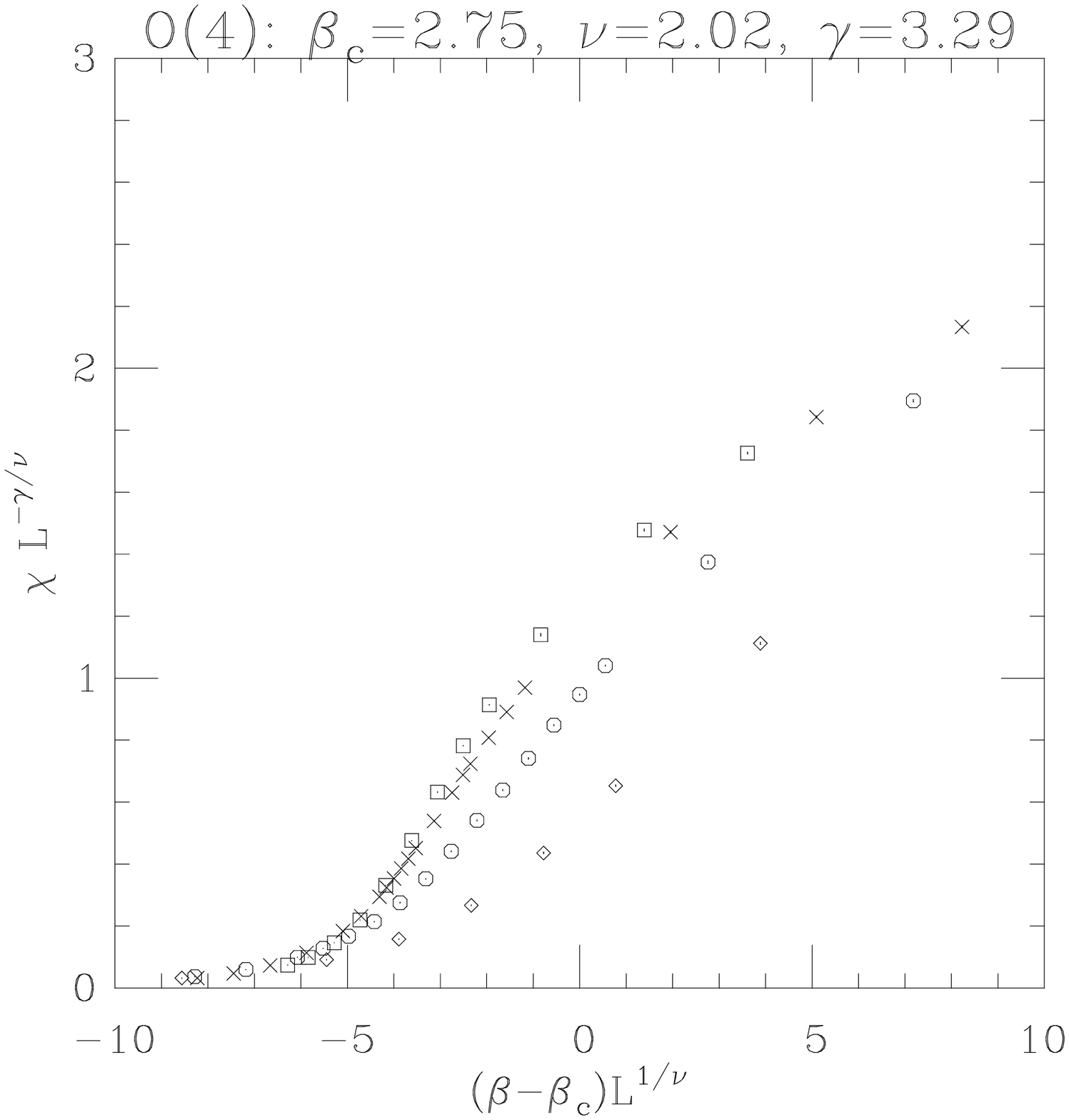} \qquad\qquad (b)
\end{center}
\vspace{0.5cm}
\caption[fig6]{  
   Finite-size-scaling plot of $\xi/L$ (a) and $\chi L^{-\gamma/\nu}$ (b)
   versus $(\beta-\beta_c) L^{1/\nu}$, for lattice sizes $L=32$ ($\Box$),
   64 ($\times$), 128 ($\bigcirc$), 256 ($\Diamond$).
   Here $\beta_c = 2.75$;
   values of $\nu=2.02$ and $\gamma=3.29$ are taken 
    from ref.~\protect\cite{Seiler_88}.
   Error bars are in most cases invisible.
}
\label{mgmco4_chi_seiler_scaling_2.75_2.02_3.29} 
\end{figure}

\begin{figure}
\begin{center}
\epsfxsize=0.5\textwidth
\leavevmode\epsffile{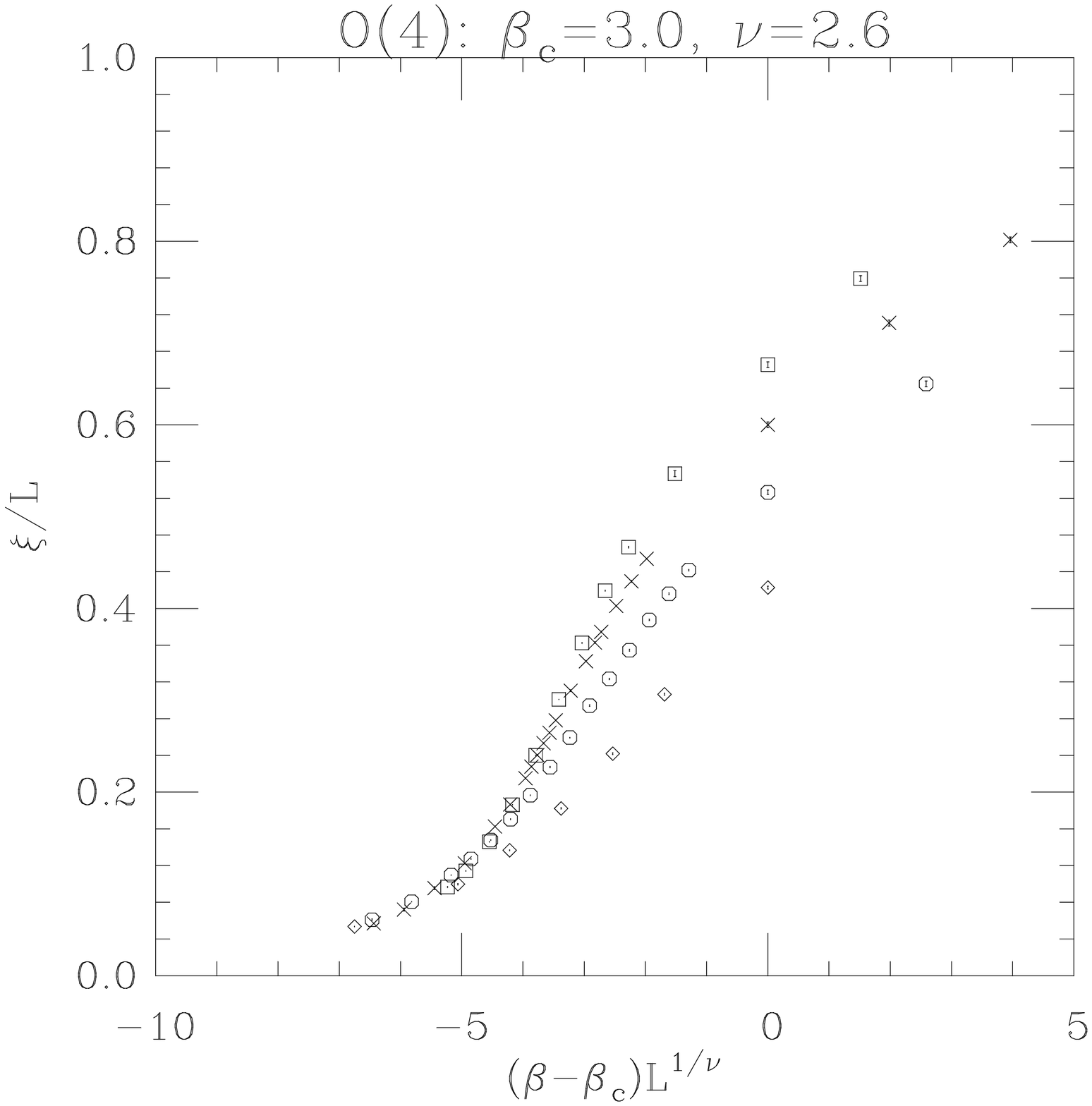} \qquad\qquad (a)   \\
\vspace{2.5cm}
\epsfxsize=0.5\textwidth
\leavevmode\epsffile{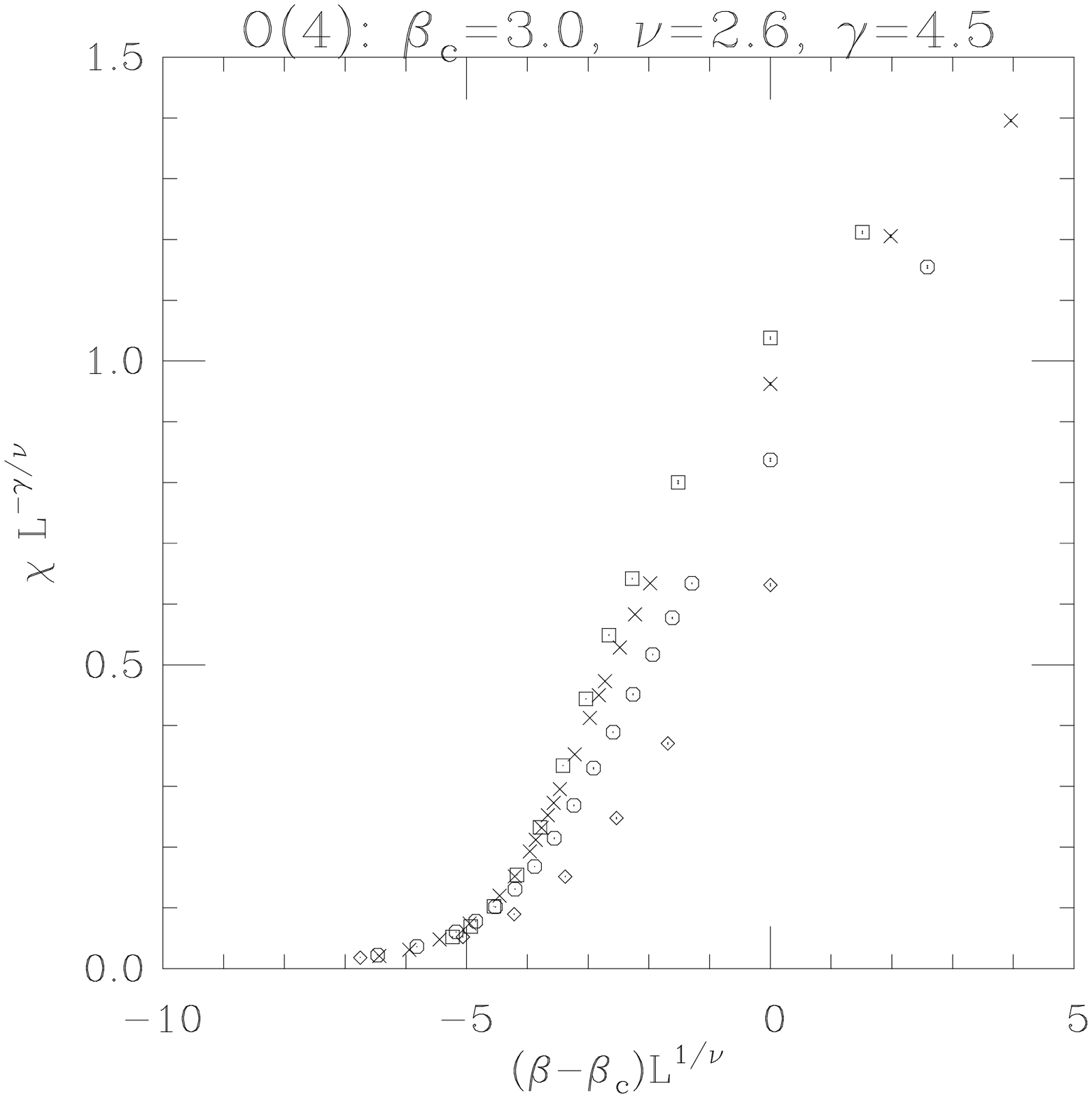} \qquad\qquad (b)
\end{center}
\vspace{0.5cm}
\caption[fig7]{
   Finite-size-scaling plot of $\xi/L$ (a) and $\chi L^{-\gamma/\nu}$ (b)
   versus $(\beta-\beta_c) L^{1/\nu}$, for lattice sizes $L=32$ ($\Box$),
   64 ($\times$), 128 ($\bigcirc$), 256 ($\Diamond$).
   Here $\beta_c = 3.0$, $\nu = 2.6$ and $\gamma = 4.5$.
   Error bars are in most cases invisible.
}
\label{mgmco4_xi-and-chi_2nd_order_scaling_betac_3} 
\end{figure}

\begin{figure}
\begin{center}
\epsfxsize=0.5\textwidth
\leavevmode\epsffile{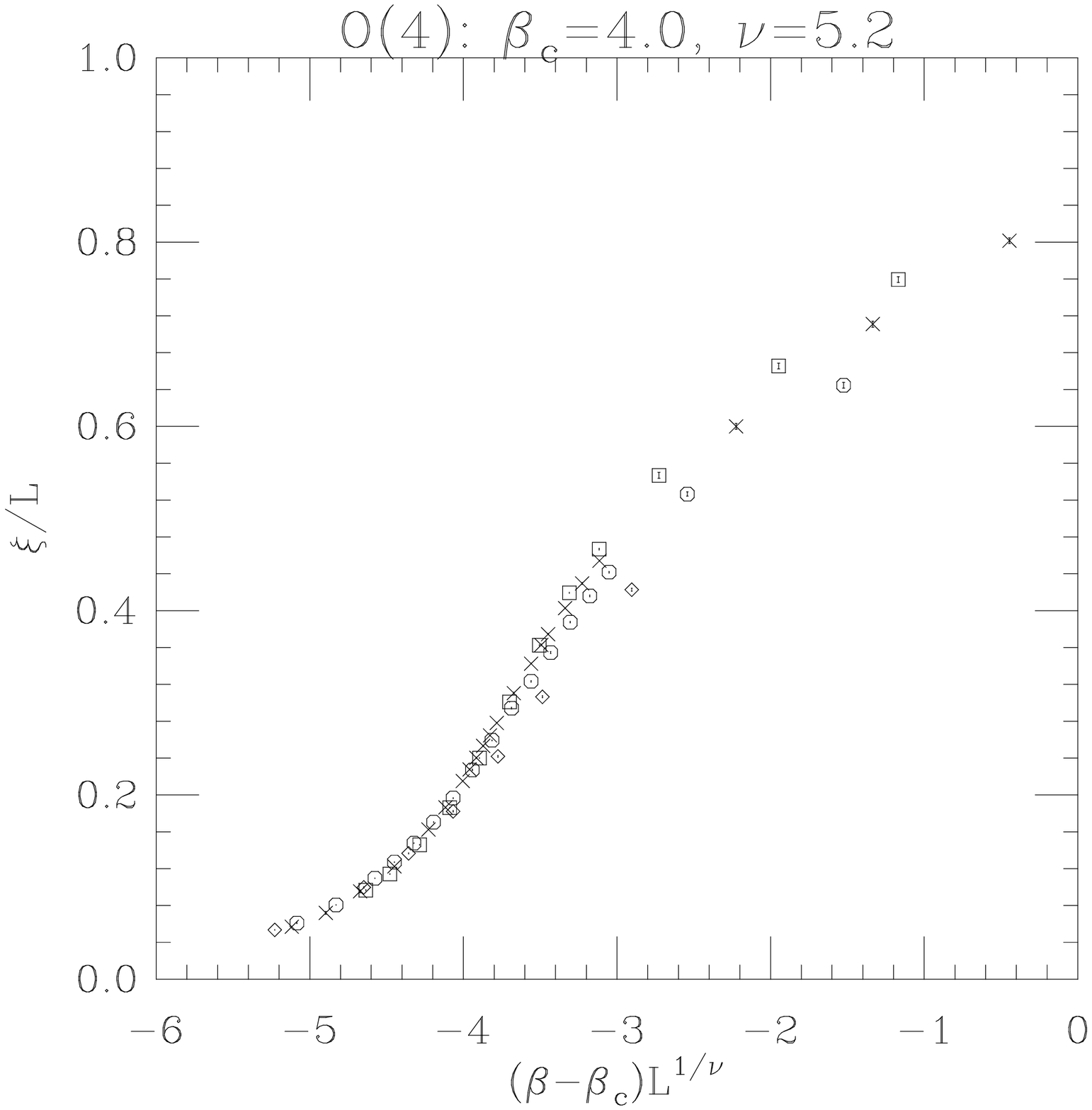} \qquad\qquad (a)   \\
\vspace{2.5cm}
\epsfxsize=0.5\textwidth
\leavevmode\epsffile{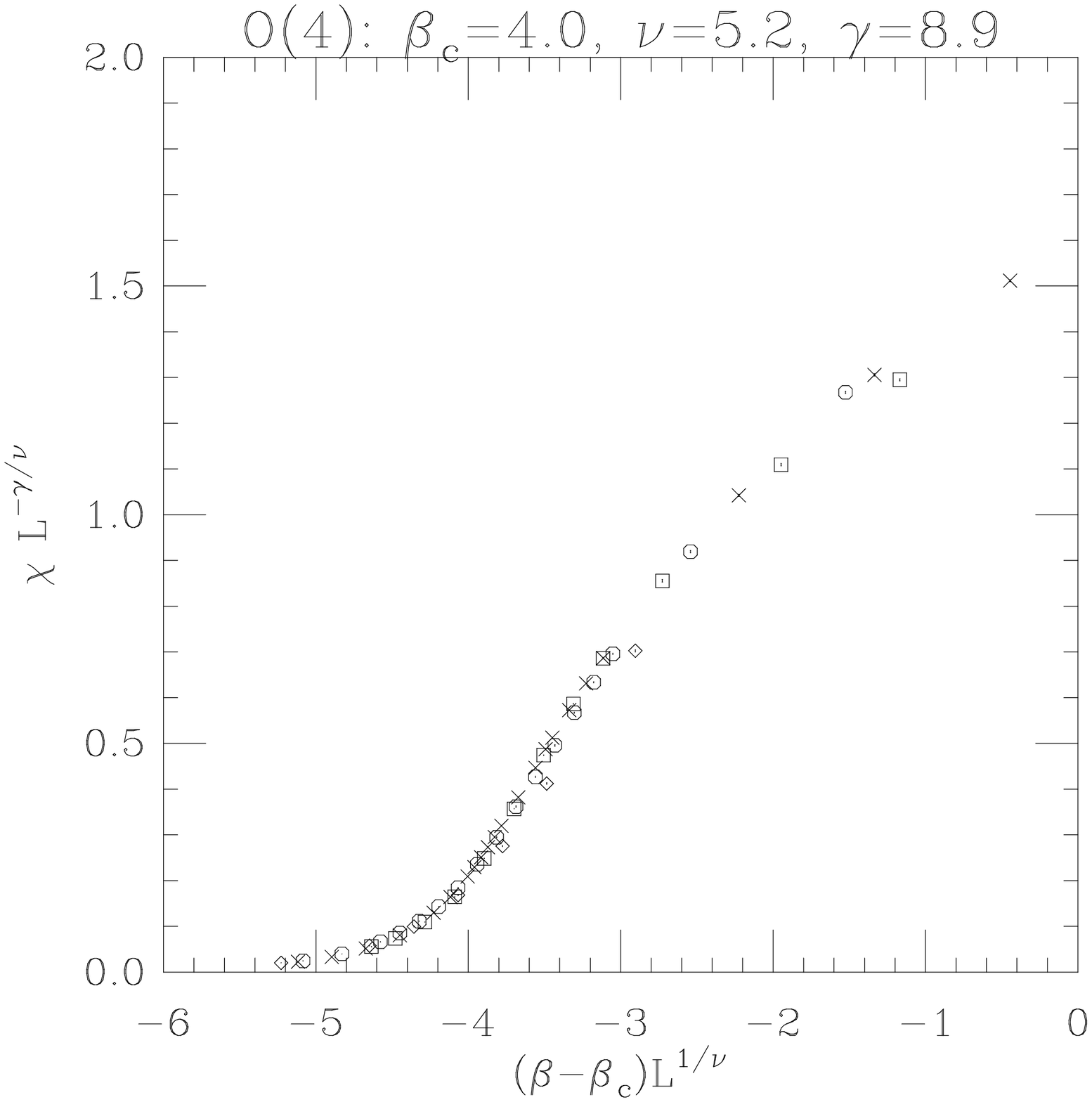} \qquad\qquad (b)
\end{center}
\vspace{0.5cm}
\caption[fig8]{
   Finite-size-scaling plot of $\xi/L$ (a) and $\chi L^{-\gamma/\nu}$ (b)
   versus $(\beta-\beta_c) L^{1/\nu}$, for lattice sizes $L=32$ ($\Box$),
   64 ($\times$), 128 ($\bigcirc$), 256 ($\Diamond$).
   Here $\beta_c = 4.0$, $\nu = 5.2$ and $\gamma = 8.9$.
   Error bars are in most cases invisible.
}
\label{mgmco4_xi-and-chi_2nd_order_scaling_betac_4} 
\end{figure}

\begin{figure}
\begin{center}
\epsfxsize=0.5\textwidth
\leavevmode\epsffile{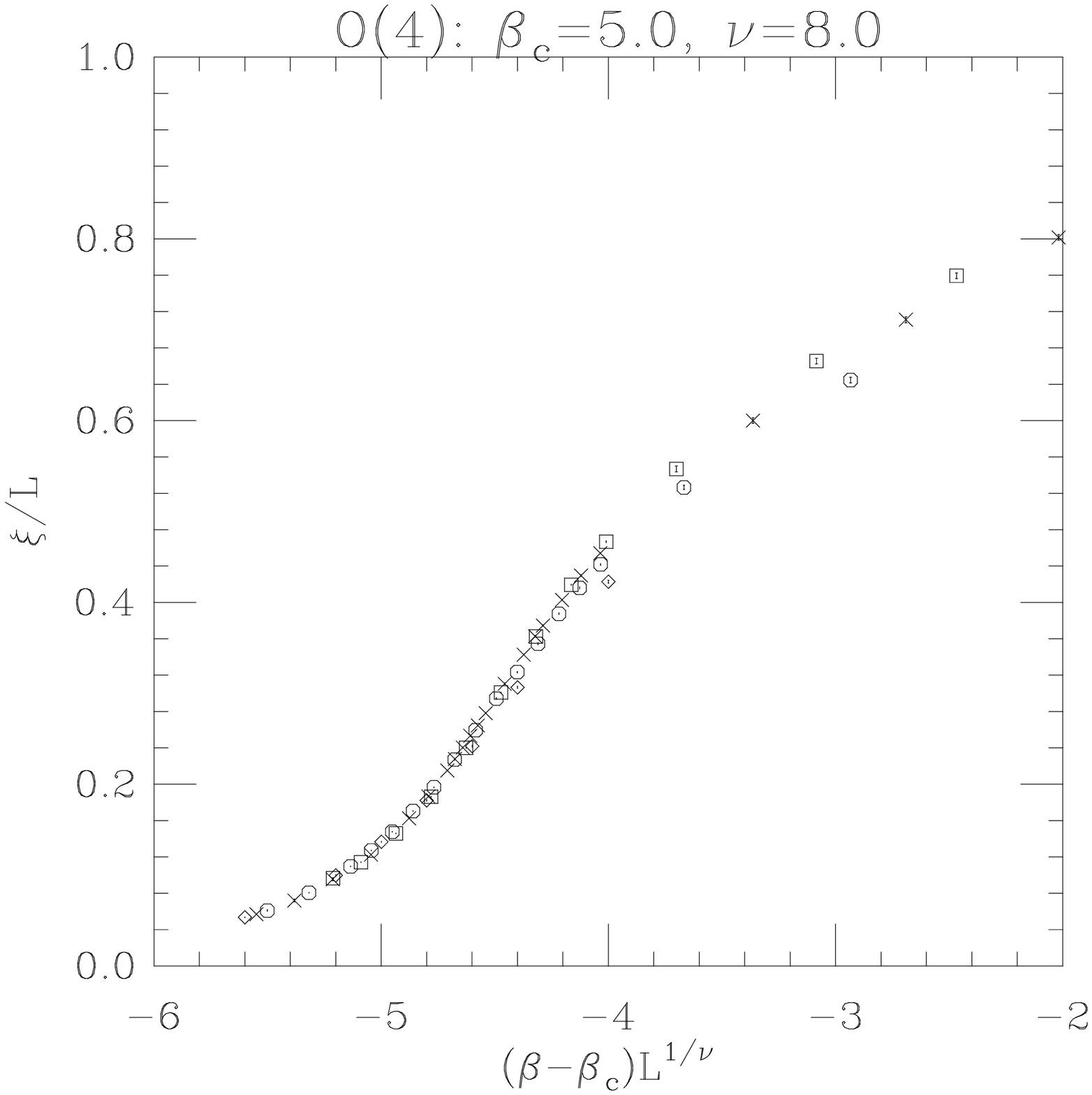} \qquad\qquad (a)   \\
\vspace{2.5cm}
\epsfxsize=0.5\textwidth
\leavevmode\epsffile{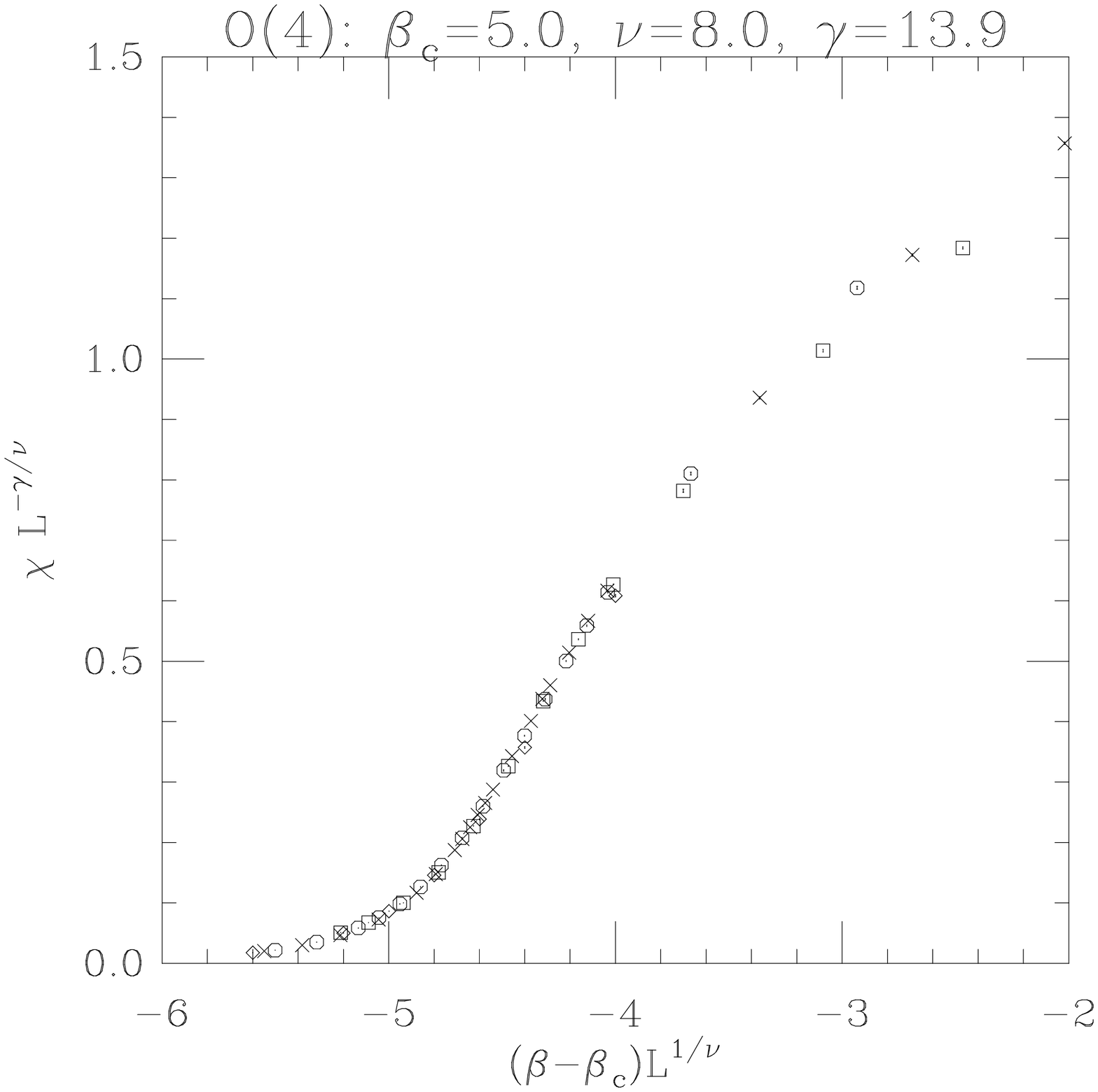} \qquad\qquad (b)
\end{center}
\vspace{0.5cm}
\caption[fig9]{
   Finite-size-scaling plot of $\xi/L$ (a) and $\chi L^{-\gamma/\nu}$ (b)
   versus $(\beta-\beta_c) L^{1/\nu}$, for lattice sizes $L=32$ ($\Box$),
   64 ($\times$), 128 ($\bigcirc$), 256 ($\Diamond$).
   Here $\beta_c = 5.0$, $\nu = 8.0$ and $\gamma = 13.9$.
   Error bars are in most cases invisible.
}
\label{mgmco4_xi-and-chi_2nd_order_scaling_betac_5} 
\end{figure}

\begin{figure}
\begin{center}
\epsfxsize=0.5\textwidth
\leavevmode\epsffile{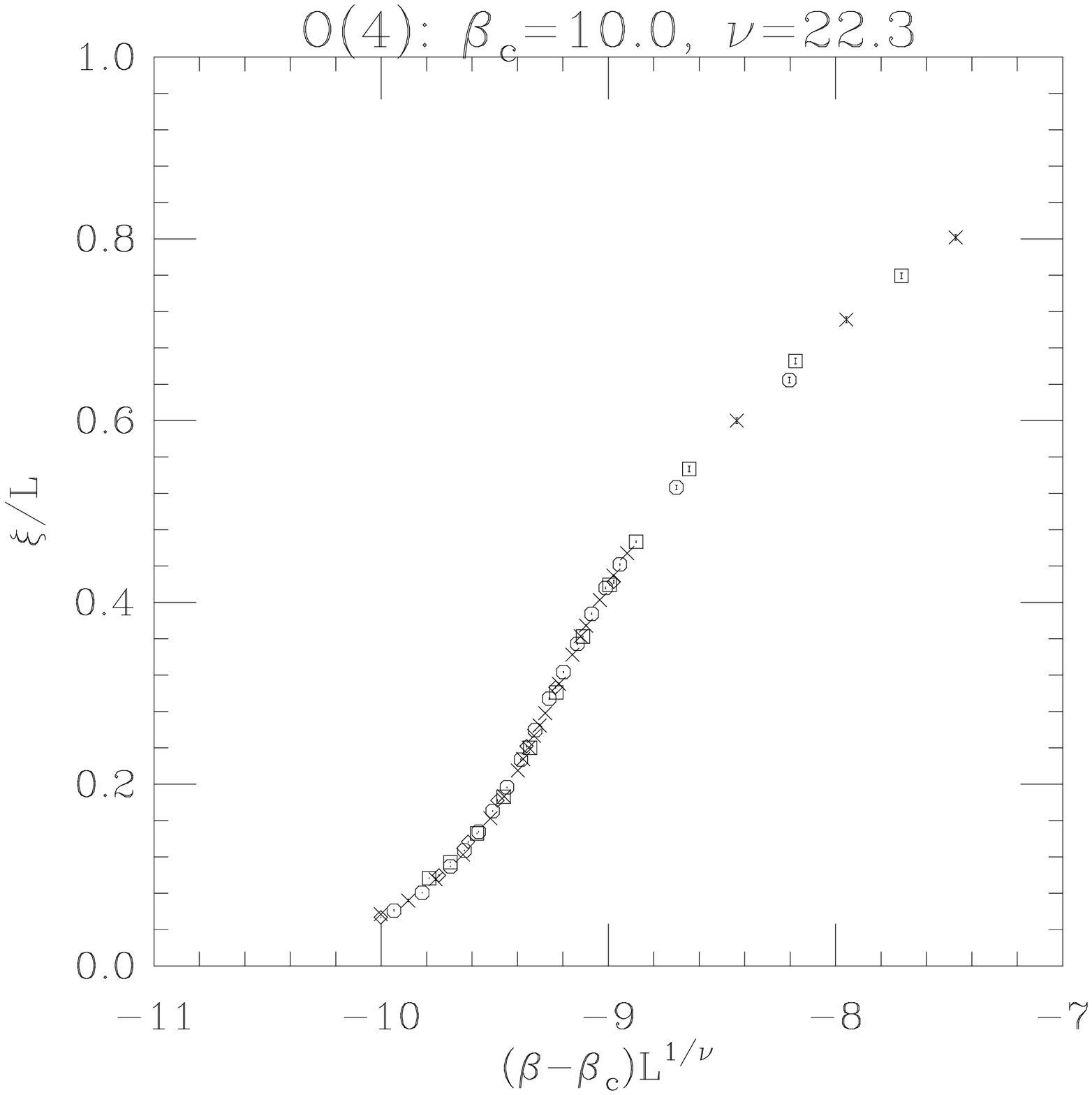} \qquad\qquad (a)   \\
\vspace{2.5cm}
\epsfxsize=0.5\textwidth
\leavevmode\epsffile{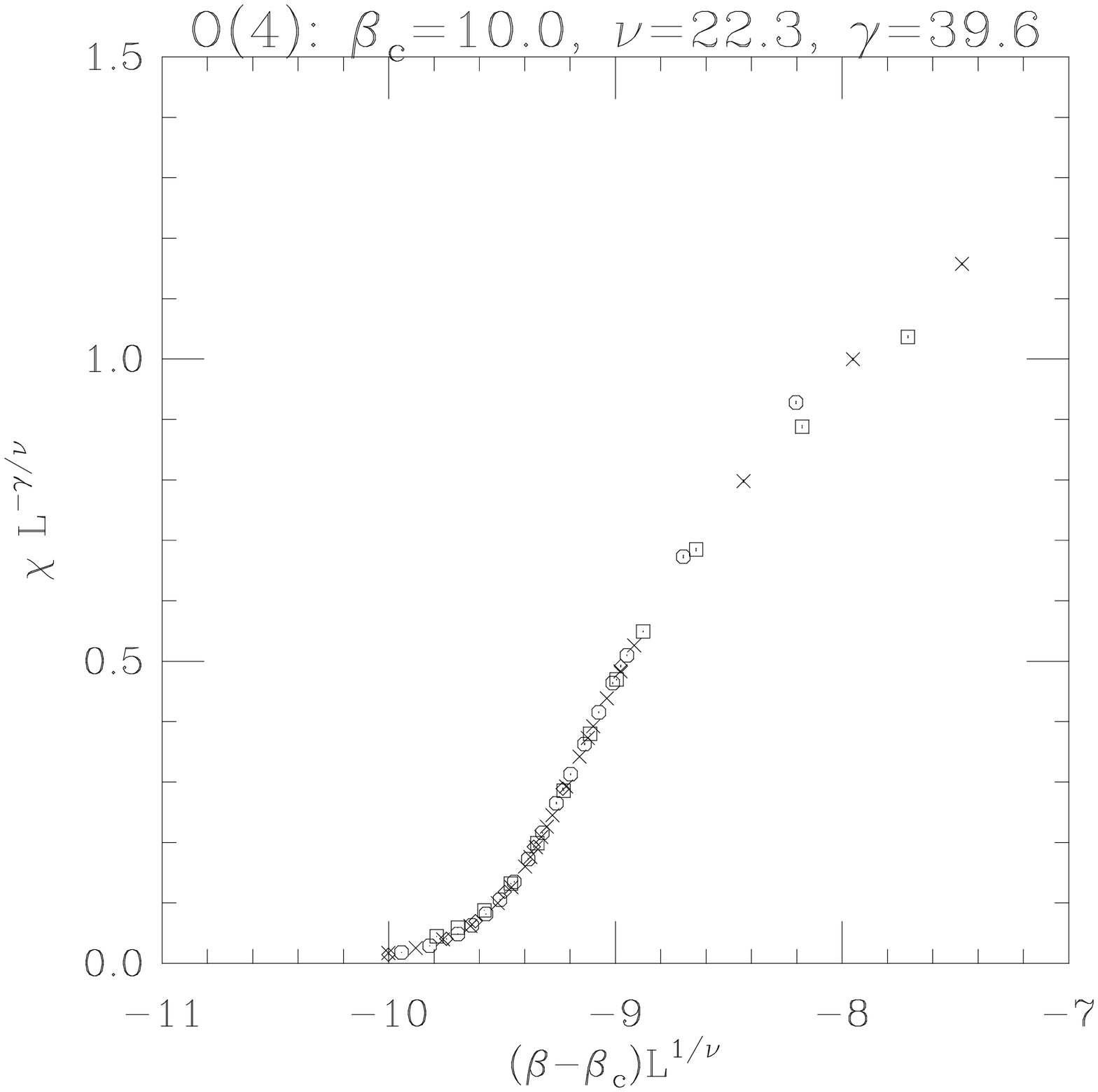} \qquad\qquad (b)
\end{center}
\vspace{0.5cm}
\caption[fig10]{
   Finite-size-scaling plot of $\xi/L$ (a) and $\chi L^{-\gamma/\nu}$ (b)
   versus $(\beta-\beta_c) L^{1/\nu}$, for lattice sizes $L=32$ ($\Box$),
   64 ($\times$), 128 ($\bigcirc$), 256 ($\Diamond$).
   Here $\beta_c = 10.0$, $\nu = 22.3$ and $\gamma = 39.6$.
   Error bars are in most cases invisible.
}
\label{mgmco4_xi-and-chi_2nd_order_scaling_betac_10} 
\end{figure}

Next we attempted fits to \reff{2nd_order_FSS_xi} and \reff{2nd_order_FSS_chi},
which have two and three free parameters, respectively.
We began by imposing the Patrascioiu-Seiler values
$\nu = 2.02$ and $\gamma = 3.29$ 
and varying $\beta_c$ until we obtained the best fit;
the results are shown in Figure~\ref{mgmco4_chi_seiler_scaling_2.75_2.02_3.29}.
We see that at $\beta \ltapprox 2.1 - 2.3$
(corresponding to $\xi \ltapprox 10 - 19$),
the fits are reasonable.
However, there are very large deviations at larger values of $\beta$,
which cannot be removed simply by varying $\beta_c$.
Therefore we tried an unbiased study:
at each of various trial values of $\beta_c$,
we first varied $\nu$ until we obtained the best fit to
\reff{2nd_order_FSS_xi}; 
then we fixed this value of $\nu$
and varied $\gamma$ until we obtained the best fit to \reff{2nd_order_FSS_chi}.
Some typical results are shown in
Figures~\ref{mgmco4_xi-and-chi_2nd_order_scaling_betac_3} ($\beta_c = 3$),
\ref{mgmco4_xi-and-chi_2nd_order_scaling_betac_4} ($\beta_c = 4$),
\ref{mgmco4_xi-and-chi_2nd_order_scaling_betac_5} ($\beta_c = 5$), and
\ref{mgmco4_xi-and-chi_2nd_order_scaling_betac_10} ($\beta_c = 10$).
%
For the smaller values of $\beta_c$, the fit is still rather poor;
as $\beta_c$ is raised, the fit remains good out to larger values of $\beta$.
Moreover, the optimal values of $\nu$ and $\gamma$ grow rapidly
with $\beta_c$:  they are in fact very roughly proportional to $\beta_c$
(or rather, to $\beta_c - {\rm const}$ with ${\rm const} \approx 2$).
This is exactly the behavior one would expect if the asymptotic-freedom
Ansatz \reff{O(N)_xi_predicted}--\reff{O(N)_chi-versus-xi_FSS_predicted2}
holds, since a power-law critical point at large $\beta_c$ and with
an exponent proportional to $\beta_c$ can mimic
(for $\beta$ considerably smaller than $\beta_c$)
an exponential growth:
\be
  \left( 1 - {\beta \over \beta_c} \right) ^{-C \beta_c}
  \;=\;
  e^{C \beta} \,
    \left[ 1 + {\beta \over 2 \beta_c} +
           O \!\left( \left( {\beta \over \beta_c} \right) ^2 \right)
    \right]   \;.
\ee
%
%

%
%

We believe that our data rule out a power-law critical point for
$\beta_c \ltapprox 5$ (a value well beyond the betas at which we ran);
and our data are compatible with a power-law critical point at
$\beta_c \gtapprox 5$ only if one assumes
implausibly large values of $\nu$ and $\gamma$.
We conclude that the two-dimensional $O(4)$ model almost certainly
does {\em not}\/ have a power-law critical point at finite $\beta$.

\subsection{Finite-Size-Scaling Analysis: Dynamic Quantities}
\label{subsection:finite-size-scaling_dynamic}
\vspace{-0.3cm}\quad\par

Next we apply finite-size scaling to the dynamic quantities $\taum$ and $\taux$.
We use the Ansatz
\be
  \tau_{int,A}(\beta,L)  \;\sim\;
     \xi(\beta,L)^{z_{int,A}} \, g_A \Bigl( \xi(\beta,L)/L  \Bigr)
 \label{dyn_FSS_Ansatz}
\ee
for $A = \scrmvec, \scrm^2$.
Here $g_A$ is an unknown scaling function, and
$g_A(0) = \lim_{x \downarrow 0} g_A(x)$
is supposed to be finite and nonzero.\footnote{
   It is of course equivalent to use the Ansatz
   $$\tau_{int,A}(\beta,L)  \;\sim\;
     L^{z_{int,A}} \, h_A \Bigl( \xi(\beta,L)/L  \Bigr)  \;,$$
   and indeed the two Ans\"atze are related by $h_A(x) = x^{z_{int,A}} g_A(x)$.
   However, to determine whether
   $\lim_{x \downarrow 0} g_A(x) = \lim_{x \downarrow 0} x^{-z_{int,A}} h_A(x)$
   is nonzero, it is more convenient to inspect a graph of $g_A$
   than one of $h_A$.
}
We determine $z_{int,A}$ by plotting $\tau_{int,A} \xi^{-z_{int,A}}$
versus $\xi/L$ and adjusting $z_{int,A}$ until the points fall as closely
as possible onto a single curve (with priority to the larger $L$ values).
We emphasize that the dynamic critical exponent $z_{int,A}$
is in general {\em different}\/ from the exponent $z_{exp}$
associated with the exponential autocorrelation time $\tau_{exp}$
\cite{Sokal_Lausanne,CPS_90,Sokal_LAT90}.

\begin{figure}
\epsfxsize=\textwidth
\epsffile{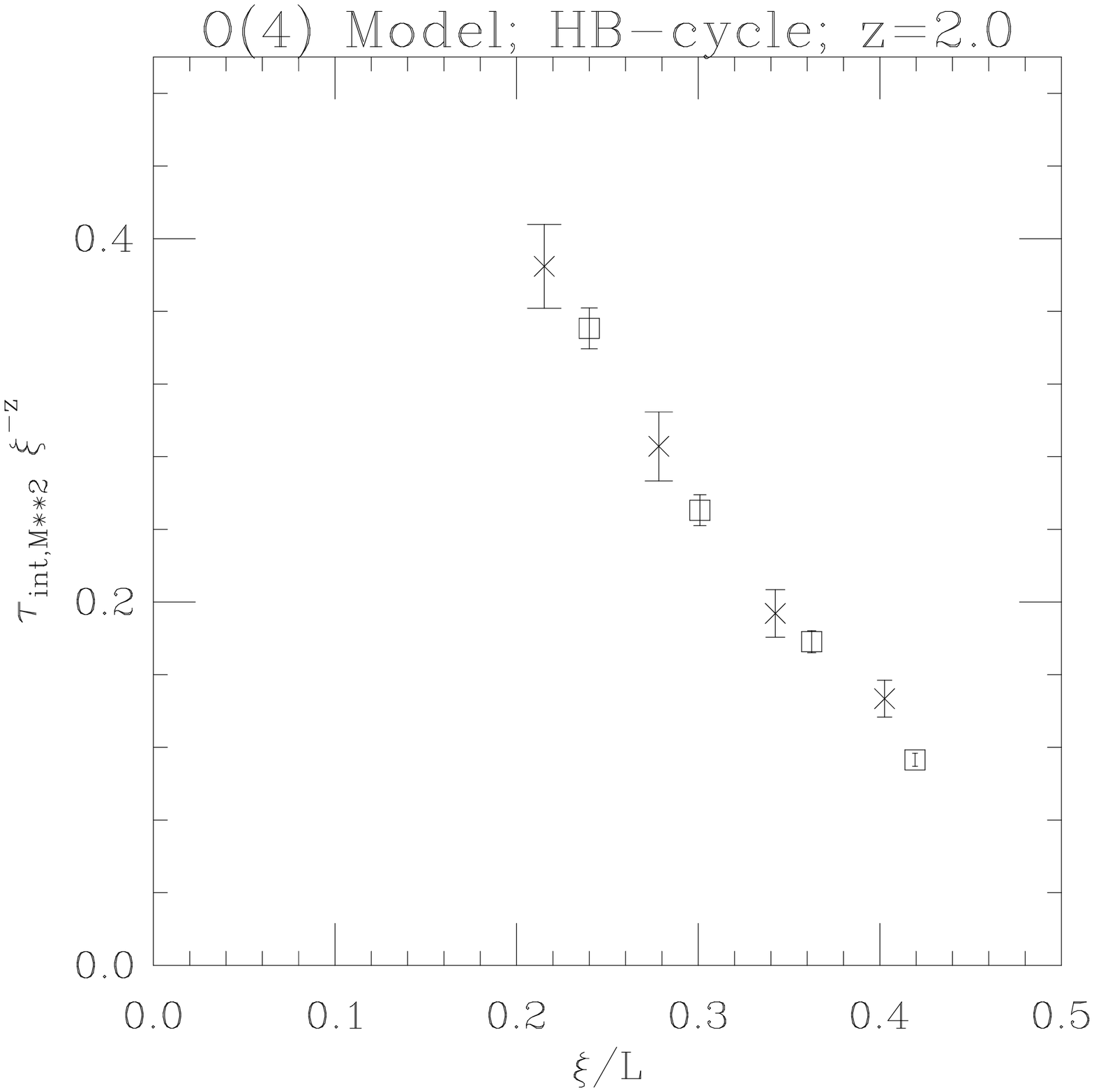}
\caption[fig11]{
   Finite-size-scaling plot of $\taux \xi^{-z_{int,\scrm^2}}$
   versus $\xi/L$ for the heat-bath algorithm,
   for lattice sizes $L=32$ ($\Box$) and 64 ($\times$).
   Here $z_{int,\scrm^2}=2.0$.
}
\label{fig_dynamic_fss_M2_HB}
\end{figure}

\begin{figure}
\epsfxsize=\textwidth
\epsffile{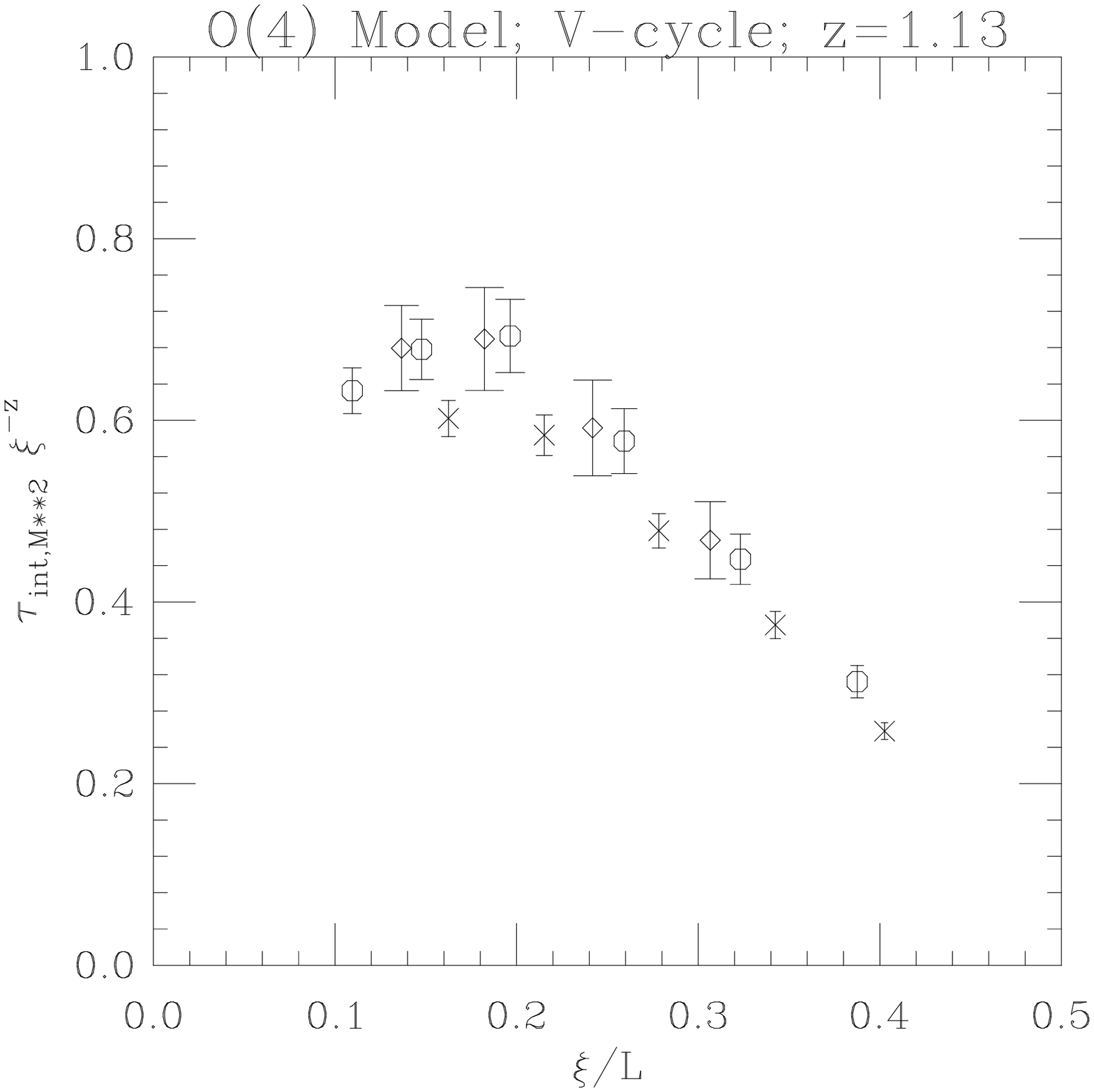}
\caption[fig12]{
   Finite-size-scaling plot of $\taux \xi^{-z_{int,\scrm^2}}$
   versus $\xi/L$ for the V-cycle algorithm,
   for lattice sizes $L=64$ ($\times$), 128 ($\bigcirc$) and 256 ($\Diamond$).
   Here $z_{int,\scrm^2}=1.13$.
}
\label{fig_dynamic_fss_M2_V}
\end{figure}

\begin{figure}
\epsfxsize=\textwidth
\epsffile{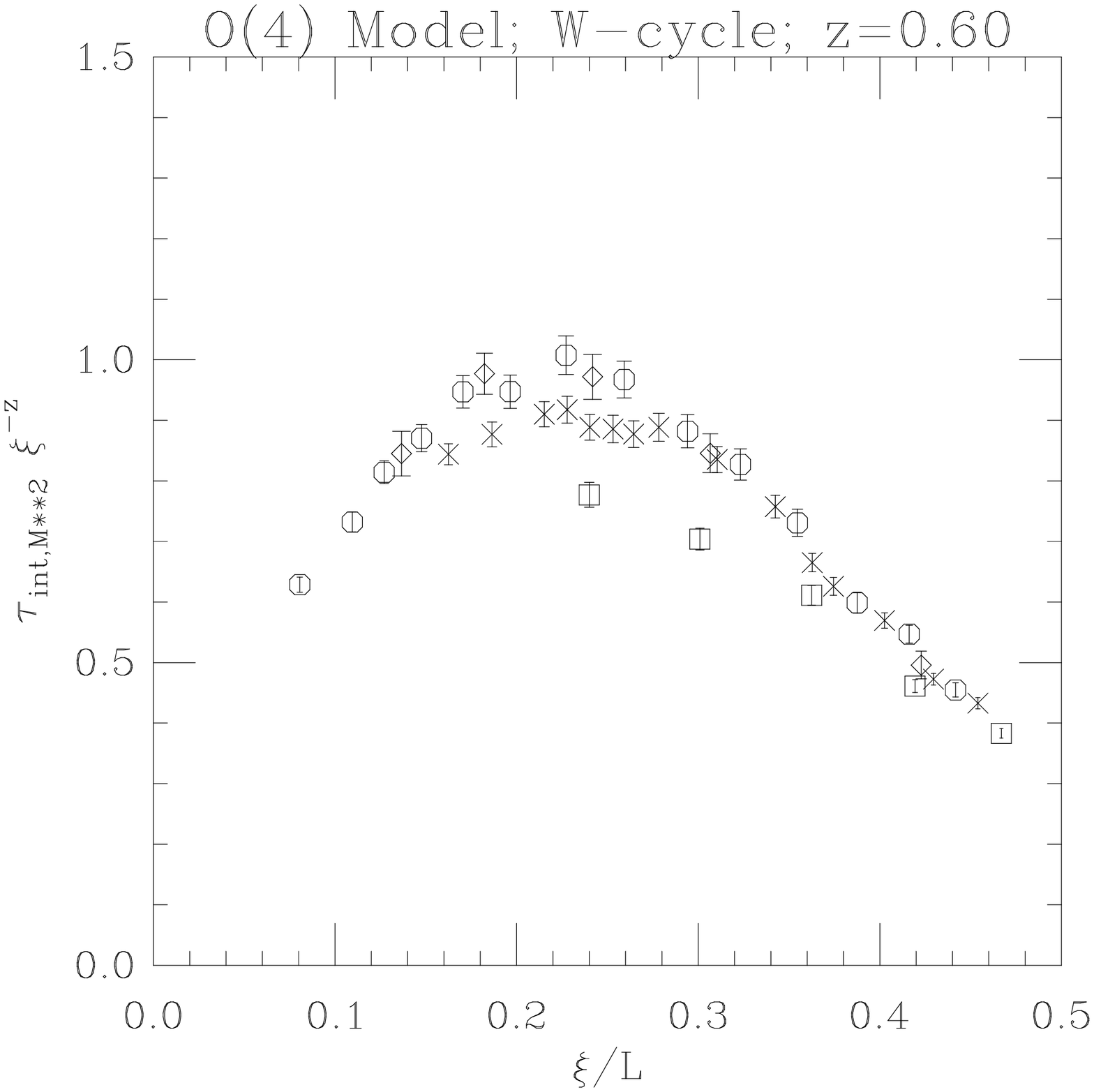}
\caption[fig13]{
   Finite-size-scaling plot of $\taux \xi^{-z_{int,\scrm^2}}$
   versus $\xi/L$ for the W-cycle algorithm,
   for lattice sizes $L=32$ ($\Box$),
   64 ($\times$), 128 ($\bigcirc$) and 256 ($\Diamond$).
   Here $z_{int,\scrm^2}=0.60$.
}
\label{fig_dynamic_fss_M2_W}
\end{figure}

\begin{figure}
\epsfxsize=\textwidth
\epsffile{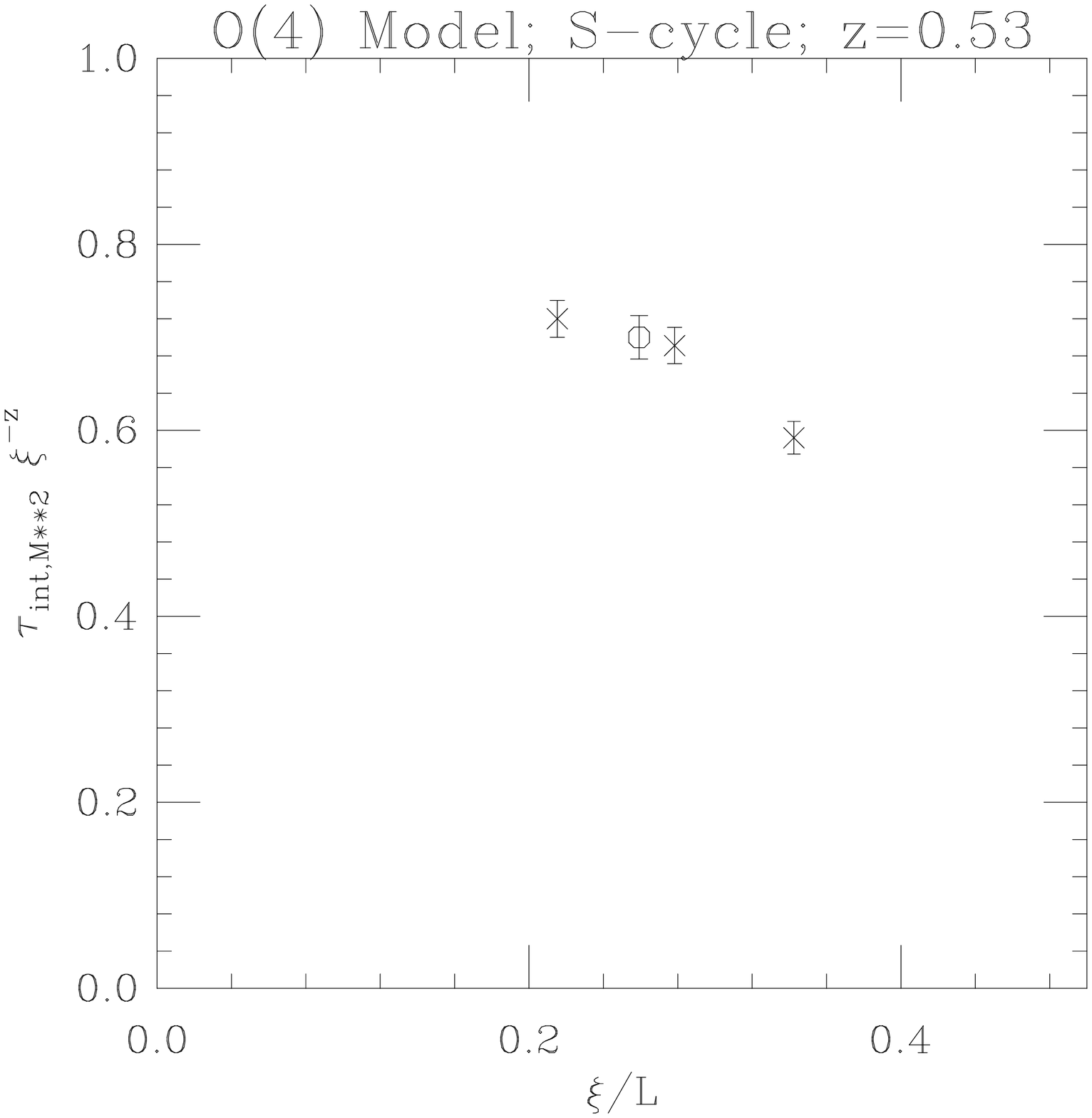}
\caption[fig14]{
   Finite-size-scaling plot of $\taux \xi^{-z_{int,\scrm^2}}$
   versus $\xi/L$ for the super-W-cycle algorithm,
   for lattice sizes $L=64$ ($\times$) and 128 ($\bigcirc$).
   Here $z_{int,\scrm^2}=0.53$.
}
\label{fig_dynamic_fss_M2_S}
\end{figure}

Let us begin with $\taux$.
Using the procedure just described, we find
\be
   z_{int,\scrm^2}   \;=\;
      \cases{ 2.0 \pm 0.15   & for the heat-bath  \cr
              1.13 \pm 0.11  & for the V-cycle    \cr
              0.60 \pm 0.07  & for the W-cycle    \cr
              0.53 \pm 0.10  & for the S-cycle    \cr
            }
 \label{z_estimates}
\ee
(subjective 68\% confidence limits).
In Figures~\ref{fig_dynamic_fss_M2_HB}--\ref{fig_dynamic_fss_M2_S}
we show the ``best'' finite-size-scaling plots for each case.
Note that the corrections to scaling are quite strong:
the $L=32$ points deviate considerably from the asymptotic scaling curve,
and even for $L=64$ the deviation is noticeable
(see Figures \ref{fig_dynamic_fss_M2_V} and \ref{fig_dynamic_fss_M2_W}).
However, it is reasonable to hope that these corrections to scaling
will have decayed to a small value by $L=128$;
if so, our estimates of $z_{int,\scrm^2}$ for the V- and W-cycles ---
which attempt to place the $L=128$ and $L=256$ points on top of one another
--- will be afflicted by only a small systematic error.
In any case, the error bars in \reff{z_estimates} take account of this
potential systematic error.

%
%
%
%
%

%

Note also that the finite-size effects on dynamic quantities
are {\em extremely strong}\/,
even at $\xi/L \ltapprox 0.2$ where the finite-size effects on static
quantities are very weak:
compare Figures~\ref{fig_dynamic_fss_M2_HB}--\ref{fig_dynamic_fss_M2_S}
with Figures~\ref{mgmco4_xi_scaling}--\ref{mgmco4_chi_scaling}.
Indeed, in the heat-bath case (Figure~\ref{fig_dynamic_fss_M2_HB})
it is far from clear that
$g_{\scrm^2}(0) = \lim_{x \downarrow 0} g_{\scrm^2}(x)$
is finite as it should be;
additional data at smaller values of $\xi/L$ would be needed to verify this.
In the W-cycle case it does seem likely that
$g_{\scrm^2}(0)$ is finite and nonzero ($\approx 0.3$?);
but the scaling function $g_{\scrm^2}(x)$ may well vary
by as much as a factor of 2 in the range $x \equiv \xi/L \ltapprox 0.1$.
We conclude that finite-size corrections to dynamic critical behavior
can be surprisingly strong;
therefore, serious studies of dynamic critical phenomena
{\em must}\/ include a finite-size-scaling analysis.
It can be very misleading to assume that the finite-size corrections
to dynamic quantities are small simply because $\xi/L$ is small,
or because the finite-size corrections to {\em static}\/ quantities
are small.

\begin{figure}
\epsfxsize=\textwidth
\epsffile{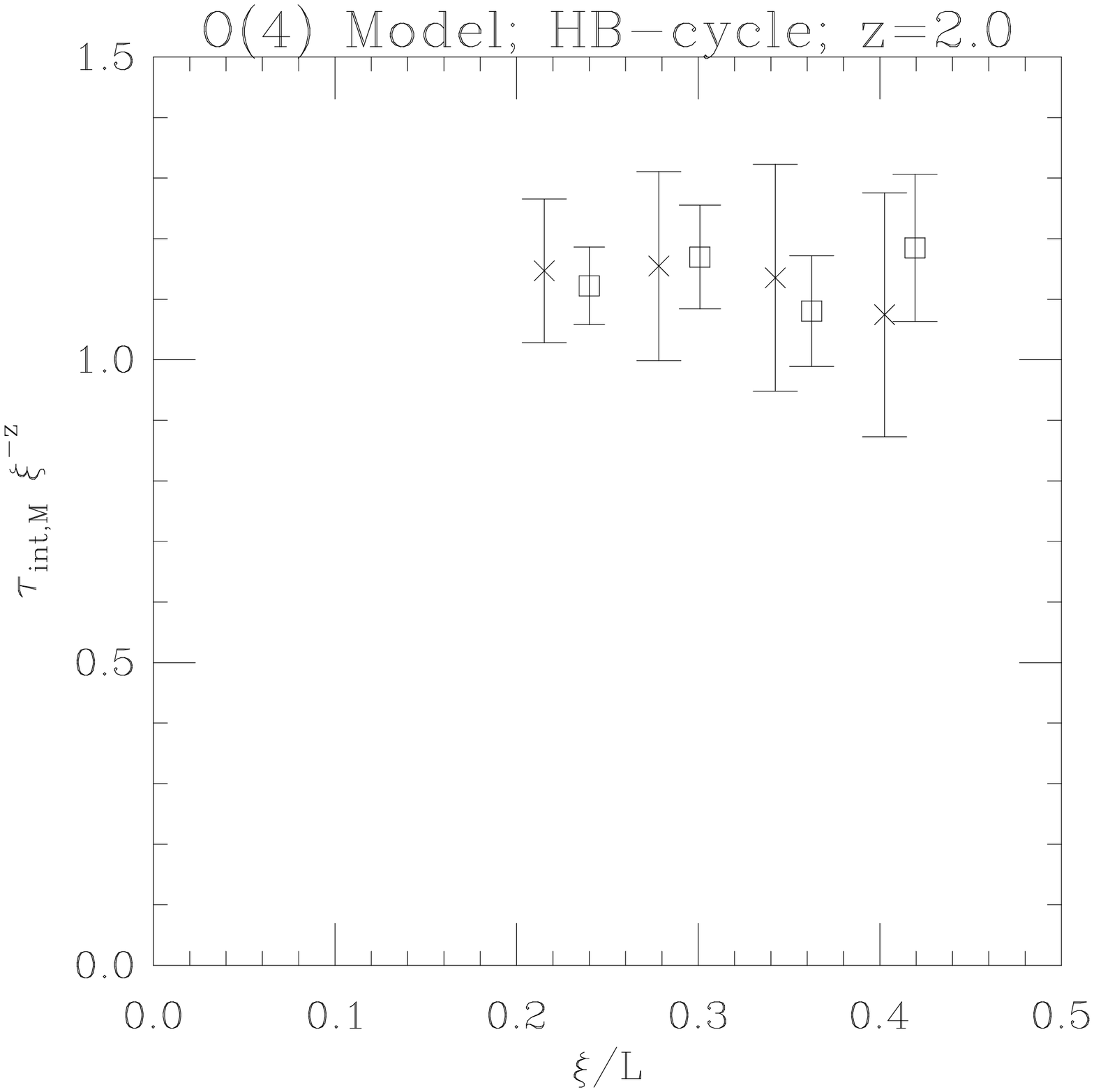}
\caption[fig15]{
   Finite-size-scaling plot of $\taum \xi^{-z_{int,\vec{\cal M}}}$
   versus $\xi/L$ for the heat-bath algorithm,
   for lattice sizes $L=32$ ($\Box$) and 64 ($\times$).
   Here $z_{int,\vec{\cal M}}=2.0$.
}
\label{fig_dynamic_fss_M_HB}
\end{figure}

\begin{figure}
\epsfxsize=\textwidth
\epsffile{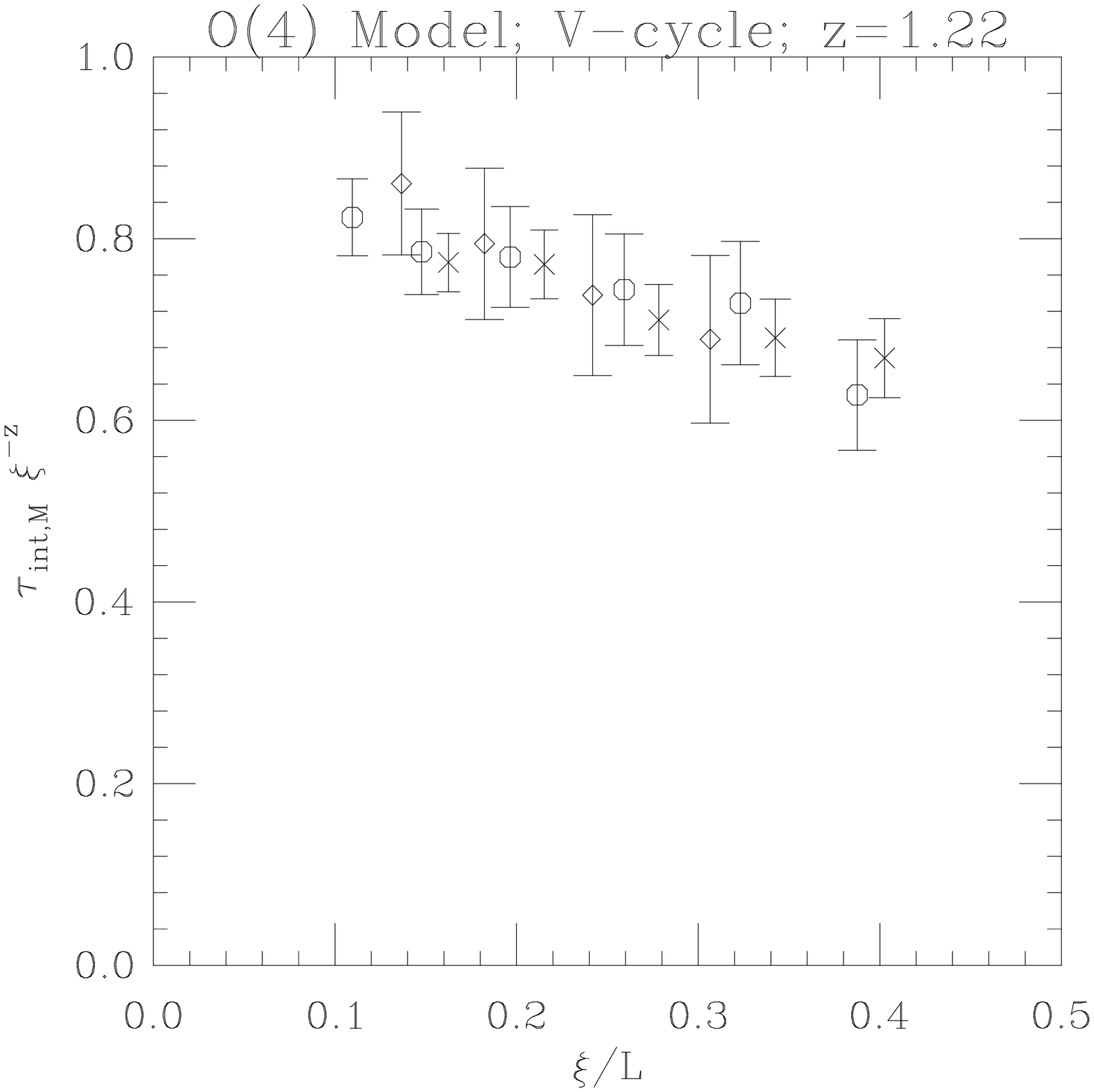}
\caption[fig16]{
   Finite-size-scaling plot of $\taum \xi^{-z_{int,\vec{\cal M}}}$
   versus $\xi/L$ for the V-cycle algorithm,
   for lattice sizes $L=64$ ($\times$), 128 ($\bigcirc$) and 256 ($\Diamond$).
   Here $z_{int,\vec{\cal M}}=1.22$.
}
\label{fig_dynamic_fss_M_V}
\end{figure}

Applying the same procedure to $\taum$, we obtain
\be
   z_{int,\scrmvec}   \;=\;
      \cases{ 2.0 \pm 0.20   & for the heat-bath  \cr
              1.22 \pm 0.09  & for the V-cycle    \cr
            }
\ee
(subjective 68\% confidence limits).
In Figures~\ref{fig_dynamic_fss_M_HB} and \ref{fig_dynamic_fss_M_V}
we show the ``best'' finite-size-scaling plots for each case.
For this observable the finite-size corrections are much weaker.

%
%

%
%

\subsection{Computational Work}
\vspace{-0.3cm}\quad\par

In Table~\ref{mgmco4_cpu_timings}, we show the CPU time per iteration
for our MGMC program running on one processor of a Cray Y-MP 8/832,
as a function of $L=32,64,128,256$ and $\gamma=0,1,2,3$.
The data deviate from the predicted behavior \reff{labor}
for two reasons:
Firstly, the fine-grid Hamiltonian \reff{eqn1b} is simpler than
the coarse-grid Hamiltonians \reff{eq:new_ham},
and we have exploited this by writing special streamlined versions of the
heat-bath and ``compute $H_{l-1}$'' subroutines for the finest grid alone.
Therefore, the CPU time on each grid other than the finest is a factor
$C \approx 2$ greater than the estimate \reff{labor} assumes.
Secondly, the vectorization is more effective on the larger lattices
because the vectors are longer, so the CPU time grows less than
proportionately to the volume, at least through $L=256$.

\clearpage
%
%
\begin{table}
\begin{center}
\begin{tabular}{|c|r|r|r|r|} \hline
\multicolumn{5}{|c|}{CPU timings for $O(4)$ MGMC} \\ \hline
\multicolumn{1}{|c|}{$L$ }
  &\multicolumn{1}{|c|}{$\gamma=0$}
  &\multicolumn{1}{|c|}{$\gamma=1$}
  &\multicolumn{1}{|c|}{$\gamma=2$}
  &\multicolumn{1}{|c|}{$\gamma=3$} \\ \hline
32 &   5.9 & 12.2 & 33.0 & 92.4 \\
64 &  18.7 & 33.8 & 88.9 & 301.2 \\
128 & 66.3 & 109.6 & 256.7 & 986.2 \\
256 & 252.8 & 397.6 & 815.3 & 3305.2 \\
\hline
\end{tabular}
\end{center}
\caption[tab7]{
  CPU times in milliseconds per iteration for MGMC algorithm for
  the two-dimensional $O(4)$ model with $m_1 = m_2 =1$ and $\gamma=0,1,2,3$,
  on a Cray Y-MP 8/832.
}
\label{mgmco4_cpu_timings}
\end{table}

The running speed for our program at $L=256$ was about 100 MFlops.
The total CPU time for the runs reported here was about 650 hours.

The statistical efficiency of a Monte Carlo algorithm
is inversely proportional
to the integrated autocorrelation time for the observable(s) of interest,
{\em measured in CPU units}\/.
For the W-cycle at $L=256$, $\beta=2.6$
(corresponding to $\xi \approx 47$),
we have $\taux \approx 8$ Cray seconds for a W-cycle,
compared to $\approx 43$ seconds for a V-cycle,
and $\approx 280$ seconds for a heat bath.
(The heat-bath value is an extrapolation from
the $L=64$, $\beta=2.2$ point, using finite-size-scaling
together with our result $z_{int,\scrm^2,HB} \approx 2$.)
Thus, on a $256 \times 256$ lattice the W-cycle MGMC algorithm
is about 35 times as efficient as a heat-bath algorithm.

\section{Discussion} \label{section:discussion}
\setcounter{equation}{0}

\subsection{Heuristic Predictions for $z$}   \label{sec4.1}
\vspace{-0.3cm}\quad\par

In paper I \cite[Section IX.C]{Goodman:Multigrid1}
we predicted that for an asymptotically free theory
with a continuous symmetry group,
such as the $O(N)$ $\sigma$-model ($N > 2$),
the MGMC algorithm using piecewise-constant interpolation and a W-cycle
would completely eliminate critical slowing-down
except for a possible logarithm.
Our data in Section \ref{subsection:finite-size-scaling_dynamic}
show clearly that this prediction is {\em incorrect}\/:
the exponent $z_{W-cycle} = 0.60 \pm 0.07$
is manifestly {\em not}\/ equal to zero.
It is obviously of great importance to understand heuristically
what is going on here.

Our argument in \cite{Goodman:Multigrid1} went as follows:
For an asymptotically free theory with a continuous symmetry group,
the important excitations at short wavelengths are weakly-interacting
spin waves, i.e.\ the theory at short wavelengths is {\em almost Gaussian}\/.
Now, we know   
\cite{GS_unpublished}
that MGMC with piecewise-constant interpolation and a W-cycle
{\em completely eliminates}\/ critical slowing-down in a Gaussian model.
Therefore, we expect that MGMC will {\em almost}\/ perfectly handle
the short-wavelength modes of an asymptotically free theory.
On the other hand, we know that the long-wavelength modes of an
asymptotically free theory are strongly non-Gaussian
(e.g.\ there is strong scattering \cite{Zamolodchikov_79}),
and we do not expect MGMC to handle these modes well.
Suppose, therefore,
that the modes of wavelength $\ltapprox \ell_{max}$ are well handled by
MGMC moves on the corresponding grids
(in the sense that the autocorrelation time for these modes is of order 1),
but that coarse-grid moves on scales $\gtapprox \ell_{max}$ are ineffective.
In this case, the correlation length ``seen'' on the grid of scale $\ell_{max}$
will be of order $\xi/\ell_{max}$;
and since further coarse grids are ineffective,
we expect the autocorrelation time of the MGMC algorithm
to be of order $\tau_\MGMC \sim (\xi/\ell_{max})^{z_\HB}$,
where $z_\HB$ is the dynamic critical exponent of the single-site heat-bath
algorithm.

The question is: {\em how big is $\ell_{max}$?}\/
In \cite{Goodman:Multigrid1} we assumed that $\ell_{max} \sim \xi$
(for example, $\ell_{max} \approx \xi/10$) or at worst
$\ell_{max} \sim \xi/\log^p \xi$ --- in which case we would have
$\tau_\MGMC \sim 1$ or $\tau_\MGMC \sim \log^q \xi$, respectively.
But this argument is, in retrospect, much too sloppy.
An asymptotically free theory is indeed ``almost'' Gaussian at short length
scales, but this Gaussianness is approached {\em very slowly}\/:
on scale $\ell$, the coupling strength
(defined e.g.\ by the truncated 4-point function at momentum $p \sim 1/\ell$)
behaves like $g(\ell) \sim 1/\log(\xi/\ell)$.
Now, suppose that this non-Gaussianness causes MGMC
to make small ``mistakes''\footnote{
   By ``mistakes'' we do {\em not}\/ mean to imply a failure to simulate
   the correct Gibbs measure;  of course, by construction, the unique
   invariant measure of the MGMC algorithm is the correct Gibbs measure.
   Rather, we mean a failure to generate {\em independent}\/ samples
   from this Gibbs measure in a time of order 1.
};
and suppose that these ``mistakes'' accumulate (additively)
over successive length scales until they reach order 1,
at which point MGMC becomes ineffective.
The length scales (i.e.\ grids) are indexed by $\log_2 \ell$.
This hand-waving argument suggests that $\ell_{max}$
should be given by the relation
\be
   \int\limits_{1}^{\ell_{max}} {1 \over \log(\xi/\ell)} \, d(\log\ell)
   \;\approx\;  C  \;,
\ee
where $C$ is a constant of order 1.  It follows that
\be
   \ell_{max}   \;\sim\;   \xi^p   \;,
\ee
where
\be
   p   \;=\;  1 - e^{-C}
\ee
lies strictly between 0 and 1 (but we are unable to say more than this).
Hence we predict that $z_\MGMC = p z_\HB$ lies strictly between 0 and $z_\HB$
--- a very weak prediction, but one that seems at least to be true.
 
(We remark that this heuristic argument is very unstable to small
perturbations.  Suppose, for example, that we take instead
$g(\ell) \sim 1/\log^p(\xi/\ell)$ with some exponent $p \neq 1$.
Then if $p < 1$ we get $\ell_{max} \sim O(1)$,
while if $p > 1$ we get $\ell_{max} \sim \xi$.
So $p=1$ is a very delicate borderline case.
Thus, our heuristic argument, if in fact it is correct,
depends crucially on the particular measure of ``non-Gaussianness''
that we have chosen.)

On the other hand, for the V-cycle the observed exponents
$z_{int,\scrm^2,V-cycle} = 1.13 \pm 0.11$
and $z_{int,\scrmvec,V-cycle} = 1.22 \pm 0.09$
are very close to the believed behavior $z_{V-cycle} = 1$
of the Gaussian model
[see the discussion following \reff{eq2.10}].
This suggests that in V-cycle MGMC (with piecewise-constant interpolation)
for an asymptotically free model, the slowest modes are indeed
the Gaussian-like long-wavelength spin waves.
It would be interesting to know whether $z_{V-cycle}$ for the $O(4)$ model
is exactly equal to 1 (possibly modulo a logarithm),
or is slightly higher as our data suggest.

\clearpage
\subsection{Piecewise-Constant vs.\ Smooth Interpolation}   \label{sec4.2}
\vspace{-0.3cm}\quad\par
  
As noted at the end of Section \ref{sec2},
Mack and collaborators
\cite{Mack_87,Mack-Meyer_90,Hasenbusch_LAT90,Hasenbusch-Meyer_91}
have advocated the use of a smooth interpolation
(e.g.\ piecewise-linear or better)
in place of piecewise-constant.
The key question is this:
Does the MGMC algorithm have a {\em different dynamic critical exponent}\/
depending on whether smooth or piecewise-constant interpolation is used?

In the Gaussian case, the answer is ``yes'' for a V-cycle
but ``no'' for a W-cycle, as discussed in Section \ref{sec2}.
For non-Gaussian asymptotically free models, we conjecture that the answer
is the same, but to date no full-scale test has been made.
Hasenbusch, Meyer and Mack \cite{Hasenbusch_LAT90}
have reported preliminary data for the $O(3)$ model suggesting that
$z_{int,\scrm^2} \approx 1.2$ for a V-cycle with piecewise-constant
interpolation, compared to $z_{int,\scrm^2} \approx 0.2$ for a V-cycle with
smooth interpolation.
(These authors have not studied a W-cycle, as this would be extremely costly
in their ``unigrid'' approach, but it is safe to assume that
$z_{W-cycle} \le z_{V-cycle}$.)
For the piecewise-constant V-cycle, this estimate is very close to ours.
But the estimate for the smooth-interpolation V-cycle
(and hence also the smooth-interpolation W-cycle)
is considerably lower than our estimate $z_{int,\scrm^2} \approx 0.6$
for the piecewise-constant W-cycle ---
suggesting that the smooth interpolation might indeed lead to a
smaller dynamic critical exponent than piecewise-constant interpolation,
even for a W-cycle.
However, this estimate is based on runs at only four $(\beta,L)$ pairs,
which moreover have different values of $\xi_\infty/L$,
so it is very difficult to perform a correct finite-size-scaling analysis.
It would be very useful to have data at additional $(\beta,L)$ pairs,
and with higher statistics.

It should also be noted that since the dynamic critical exponent
for MGMC in asymptotically free models is apparently nontrivial
(i.e.\ $z \neq 0,1,2$), it is reasonable to expect this exponent to be
{\em different}\/ for each asymptotically free model:
for example, in the $O(N)$ models the exponent would depend on $N$
--- albeit probably weakly --- just as static exponents in
nontrivial $N$-vector models (such as those in $d=3$) depend on $N$.
For this reason, it would be very useful to have data comparing
the two interpolations at the {\em same}\/ value of $N$.

A very recent preprint by Hasenbusch and Meyer \cite{Hasenbusch-Meyer_91}
gives preliminary data for the $SU(3)$ and $\CP^3$ $\sigma$-models
in two dimensions, using a V-cycle with piecewise-linear interpolation.
For $\CP^3$ the data are consistent with the dynamic finite-size-scaling
Ansatz \reff{dyn_FSS_Ansatz} with $z_{int,\scrm^2} \approx 0.3 \pm 0.1$.
For $SU(3)$, however, the data are very erratic,
and do not appear to be consistent with \reff{dyn_FSS_Ansatz}.
More data would be very useful here too.

\subsection{Comparison with Other Collective-Mode Algorithms}   \label{sec4.3}
\vspace{-0.3cm}\quad\par

It is of great interest to compare the behavior of MGMC
with that of Fourier acceleration
\cite{Parisi_84,Batrouni_85,Dagotto_87:XY,Dagotto_87:SU(3)}.
The two algorithms are very similar in spirit:
both are motivated by the Gaussian model
(for which they eliminate critical slowing-down entirely);
both make additive updates of a fixed-shape collective mode.
The principal difference is the choice of this collective mode:
for Fourier acceleration it is a sine wave,
while for MGMC it is a square wave (or triangular wave
if one uses piecewise-linear interpolation).
Of course, the ``correct'' collective modes in the Gaussian model
are sine waves;  but the MGMC convergence proof 
\cite{GS_unpublished}
shows that square waves are ``close enough'' (for a W-cycle).
There is one additional difference:
Fourier acceleration is based on small-amplitude updates
(i.e.\ a Langevin or hybrid algorithm),
while MGMC is based on arbitrary-amplitude updates
(a heat-bath algorithm).

These close analogies between MGMC and Fourier acceleration suggest
that their behavior
may be qualitatively similar, in the sense that they
work well for the same models and work badly for the same models.
More precisely, we expect that on length scales where
the dominant collective modes are very weakly interacting spin waves,
both MGMC and Fourier acceleration will handle such modes well.
On the other hand, on length scales where the dominant collective modes
are strongly non-Gaussian --- either because the spin waves are strongly
interacting, as in the $\sigma$-model at $\ell \gtapprox \ell_{max}$,
or because these modes are inherently discrete,
as in the one-component $\varphi^4$ model
or the high-temperature phase of the two-dimensional $XY$ model ---
we expect that {\em neither}\/ MGMC nor Fourier acceleration
will work particularly well:  they will gain either a constant factor
in efficiency over the local algorithms,
or perhaps a small reduction in the dynamic critical exponent.
(For some nonlinear models, Fourier acceleration might work significantly
worse than MGMC, if barrier penetration by the small-amplitude updates
becomes a limiting factor.  This difficulty is unrelated
to {\em critical}\/ slowing-down;  for example, it occurs already for
the double-well $\varphi^4$ model with {\em one}\/ lattice site.)

To test this hypothesis, we would like to compare the dynamic critical
behavior of MGMC and Fourier acceleration for the $\sigma$-model.
Unfortunately, there is at present very little published
data on the dynamic critical behavior of Fourier acceleration.
Studies of the two-dimensional $XY$ model
\cite{Batrouni_85,Dagotto_87:XY}
were discussed in paper II \cite{Goodman:Multigrid2}.
The only other published test of Fourier acceleration of which we are aware
is a study of the two-dimensional $SU(3)$ principal chiral model by
Dagotto and Kogut \cite{Dagotto_87:SU(3)}.
This article reported $\taux$ and $\taue$
for Langevin and hybrid algorithms with and without Fourier acceleration,
on lattices of size $16^2$ and $32^2$, for several temperatures in
the scaling region.
The results indicate qualitatively that $z \ll 2$ for the hybrid algorithm
with Fourier acceleration, but the data are unfortunately too crude
to allow quantitative conclusions to be drawn.

The available data are in any case
at least roughly consistent with the conjecture
that, for each model, Fourier acceleration and MGMC have
qualitatively similar, and possibly even identical, dynamic critical exponents.
However, it would be desirable to have more precise and comprehensive
measurements of the dynamic critical behavior of the Fourier-accelerated
Langevin and hybrid algorithms, so as to test this conjecture
of dynamic (quasi-)universality.\footnote{
   We emphasize that a fair test must use the {\em exact}\/ versions
   of the Langevin \cite{exact_Langevin} and hybrid \cite{exact_hybrid}
   algorithms, and should tune the step-size in these algorithms
   (separately for each $\beta$ and $L$)
   so as to optimize the autocorrelation time.
   We emphasize also that Fourier acceleration most likely encompasses
   a {\em continuously infinite family}\/ of dynamic universality classes,
   according to the long-distance (low-momentum) properties of the
   acceleration kernel.
}
It would also be desirable to have precise measurements of the dynamic
critical exponent of MGMC for another asymptotically free model,
such as the $SU(3)$ principal chiral model,
in order to determine whether $z$ varies from one such model to another.

A very different type of collective-mode algorithm for $N$-vector models
was proposed very recently by Wolff \cite{Wolff_89a},
and independently by Hasenbusch \cite{Hasenbusch_90}.
This algorithm is based on embedding a field of Ising variables
into the $N$-vector model, and then simulating the induced Ising model
by the Swendsen-Wang \cite{Swendsen_87} algorithm
(or a variant thereof \cite{Wolff_89a}).
It can be argued heuristically \cite{Edwards_89} that this algorithm
is effective at creating long-wavelength spin waves,
and extensive numerical tests show
that critical slowing-down is entirely (or almost entirely) eliminated
in the two-dimensional $N$-vector models for $N = 2,3,4$
\cite{Wolff_89a,Edwards_89,Wolff_90,CEPS_swwo4c2}.\footnote{
   For $N=2$ this requires that the algorithm also be effective at
   creating vortex-antivortex pairs; the mechanism for this is
   explained in \cite{Edwards_89}.
}

The principal difference between MGMC and Fourier acceleration on the
one hand, and algorithms of Swendsen-Wang and Wolff type on the other,
is how proposals are made for updating the long-wavelength modes.
MGMC and Fourier acceleration propose additive updates
of {\em fixed}\/ shape (sinusoidal, triangular or square)
and variable amplitude, while
cluster algorithms like Swendsen-Wang and Wolff allow the system,
in some sense, to {\em choose its own collective modes}\/.

At present, the Wolff-Swendsen-Wang algorithm is clearly the best
algorithm for simulating $N$-vector models (at least in an unbroken phase):
it is as efficient as MGMC for the two-dimensional $XY$ model
in the spin-wave regime,
it is somewhat more efficient for two-dimensional asymptotically free
$\sigma$-models,
and it is far more efficient for the two-dimensional $XY$ model
in the vortex regime ---
and it is much simpler to program.
On the other hand, the MGMC algorithm has a natural extension
to $\sigma$-models taking values in an arbitrary homogeneous space $G/H$,
and at least in principle to lattice gauge theories
\cite[Sections IV and V]{Goodman:Multigrid1};
but it is as yet far from clear whether an efficient Wolff-type embedding
algorithm can be devised for these models.
Indeed, in a separate work \cite{CEPS_swwo4c2}
we argue that the generalized Wolff-type embedding algorithm
can have dynamic critical exponent $z \ll 2$
{\em only}\/ if the target manifold is a sphere (as in the $N$-vector model),
a product of spheres,
or the quotient of such a space by a discrete group
(e.g.\ real projective space $RP^{N-1}$).
Finally, Fourier acceleration applies to all continuous-spin models,
even in the presence of dynamical fermions
(which is its greatest advantage);
but the restriction to small step size is a disadvantage,
and the need to optimize several tunable parameters
makes it more complicated to test and to use.
We therefore believe that all three classes of collective-mode algorithms
--- Fourier acceleration, MGMC and embedding algorithms ---
should be developed in parallel, in order to determine the conditions
under which each one works well or badly.
In particular, we advocate that a standard test procedure be developed
(which autocorrelation times to measure, what statistical procedures to use)
so that test results obtained by diverse research groups
can be readily compared.

\section*{Acknowledgments}

We want to thank Achi Brandt, Richard Brower, Martin Hasenbusch,
Steffen Meyer, Claudio Parrinello and Ulli Wolff
for several helpful suggestions.
The computations reported here were carried out
on the Cray Y-MP at the Pittsburgh Supercomputing Center (PSC).
This work was supported in part 
by the U.S.\ National Science Foundation grants DMS-8705599 and DMS-8911273
(J.G. and A.D.S.),
by the U.S.\ Department of Energy through contracts
DE-FC05-85ER250000 (R.G.E.) and DE-FG02-90ER40581 (A.D.S.),
and by PSC grant PHY890025P.
One of the authors (S.J.F.) was supported by the Conselho Nacional de Pesquisas
(Brazil),
and another of the authors (J.G.) was partially supported by a Sloan Foundation 
Fellowship.

\clearpage

\appendix
\section{Heat-Bath Algorithm for the $N$-Vector Model} \label{appendix_A}
\setcounter{equation}{0}
\vspace{-0.3cm}\quad\par

In implementing the heat-bath step of the MGMC algorithm for
the $SU(2)$ principal chiral model,
we need a subroutine that generates a random matrix $U \in SU(2)$
from the probability distribution
\be
   d\mu(U)   \;=\;   {\rm const} \times \exp(\tr M^\dagger U) \, dU   \;,
\ee
where $M$ is a fixed matrix lying in the linear span of $SU(2)$,
and $dU$ is Haar measure.  Since $M = \lambda V$ with
$\lambda = (\det M)^{1/2} \ge 0$ and $V \in SU(2)$,
and $U$ can be parametrized in the form \reff{su2},
it clearly suffices to generate a random unit vector
$\bsigma = (\sigma^0,\sigma^1,\sigma^2,\sigma^3) \in \R^4$
from the probability distribution
\be
   d\nu(\bsigma)   \;=\;   {\rm const} \times \exp(h\sigma^0) \, d\bsigma   \;,
\ee
where $d\bsigma$ is uniform measure on the unit sphere in $\R^4$
(here $h = 2\lambda$).
More generally, we can pose this latter problem for unit vectors in $\R^N$
($N \ge 2$); this problem arises also in implementing the heat-bath algorithm
for an $N$-vector model.  The problem then divides naturally into two parts:
\begin{itemize}
  \item[(a)]  Generate the component $t \equiv \sigma^0$
      with probability density proportional to
\be
   f(t)   \;=\;   e^{ht} \, (1-t^2)^{(N-3)/2} \, dt
 \label{eqA.3}
\ee
      restricted to the interval $[-1,1]$.  We can assume without loss of
      generality that $h \ge 0$.
  \item[(b)]  Generate $\bsigma^\perp \equiv (\sigma^1,\ldots,\sigma^{N-1})$
     to be a random vector on the sphere of radius $(1-t^2)^{1/2}$ in
     $\R^{N-1}$.
\end{itemize}
The principal purpose of this Appendix is to discuss three methods for
solving problem (a).\footnote{
  For the special case $N=2$, one additional method (rejection from a Gaussian
  in the angular variables) was discussed
  in the Appendix of \cite{Goodman:Multigrid2},
  and several different methods are discussed in \cite{Best-Fisher}.
  See also \cite[pp.\ 473--476]{Devroye_86}.
}
At the end we shall make some brief remarks about the much easier problem (b).

All of our methods are based on the von Neumann rejection technique
\cite{Knuth_81,Devroye_86}:  given a function $g(t) \ge f(t)$,
we generate a random variable $T$ with density
proportional to $g(t)$ and then accept it with probability $f(T)/g(T)$;
we keep trying until success.  The acceptance probability is 
\be
A   \;=\;   {{\int_{-\infty}^{\infty} f(t)\, dt}\over
             {\int_{-\infty}^{\infty} g(t)\,dt}}    \;,
\ee
and the expected number of trials is $A^{-1}$. 
Obviously, $g$ must be simple enough that we can conveniently generate
a random variable with density proportional to $g$;
but subject to this constraint we want $g$ to be as similar to $f$ as possible.
Our three methods are based on different choices of the trial distribution $g$:

\medskip

{\bf Method 1} (valid only for $N \ge 3$).
$g(t) = e^{ht}$ restricted to $[-1,1]$.
Such a random variable can be generated by transformation:
\be
   T  \;=\;   h^{-1} \log[e^{-h} + (e^h - e^{-h})U]   \;,
\ee
where $U$ is uniform on $[0,1]$.
The proposal is accepted with probability $(1-t^2)^{(N-3)/2}$.
The acceptance probability is
\begin{eqnarray}
   A   & = &   2^{(N-2)/2} \, \sqrt{\pi} \, \Gamma\!\left( {N-1 \over 2} \right)
                 \, h^{2-(N/2)} \, (e^h - e^{-h})^{-1} \, I_{(N-2)/2}(h)
                                                                   \\[0.2cm]
 & = &   \cases{ {\sqrt{\pi} \, \Gamma\!\left( {N-1\over 2}\right)
                  \over
                  2 \Gamma\!\left( {N \over 2}\right)
                 }
                 \left[ 1 \,-\, {N-3 \over 6N} h^2  \,+\, O(h^4) \right]
                                                          &  as $h \to 0$   \cr
               \noalign{\vskip 4pt}
                 2^{(N-3)/2} \, \Gamma\!\left( {N-1\over 2}\right) \,
                   h^{-(N-3)/2} \, \Bigl[ 1 + O(h^{-1}) \Bigr] 
                                                       &  as $h \to \infty$ \cr
               }
\end{eqnarray}
By construction this algorithm works perfectly for $N=3$.
For $N > 3$ it works decently but not perfectly for small $h$
(and deteriorates there when $N$ is very large),
and badly at large $h$.
For the case $N=4$, this method was proposed previously by Creutz
\cite{Creutz_80}.
   
\medskip

{\bf Method 2} (valid only for $N \ge 3$).
$g(t) = e^{ht} [2(1-t)]^{(N-3)/2}$ restricted to $(-\infty,1]$.
This is expressed more conveniently by making the change of variables
$s = 1-t$.  Then the desired distribution is
\be
   \widetilde{f}(s)   \;=\;   e^{-hs} \, s^{(N-3)/2} \, (2-s)^{(N-3)/2}
\ee
restricted to $[0,2]$, the trial distribution is
\be 
   \widetilde{g}(s)   \;=\;   e^{-hs} \, (2s)^{(N-3)/2}
\ee 
on $[0,\infty)$,
and the proposal is accepted with probability
\be
   {\widetilde{f}(s)  \over  \widetilde{g}(s)}   \;=\;
   \cases{ \left(1 - {s \over 2} \right) ^{(N-3)/2}   &  if $0 \le s \le 2$ \cr
           \noalign{\vskip 4pt}
           0   &  if $s > 2$   \cr
         }
\ee
The trial $S$ is a Gamma$\left( {N-1 \over 2}, h^{-1} \right)$
random variable;  for $N$ a positive integer,
such a random variable can be generated by combining exponential
and Gaussian random variables \cite[p.\ 405]{Devroye_86},
and this algorithm is efficient if $N$ is not too large.
The acceptance probability is Method 2 is
\begin{eqnarray}
   A   & = &   (2\pi h)^{1/2} \, e^{-h} \, I_{(N-2)/2}(h)        \\[0.2cm]
       & = &   \cases{ {2 \sqrt{\pi} \over \Gamma(N/2)} \,
                          \left( {h \over 2} \right) ^{(N-1)/2} \,
                          \Bigl[ 1 + O(h) \Bigr]       &  as $h \to 0$   \cr
                      \noalign{\vskip 4pt}
                       1 \,-\, {(N-3)(N-1) \over 8h} \,+\, O(h^{-2})
                                                       &  as $h \to \infty$ \cr
                     }
\end{eqnarray}
where $I_\nu$ is a modified Bessel function.
Method 2 is extremely bad for small $h$ (especially for large $N$),
but it is quite efficient for large $h$.
For the case $N=4$, this method was proposed previously by several authors
\cite{Fabricius_84,Kennedy_85,DeForcrand_86}.

\medskip

{\bf Method 3.}  $g(t) = e^h (1-t^2)^{(N-3)/2}$ restricted to $[-1,1]$.
This is a Beta$\left( {N-1 \over 2}, {N-1 \over 2} \right)$
distribution, which can be generated from a pair of
Gamma$\left( {N-1 \over 2} \right)$ variates or in many other ways
\cite[pp.\ 428--445]{Devroye_86}.
The proposal is accepted with probability $e^{h(t-1)}$.
The acceptance probability is
\begin{eqnarray}
   A   & = &   \Gamma\!\left( {N \over 2}\right) \, (h/2)^{-(N-2)/2} \,
               e^{-h}  \,  I_{(N-2)/2}(h)                        \\[0.2cm]
 & = &   \cases{ 1 \,-\, h  \,+\, O(h^2)               &  as $h \to 0$   \cr
               \noalign{\vskip 4pt}
                 2^{(N-3)/2} \, \pi^{-1/2} \,
                 \Gamma\!\left( {N\over 2}\right) \,
                 h^{-(N-1)/2} \, \Bigl[ 1 + O(h^{-1}) \Bigr]  
                                                       &  as $h \to \infty$ \cr
               }                                   
\end{eqnarray} 
This works excellently for small $h$, and badly for large $h$.
For the case $N=2$, this method was studied in \cite{Goodman:Multigrid2}:
the trial variable $T$ is produced by
generating a uniform angle $\theta \in [0,2\pi]$ and setting $T = \cos\theta$.

\medskip

%

%
%
\begin{table}
\begin{center}
\begin{tabular}{|r|c|c|c|} \hline
\multicolumn{1}{|c|}{\ }  & \multicolumn{3}{|c|}{Acceptance Probability}   \\
\cline{2-4}
\multicolumn{1}{|c|}{$h$}  &  \multicolumn{1}{|c|}{Method 1} &
   \multicolumn{1}{|c|}{Method 2}  &  \multicolumn{1}{|c|}{Method 3}   \\
\hline
0.0000  &  0.7854  &  0.0000  &  1.0000  \\
0.1000  &  0.7851  &  0.0359  &  0.9060  \\
0.2000  &  0.7841  &  0.0922  &  0.8228  \\
0.2521  &  0.7833  &  0.1243  &  0.7834  \\
0.3000  &  0.7825  &  0.1543  &  0.7492  \\
0.4000  &  0.7802  &  0.2168  &  0.6838  \\
0.5000  &  0.7774  &  0.2772  &  0.6257  \\
0.6000  &  0.7740  &  0.3343  &  0.5739  \\
0.7000  &  0.7700  &  0.3873  &  0.5276  \\
0.8000  &  0.7656  &  0.4361  &  0.4862  \\
0.8603  &  0.7627  &  0.4634  &  0.4634  \\
0.9000  &  0.7607  &  0.4806  &  0.4491  \\
1.0000  &  0.7554  &  0.5212  &  0.4158  \\
1.5000  &  0.7242  &  0.6724  &  0.2921  \\
1.6847  &  0.7114  &  0.7114  &  0.2596  \\
2.0000  &  0.6889  &  0.7631  &  0.2153  \\
3.0000  &  0.6199  &  0.8545  &  0.1312  \\
4.0000  &  0.5618  &  0.8961  &  0.0894  \\
5.0000  &  0.5152  &  0.9191  &  0.0656  \\
10.0000  &  0.3810  &  0.9612  &  0.0243  \\
20.0000  &  0.2749  &  0.9809  &  0.0088  \\
30.0000  &  0.2259  &  0.9874  &  0.0048  \\
40.0000  &  0.1963  &  0.9906  &  0.0031  \\
50.0000  &  0.1759  &  0.9925  &  0.0022  \\
100.0000  &  0.1249  &  0.9962  &  0.0008  \\
200.0000  &  0.0885  &  0.9981  &  0.0003  \\
500.0000  &  0.0560  &  0.9992  &  0.0001  \\
\multicolumn{1}{|c|}{$\infty$} &    0.0000  &  1.0000  &  0.0000  \\
\hline
\end{tabular}
\caption[tab8]{
   Acceptance probability for generating a random variable from
   (\protect\ref{eqA.3}) with $N=4$,
   using rejection from $e^{ht}$ (method 1),
   rejection from gamma (method 2)
   or rejection from beta (method 3).
}
\label{accprob_rvo4}
\end{center}
\end{table}

In Table \ref{accprob_rvo4} we show the acceptance probability
as a function of $h$ for each of the three methods, for the case $N=4$.
(However, the relevant issue in practice
is not acceptance probability {\em per se\/}, but CPU time;
the three methods have slightly different CPU time per trial.)
Clearly it is necessary to use a combination of methods,
in order to obtain a generator whose acceptance probability is
bounded away from zero uniformly in $h$.\footnote{
   In a Monte Carlo application, the $h$ are themselves random variables,
   and the CPU time is proportional to $\< A(h)^{-1} \>$
   ({\em not}\/ $\< A(h) \> ^{-1}$, as we once learned the hard way!).
   It is thus important to use a generator in which $A(h)$ stays away from
   zero, {\em even for ``rare'' values of $h$}\/.
   This is especially important in the MGMC application,
   in which an extremely wide range of values of $h$ are observed
   on the different grids.
}
In our Monte Carlo studies of the $O(4)$ $\sigma$-model,
we used Method 1 when $h \le$ some $h_{\hbox{\scriptsize\em cutoff}}$,
and Method 2 when $h > h_{\hbox{\scriptsize\em cutoff}}$.
We vectorized the algorithm as follows:
Gather all sites with $h \le h_{\hbox{\scriptsize\em cutoff}}$,
and compress them into a single vector;  make one trial of Method 1;
scatter the ``successful'' $T$ values;
gather and recompress the ``failures'';
and repeat until all sites are successful.
Then do the same for sites with $h > h_{\hbox{\scriptsize\em cutoff}}$,
using Method 2.
Empirically we found that for runs in the physically interesting range
of $\beta$ values (i.e.\ $\beta \approx 2$),
the CPU time was almost completely insensitive to
 $h_{\hbox{\scriptsize\em cutoff}}$
in the range $0.5 \ltapprox h_{\hbox{\scriptsize\em cutoff}} \ltapprox 5$.
We used $h_{\hbox{\scriptsize\em cutoff}} = 1.68474$
in our production runs.

\medskip

Finally, let us make some remarks about the problem of generating
a random vector $(\sigma^1,\ldots,\sigma^{N-1})$
on the unit sphere of $\R^{N-1}$.
For $N=3$ this is easy: generate a uniform random angle $\theta \in [0,2\pi]$
and take the cosine and sine.
For $N=4$ it is also easy:  generate a uniform random variable
$\sigma^1 \in [-1,1]$, and then generate a random vector $(\sigma^2,\sigma^3)$
on the circle of radius $[1- (\sigma^1)^2]^{1/2}$ as just described.
(This is the method we employed.)
For $N>4$ one can generate first $\sigma^1$ from a
Beta$\left( {N-2 \over 2}, {N-2 \over 2} \right)$ distribution,
then $\sigma^2$ from a Beta$\left( {N-3 \over 2}, {N-3 \over 2} \right)$
distribution scaled by $[1- (\sigma^1)^2]^{1/2}$, and so forth.
Numerous other methods are described in \cite[Section V.4]{Devroye_86}.


\end{document}